\documentclass[reprint,longbibliography,aps,prd,superscriptaddress,amsmath,amssymb,amsfonts,nobibnotes,floatfix]{revtex4-2}

\usepackage{breqn}
\usepackage{mathrsfs}
\usepackage{ytableau}
\usepackage{tikz-cd}
\usepackage{hyperref}
\hypersetup{colorlinks=true,allcolors=blue,linktocpage}

\newtheorem{definition}{Definition}

\newcommand{\diag}{\operatorname{diag}}

\newcommand{\tr}{\mathop{\mathrm{tr}}}
\newcommand{\we}{\mathop{\scriptstyle\wedge}}

\newcommand\UTFSM{Departamento de F\'isica, Universidad T\'{e}cnica Federico Santa Mar\'\i a\\ Casilla 110-V, Valpara\'iso, Chile}

\newcommand\CCTVal{Centro Cient\'ifico Tecnol\'ogico de Valpara\'iso\\ Casilla 110-V, Valpara\'\i so, Chile}
\newcommand\PUCV{Instituto de F\'isica, Pontificia Universidad Cat\'olica de Valpara\'iso\\ Casilla 4059, Valpara\'iso, Chile}
\newcommand{\UdelaR}{Instituto de F\'isica, Facultad de Ciencias\\Igu\'a 4225, esq. Mataojo, 11400 Montevideo, Uruguay.}
\newcommand{\espol}{Departamento de Ciencias F\'isicas, Facultad de Ciencias Naturales y Matem\'aticas\\Escuela Superior Polit\'ecnica del Litoral, km 30,5 Vía Perimetral, Guayaquil, Ecuador.}

\begin{document}
\title{A polynomial affine model of gravity: after ten years}
\author{Oscar \surname{Castillo-Felisola}}
\affiliation{\UTFSM}
\affiliation{\CCTVal}
\email{o.castillo.felisola:at:proton.me}
\author{Bastian \surname{Grez}}
\affiliation{\UTFSM}
\affiliation{\PUCV}
\author{Manuel \surname{Morocho-L\'opez}}
\affiliation{\espol}
\author{Jose \surname{Perdiguero}}
\noaffiliation
\author{Aureliano \surname{Skirzewski}}
\affiliation{\UdelaR}
\author{Jefferson \surname{Vaca}}
\affiliation{\UTFSM}
\affiliation{\PUCV}
\author{Nicolas \surname{Zambra-G\'omez}}
\affiliation{\UTFSM}
\affiliation{\PUCV}

\begin{abstract}
  The polynomial affine model of gravity was proposed as an alternative to metric and metric-affine gravitational models. What at the beginning was thought as a source of unpredictability, the presence of many terms in the action, turned out to be a milestone, since it contains all possible combinations of the fields compatible with the covariance under diffeomorphisms. Here, we present a review of the advances in the analysis of the model after ten years of its proposal, and sketch the guideline of our future perspectives.
\end{abstract}

\maketitle

\section{Introduction\label{sec:intro}}

General relativity was the model proposed by Einstein in response to the need to make the constancy of the velocity of light and the Newtonian theory of gravity compatible \cite{einstein15_zur_allgem_relat}. The equations of gravitational interaction, described as a field theory for the metric tensor field, were presented almost simultaneously by Einstein and Hilbert \cite{hilbert15_grund_physik}. However, the latter obtained the field equations based on an optimisation problem, aligned with the Lagrangian formalism of classical mechanics. Within the following months the first (non-trivial) exact solution to these equations was obtained by Schwarzschild \cite{schwarzschild16_on_the_gravit_fiel}, but also the first phenomenological predictions of the model were derived: (i) Perihelion precession of Mercury's orbit; (ii) Deflection of light by the Sun, and; (iii) The gravitational redshift of light. A modern comparison between the predictions of General Relativity and experimental observations can be found in Refs. \cite{will14_confr_between_gener_relat,will18_theor_exper_gravit_physic}. The amount of evidence supporting the validity of General Relativity is vast; however, nowadays it is presumed that Einstein's theory is an effective model, mostly due to the unsuccessful attempts to renormalise and quantise the model \cite{arnowitt59_dynam_struc_defin_energ_gener_relat,arnowitt60_canon_variab_gener,wheeler64_relat,dewitt67_quant_theor_gravit_i,dewitt67_quant_theor_gravit_ii,dewitt67_quant_theor_gravit_iii,hooft73_algor_poles_at_dimen_four,hooft74_one_loop_diver_theor,hooft76_comput_quant_effec_due_to,hooft76_symmet_break_throug_bell_jackiw_anomal,hooft78_errat}, and the necessity of hypothesising the existence of an extensive dark sector to commensurate the cosmological observations with the predictions from the model \cite{papantonopoulos07_invis_univer,weinberg08_cosmol,dodelson20_moder}. Such a belief encourages the inquiry of models of gravity that extend General Relativity.

Einstein formulated the theory of General Relativity under the following precepts: the fundamental field is the metric tensor field, the theory is covariant under the group of diffeomorphism, and the field equations are second-order differential equations. The \emph{uniqueness} of General Relativity was shown by Lovelock \cite{lovelock69_uniquen_einst_field_equat_four_dimen_space}, and in addition he proposed gravitational models in diverse dimensions that satisfy the same axioms \cite{lovelock71_einst_tensor_its_gener}.\footnote{Lanczos--Lovelock models of gravity might be inspired by Sakharov's proposal that General Relativity might be an effective model that receives higher curvature corrections \cite{sakharov00_vacuum_quant_fluct_curved}.} These models are known as Lanczos--Lovelock models of gravity. In order to build extensions of General Relativity one needs to relax the axioms stated above.

One of the most interesting formulations of gravitational models comes with the independence of the connection from the metric tensor. These are dubbed metric-affine models of gravity (for a review, see Ref. \cite{hehl95_metric_affin_gauge_theor_gravit}). The first metric-affine models of gravity were proposed by Weyl \cite{weyl21_zur_infin,weyl22_space} and Cartan \cite{cartan22_sur_une_de_la_notion,cartan23_sur_les_connex_affin_et,cartan24_sur_les_connex_affin_et,cartan25_sur_les_connex_affin_et}, who consider connections with nonmetricity and torsion, respectively. There were other formulations of gravity, for example, based on projective transformations \cite{veblen33_projec_gener_relat}, or affine formulations \cite{einstein23_zur_affin_feldt,einstein23_theor_affin_field,eddington23,schroedinger46_general_affin_field_laws,schroedinger47_final_affin_field_laws_i,schroedinger48_final_affin_field_laws_ii,schroedinger48_final_affin_field_laws_iii}, other formulations consider non-symmetric metrics \cite{hlavaty57_geomet_einst,tonnelat14_einst}, and models with extra dimensions \cite{kaluza21_probl_unity_physic,klein26_quant_theor_five_dimen_theor,einstein38_gener_kaluz_theor_elect}. Extensions that consider additional fields involved in the mediation of gravitational interaction are known as TeVeS (Tensor-Vector-Scalar) gravities. For contemporary presentations of different modifications of gravity, see Refs. \cite{clifton12_modif_gravit_cosmol,saridakis21_modif_gravit_cosmol}.

In the past decades the interest in different formulations of General Relativity and its extensions has increased, with the aspiration that some model could remediate the remaining issues of General Relativity. The teleparallel and symmetric teleparallel equivalent formulations of General Relativity (see, for example, Refs. \cite{aldrovandi13_telep_gravit,bahamonde21_telep_gravit,jimenez19_geomet_trinit_gravit}) have been fertile soil for building extended gravitational models,\footnote{In words of K. Krasnov \cite{krasnov20_formul_gener_relat},
  \begin{quote}
    There may be equivalent formulations of a theory, all leading to the same physical predictions. But such reformulations may be inequivalent if one decides to generalise.
  \end{quote}} originating a huge amount of theories \cite{sotiriou10_r_theor_of_gravit,olmo11_palat_approac_to_modif_gravit,de10_r,cai16_t_telep_gravit_cosmol,heisenberg23_review_q_gravit,jimenez20_cosmol_q_geomet,myrzakulov12_dark_energ_f_r,harko11_r_t,nakayama22_geomet_trinit_unimod_gravit}.

In the same spirit, there are modern proposals of affine gravity that extend the Einstein--Eddington--Schrödinger model. This renewed interest in affine models of gravity started with the foundational work of J. Kijiowsky \cite{kijowski78_new_variat_princ_gener_relat,kijowski79,ferraris81_gener_relat_is_gauge_type_theor,ferraris82_equiv_relat_theor_gravit,kijowski07_univer_affin_formul_gener_relat}, followed by reformulations of General Relativity by Krasnov \cite{krasnov07_non_metric_gravit,krasnov08_non_metric_gravit_i,krasnov08_non_metric_gravit_ii,krasnov11_pure_connec_action_princ_gener_relat,delfino15_pure_connec_formal_gravit_lin,delfino15_pure_connec_formal_gravit_feyn}, the affine proposal by Poplawski \cite{poplawski07_unified_purel_affin_theor_gravit_elect,poplawski07_nonsy_purel_affin,poplawski14_affin_theor_gravit}, and the polynomial affine model of gravity (target of this review) \cite{castillo-felisola15_polyn_model_purel_affin_gravit,castillo-felisola18_einst_gravit_from_polyn_affin_model}. The cosmological solutions and some phenomenological aspects of these models have been examined in Refs. \cite{poplawski10_cosmol_with_torsion_alter_to_cosmic,azri17_affin_inflat,azri18_cosmol_implic_affin_gravit,azri18_induc_affin_inflat,castillo-felisola18_cosmol,castillo-felisola20_emerg_metric_geodes_analy_cosmol,castillo-felisola24_cosmol_solut_polyn_affin}.

The affine formulations of gravity, due to the lack of a fundamental metric tensor field, do not have the flexibility of building invariant quantities offered in metric models of gravity. We call this property the \emph{rigidity} of affine gravity. Moreover, the affine connection, which in general has sixty-four components in four dimensions (much more than the ten components of the metric tensor field) has more room to accommodate additional fields, associated with the geometrical nature of the manifold, which could be interpreted as additional degrees of freedom to explain the dark sector of the Universe.

Invariants built with the affine connecting do not refer to any length scale, and therefore the group of symmetries could be naturally enhanced to the group of conformal or projective transformations. Some applications of these groups in gravitational models are found in Refs. \cite{aalbers13_confor,bailey94_thomas_struc_bundl_confor_projec_relat_struc,borthwick21_projec_differ_geomet_asymp_analy_gener_relat}.

The aim of this article is to present the state-of-the-art with regard to the development of the polynomial affine model of gravity in four dimensions. The Sec. \ref{sec:model} gives a brief overview of the polynomial affine model of gravity, highlighting the method used to obtain the action of the model [see Eq. \eqref{eq:new-action}] and listing the remarkable features of the model. Next, in Sec. \ref{sec:cov-feqs} (complemented with the content of Appendix \ref{sec:appen-a}) we find the covariant field equations of the model. We have included in Sec. \ref{sec:tor-less-sect} the field equations in the particular scenario of the torsion-less sector. In order for searching solutions to the field equations, as in any other gravitational model, we need the ansatz of the affine connection. We build the ansatz of a spherical and cosmological connection in Sec. \ref{sec:conn-ansatz}. In Sec. \ref{sec:cosm-sol}, we scan the space of solutions to the field equations of the model in the cosmological context, and we present a model to analyse cosmological perturbations in affine models of gravity in Sec. \ref{sec:coms-pert}. Issues regarding the perspectives of the model are discussed in Sec. \ref{sec:perspectives}. The content of that section is a road map of our current research interest and the reach of the model. We end the article with a brief set of concluding remarks in Sec. \ref{sec:concl}.

\section{Purely polynomial affine gravity\label{sec:model}}

Polynomial affine gravity is a model of gravitational interactions whose fundamental field is the affine connection, and does not require the existence of a (fundamental) metric tensor field to build its action functional, with the requirement of covariance under the group of diffeomorphism. In order to define an action functional on a four-dimensional manifold \(\mathcal{M}\), we write a linear combination of all possible \(4\)-forms that can be made that are linearly independent. Choosing a coordinate system \(\{x^{\mu}\}\), there is an induced basis on the tangent and cotangent bundles, \(\partial_{\mu}\big|_{x}\) and \(\mathrm{d}x^{\mu}\big|_{x}\) respectively, and on other tensor bundles. Using this notation, \(4\)-forms integrals can be written in components as follows \cite{synge78_tensor}
\begin{dmath}
  \int_{\mathcal{M}} F_{(4)}
  = \int_{\mathcal{M}} \frac{1}{4!} F_{\mu_1 \cdots \mu_4} \, \mathrm{d}x^{\mu_1} \we \cdots \we \mathrm{d}x^{\mu_4}
  = \int_{\mathcal{M}} \frac{1}{4!} F_{\mu_1 \cdots \mu_4} \, \mathfrak{E}^{\mu_1 \cdots \mu_4} \, \mathrm{d}^4 x.
  \label{eq:integral-4-form}
\end{dmath}
defined with
\begin{align}
  \mathfrak{E}^{\mu_1 \cdots \mu_4} & =
                  \begin{cases}
                    1 & \mu_1 \cdots \mu_4 \text{ an even permutation of } 0 1 2 3 \\
                    -1 & \mu_1 \cdots \mu_4 \text{ an odd permutation of } 0 1 2 3 \\
                    0 & \text{otherwise,}
                  \end{cases}
                  \label{eq:canonical-density-up}
\end{align}
the skew-symmetric tensor density of weight \(w = +1\), invariant under coordinate transformations \cite{schouten13_ricci}. Note that the term \(\mathrm{d}^4 x\) is independent of the fields, but has the role of an integration measure and has weight \(w = -1\).

The affine structure is an additional structure that could be \emph{added} to the differential manifold, and its purpose is to allow the comparison of geometric objects placed at different points of the manifold \(\mathcal{M}\). Such structure is determined by the affine connection, whose components are defined as
\begin{equation}
  \tilde{\nabla}_{\mu} \vec{e}_{\nu} = \tilde{\Gamma}_{\mu}{}^{\lambda}{}_{\nu} \vec{e}_{\lambda},
\end{equation}
where the \(\vec{e}_{\mu}\) are the basis vectors at a given point \(p \in \mathcal{M}\). Generically, an affine connection in four dimensions has sixty-four components.

The starting point for building our model would be a generic affine connection \(\tilde{\Gamma}_{\mu}{}^{\lambda}{}_{\nu}\). Without the aid of a (fundamental) metric tensor field, a connection can be decomposed into its symmetric and antisymmetric parts,
\begin{dmath}
  \tilde{\Gamma}_{\mu}{}^{\lambda}{}_{\nu}
  =
  \tilde{\Gamma}_{(\mu}{}^{\lambda}{}_{\nu)}
  +
  \tilde{\Gamma}_{[\mu}{}^{\lambda}{}_{\nu]}
  =
  {\Gamma}_{\mu}{}^{\lambda}{}_{\nu}
  +
  \frac{1}{2} \tilde{\mathcal{T}}_{\mu}{}^{\lambda}{}_{\nu}
  =
  {\Gamma}_{\mu}{}^{\lambda}{}_{\nu}
  +
  {\mathcal{B}}_{\mu}{}^{\lambda}{}_{\nu}
  +
  {\mathcal{A}}_{[\mu} {\delta}^{\lambda}_{\nu]}.
  \label{eq:conn-decomp}
\end{dmath}
In the second line of Eq. \eqref{eq:conn-decomp} we have denoted with \(\Gamma\) the symmetric part of the affine connection and with \(\tilde{\mathcal{T}}\) the torsion of the affine connection.\footnote{In our convention, the torsion is defined as the difference between the components of the connection, \(\tilde{\mathcal{T}}_{\mu}{}^{\lambda}{}_{\nu} = \tilde{\Gamma}_{\mu}{}^{\lambda}{}_{\nu} - \tilde{\Gamma}_{\nu}{}^{\lambda}{}_{\mu}\). In order to avoid the half factor in the second line of Eq. \eqref{eq:conn-decomp}, it is usually referred to as the tensor \(\tilde{\mathcal{S}}_{\mu}{}^{\lambda}{}_{\nu} = \tilde{\Gamma}_{[\mu}{}^{\lambda}{}_{\nu]}\).} In the third line, we have decomposed the torsion into its traceless and trace part, proportional to \(\mathcal{B}\) and \(\mathcal{A}\) respectively.

In four dimensions a covariant theory of the affine connection requires that it enters into the functional action as a covariant derivative, \({\nabla}\). For simplicity, let us focus on the symmetric affine connection, \(\Gamma\), because the extension to asymmetric affine connections would be straightforward.

Schematically, a generic action functional for the (symmetric) affine connection has the symbolic form,
\begin{equation}
  S[\Gamma] = \int_{\mathcal{M}} \mathrm{d}^4 x \, \mathfrak{E}^{\alpha\beta\gamma\delta} \nabla_{\alpha} \nabla_{\beta} \nabla_{\gamma} \nabla_{\delta}.
  \label{eq:schematic-action}
\end{equation}
The action in Eq. \eqref{eq:schematic-action} can be rewritten in terms of terms that are quadratic in the curvature tensor, due to the contraction of the covariant derivative with the skew-symmetric tensor density \(\mathfrak{E}\).

A typical term of the quadratic action would be of the form,
\begin{equation}
  \int_{\mathcal{M}} \mathrm{d}^4 x \, \mathfrak{E}^{\cdot\cdot\cdot\cdot}  \mathcal{R}_{\cdot\cdot}{}^{\cdot}{}_{\cdot} \mathcal{R}_{\cdot\cdot}{}^{\cdot}{}_{\cdot},
  \label{eq:typical-term-action}
\end{equation}
where the dots represent indices which have to be contracted in every possible (but inequivalent) way. A consequence of the algebraic Bianchi identity,
\begin{equation}
  \mathfrak{E}^{\mu\nu\lambda\rho} \mathcal{R}_{\mu\nu}{}^{\sigma}{}_{\lambda} = 0,
  \label{eq:bianchi1}
\end{equation}
is that each curvature tensor has to have two of it lower indices contracted with the skew-symmetric tensor density.

Using the identity in Eq. \eqref{eq:bianchi1}, also written as
\begin{equation}
  \mathcal{R}_{\mu[\alpha}{}^{\nu}{}_{\beta]} = - \frac{1}{2} \mathcal{R}_{\alpha\beta}{}^{\nu}{}_{\mu},
  \label{eq:BI1}
\end{equation}
and its contracted version,
\begin{equation}
  \mathcal{R}_{[\alpha\beta]} = - \frac{1}{2} \mathcal{R}_{\alpha\beta}{}^{\mu}{}_{\mu},
  \label{eq:cBI1}
\end{equation}
one can workout all the possible contractions admissible in the action functional, and only two of these terms are inequivalent,
\begin{equation}
  \mathfrak{E}^{\alpha\beta\gamma\delta} \mathcal{R}_{\alpha\beta}{}^{\mu}{}_{\nu} \mathcal{R}_{\gamma\delta}{}^{\nu}{}_{\mu}
  \quad
  \text{ and }
  \quad
  \mathfrak{E}^{\alpha\beta\gamma\delta} \mathcal{R}_{\alpha\beta}{}^{\mu}{}_{\mu} \mathcal{R}_{\gamma\delta}{}^{\nu}{}_{\nu},
  \label{eq:affine-action-terms}
\end{equation}
which are Pontryagin terms, which in the notation of Refs. \cite{zanelli00_chern_simon_gravit,zanelli05_lectur_notes_chern_simon,zanelli16_chern} are denoted as \(P_4\) and \((P_2)^2 = P_2 \we P_2\), which are topological terms.\footnote{Note that when the connection is metric the Pontryagin \(P_2\) vanishes explicitly, since the anti-symmetric part of the Ricci tensor is zero, \(\mathcal{R}_{[\mu\nu]} = 0\).} Therefore, the model built in this way is not very interesting from the dynamical point of view.

In order to build up more interesting models, instead of considering the whole connection, we would use the decomposition shown in Eq. \eqref{eq:conn-decomp}. We shall engage in forming the most general action that contains the fields \(\mathcal{A}\), \(\mathcal{B}\), and the symmetric connection \(\Gamma\) through the covariant derivative \(\nabla\). To keep track of all possible terms that might be included in the action functional, we used a sort of dimensional analysis of the index structure. Let us define the operator \(\mathscr{N}\), which counts the net number of indices (upper indices count \(+1\) and lower indices count \(-1\)) of a term, and also define the operator \(\mathscr{W}\), which counts the weight of a tensor density.

The action of the operators on the basic ingredients of the model is
\begin{align}
  \label{eq:operator-N}
  \mathscr{N}(\mathcal{A}) = \mathscr{N}(\mathcal{B}) = \mathscr{N}(\nabla) & = -1 & \mathscr{N}(\mathfrak{E}) & = 4,
  \\
  \label{eq:operator-W}
  \mathscr{W}(\mathcal{A}) = \mathscr{W}(\mathcal{B}) = \mathscr{W}(\nabla) & = 0 & \mathscr{W}(\mathfrak{E}) & = 1.
\end{align}
A polynomial action contains powers of the basic fields, i.e., a generic term has the form
\begin{equation}
  \mathcal{O} = \mathcal{A}^m \mathcal{B}^n \nabla^p \mathfrak{E}^{q}.
  \label{eq:generic-action-term}
\end{equation}
Since the action has to be a scalar, the action of the operators \(\mathscr{N}\) and \(\mathscr{W}\) on the term in Eq. \eqref{eq:generic-action-term} yield the restrictions
\begin{equation}
  m + n + p + q = 1, \text{ and } m + n + p = 4 q.
  \label{eq:operator-restrictions}
\end{equation}

\begin{table}[htbp]
  \caption{\label{tab:power-counting}Possible terms to be considered in the action functional, according to the indices structure analysis.}
  \centering
  \begin{ruledtabular}
    \begin{tabular}{lllll}
      \(m\) & \(n\) & \(p\) & \(q\) & \(\mathcal{O}\)\\[0pt]
      \hline
      \(4\) & \(0\) & \(0\) & \(1\) & \(\mathcal{A}\mathcal{A}\mathcal{A}\mathcal{A}\mathfrak{E}\)\\[0pt]
      \(0\) & \(4\) & \(0\) & \(1\) & \(\mathcal{B}\mathcal{B}\mathcal{B}\mathcal{B}\mathfrak{E}\)\\[0pt]
      \(0\) & \(0\) & \(4\) & \(1\) & \(\nabla\nabla\nabla\nabla\mathfrak{E}\)\\[0pt]
      \(3\) & \(1\) & \(0\) & \(1\) & \(\mathcal{A}\mathcal{A}\mathcal{A}\mathcal{B}\mathfrak{E}\)\\[0pt]
      \(3\) & \(0\) & \(1\) & \(1\) & \(\mathcal{A}\mathcal{A}\mathcal{A}\nabla\mathfrak{E}\)\\[0pt]
      \(1\) & \(3\) & \(0\) & \(1\) & \(\mathcal{A}\mathcal{B}\mathcal{B}\mathcal{B}\mathfrak{E}\)\\[0pt]
      \(0\) & \(3\) & \(1\) & \(1\) & \(\mathcal{B}\mathcal{B}\mathcal{B}\nabla\mathfrak{E}\)\\[0pt]
      \(1\) & \(0\) & \(3\) & \(1\) & \(\mathcal{A}\nabla\nabla\nabla\mathfrak{E}\)\\[0pt]
      \(0\) & \(1\) & \(3\) & \(1\) & \(\mathcal{B}\nabla\nabla\nabla\mathfrak{E}\)\\[0pt]
      \(2\) & \(2\) & \(0\) & \(1\) & \(\mathcal{A}\mathcal{A}\mathcal{B}\mathcal{B}\mathfrak{E}\)\\[0pt]
      \(2\) & \(0\) & \(2\) & \(1\) & \(\mathcal{A}\mathcal{A}\nabla\nabla\mathfrak{E}\)\\[0pt]
      \(0\) & \(2\) & \(2\) & \(1\) & \(\mathcal{B}\mathcal{B}\nabla\nabla\mathfrak{E}\)\\[0pt]
      \(2\) & \(1\) & \(1\) & \(1\) & \(\mathcal{A}\mathcal{A}\mathcal{B}\nabla\mathfrak{E}\)\\[0pt]
      \(1\) & \(2\) & \(1\) & \(1\) & \(\mathcal{A}\mathcal{B}\mathcal{B}\nabla\mathfrak{E}\)\\[0pt]
      \(1\) & \(1\) & \(2\) & \(1\) & \(\mathcal{A}\mathcal{B}\nabla\nabla\mathfrak{E}\)\\[0pt]
    \end{tabular}
  \end{ruledtabular}
\end{table}

The table \ref{tab:power-counting} shows all the solutions to the constraints in Eq. \eqref{eq:operator-restrictions}. Using the symmetries of the fields and the Bianchi identities (both algebraic and differential), the most general action, up to boundary and topological terms, is
\begin{dmath}
  \label{eq:new-action}
  S
  = \int \mathrm{d}^4x \, \mathfrak{E}^{\alpha \beta \gamma \delta} \bigg[
  B_1 \mathcal{R}_{\mu\nu}{}^{\mu}{}_{\rho} \mathcal{B}_{\alpha}{}^{\nu}{}_{\beta} \mathcal{B}_{\gamma}{}^{\rho}{}_{\delta}
  + B_2 \mathcal{R}_{\alpha\beta}{}^{\mu}{}_{\rho} \mathcal{B}_{\gamma}{}^{\nu}{}_{\delta} \mathcal{B}_{\mu}{}^{\rho}{}_{\nu}
  + B_3 \mathcal{R}_{\mu\nu}{}^{\mu}{}_{\alpha} \mathcal{B}_{\beta}{}^{\nu}{}_{\gamma} \mathcal{A}_\delta
  + B_4 \mathcal{R}_{\alpha\beta}{}^{\sigma}{}_{\rho} \mathcal{B}_{\gamma}{}^{\rho}{}_{\delta} \mathcal{A}_\sigma
  + B_5 \mathcal{R}_{\alpha\beta}{}^{\rho}{}_{\rho} \mathcal{B}_{\gamma}{}^{\sigma}{}_{\delta} \mathcal{A}_\sigma
  + C_1 \mathcal{R}_{\mu\alpha}{}^{\mu}{}_{\nu} \nabla_\beta \mathcal{B}_{\gamma}{}^{\nu}{}_{\delta}
  + C_2 \mathcal{R}_{\alpha\beta}{}^{\rho}{}_{\rho} \nabla_\sigma \mathcal{B}_{\gamma}{}^{\sigma}{}_{\delta}
  + D_1 \mathcal{B}_{\nu}{}^{\mu}{}_{\lambda} \mathcal{B}_{\mu}{}^{\nu}{}_{\alpha} \nabla_\beta \mathcal{B}_{\gamma}{}^{\lambda}{}_{\delta}
  + D_2 \mathcal{B}_{\alpha}{}^{\mu}{}_{\beta} \mathcal{B}_{\mu}{}^{\lambda}{}_{\nu} \nabla_\lambda \mathcal{B}_{\gamma}{}^{\nu}{}_{\delta}
  + D_3 \mathcal{B}_{\alpha}{}^{\mu}{}_{\nu} \mathcal{B}_{\beta}{}^{\lambda}{}_{\gamma} \nabla_\lambda \mathcal{B}_{\mu}{}^{\nu}{}_{\delta}
  + D_4 \mathcal{B}_{\alpha}{}^{\lambda}{}_{\beta} \mathcal{B}_{\gamma}{}^{\sigma}{}_{\delta} \nabla_\lambda \mathcal{A}_\sigma
  + D_5 \mathcal{B}_{\alpha}{}^{\lambda}{}_{\beta} \mathcal{A}_\sigma \nabla_\lambda \mathcal{B}_{\gamma}{}^{\sigma}{}_{\delta}
  + D_6 \mathcal{B}_{\alpha}{}^{\lambda}{}_{\beta} \mathcal{A}_\gamma \nabla_\lambda \mathcal{A}_\delta
  + D_7 \mathcal{B}_{\alpha}{}^{\lambda}{}_{\beta} \mathcal{A}_\lambda \nabla_\gamma \mathcal{A}_\delta
  + E_1 \nabla_\rho \mathcal{B}_{\alpha}{}^{\rho}{}_{\beta} \nabla_\sigma \mathcal{B}_{\gamma}{}^{\sigma}{}_{\delta}
  + E_2 \nabla_\rho \mathcal{B}_{\alpha}{}^{\rho}{}_{\beta} \nabla_\gamma \mathcal{A}_{\delta}
  + F_1 \mathcal{B}_{\alpha}{}^{\mu}{}_{\beta} \mathcal{B}_{\gamma}{}^{\sigma}{}_{\delta} \mathcal{B}_{\mu}{}^{\lambda}{}_{\rho} \mathcal{B}_{\sigma}{}^{\rho}{}_{\lambda}
  + F_2 \mathcal{B}_{\alpha}{}^{\mu}{}_{\beta} \mathcal{B}_{\gamma}{}^{\nu}{}_{\lambda} \mathcal{B}_{\delta}{}^{\lambda}{}_{\rho} \mathcal{B}_{\mu}{}^{\rho}{}_{\nu}
  + F_3 \mathcal{B}_{\nu}{}^{\mu}{}_{\lambda} \mathcal{B}_{\mu}{}^{\nu}{}_{\alpha} \mathcal{B}_{\beta}{}^{\lambda}{}_{\gamma} \mathcal{A}_\delta
  + F_4 \mathcal{B}_{\alpha}{}^{\mu}{}_{\beta} \mathcal{B}_{\gamma}{}^{\nu}{}_{\delta} \mathcal{A}_\mu \mathcal{A}_\nu
  \bigg].
\end{dmath}
In the above equation, the covariant derivative \(\nabla\) and the curvature \(\mathcal{R}\) are associated with the symmetric connection, \(\Gamma\).

The action in Eq. \eqref{eq:new-action} is much complex than the action of Einstein--Hilbert, however, it possesses very interesting features: (i) The lack of a metric tensor field endows the action with the property of rigidity, since contains all possible combinations of the fields and their derivatives; (ii) All the coupling constants are dimensionless, which might be a sign of \emph{conformal} (or projective) invariance, and also ensure that the model is power-counting renormalisable; (iii) The model has no explicit three-point graviton vertices, since all graviton self-interactions are mediated by non-Riemannian parts of the connection, allowing to bypass the general postulates supporting the no-go theorems stated in Refs. \cite{mcgady14_higher_spin_massl_s_dimen,camanho16_causal_const_correc_to_gravit}, where it was proven that generic three-point graviton interactions are highly constrained by causality and analyticity of the \(S\)-matrix; (iv) The field equations are second order differential equations for the fields, and the Einstein spaces are a subset of their solutions; (v) The supporting symmetry group is the group of diffeomorphisms, desirable for the background independence of the model; (vi) It is possible to obtain \emph{emergent} (connection-descendent) metric tensors in the space of solutions; (vii) The cosmological constant appears in the solutions as an integration constant, changing the paradigm concerning its nature;\footnote{This is similar to what happens in the unimodular model of gravity \cite{ng91_unimod_theor_gravit_cosmol_const,jirousek23_unimod_approac_to_cosmol}.} (viii) The model can be extended to be coupled with a scalar field, and the field equations are \emph{equivalent} to those of General Relativity interacting with a massless scalar field; (ix) The action possesses just first order derivative of the fields, yielding second order differential equations, this might avoid the necessity of terms analogous to the Gibbons--Hawking--York boundary term in General Relativity.

Before moving forward to the analysis of the dynamical aspects of the model, let us inspect some general facets of the model.

Firstly, note that all the terms in the action of the model, Eq. \eqref{eq:new-action}, contain powers of the torsional fields, that is, \(\mathcal{A}\) and \(\mathcal{B}\). Therefore, it is not possible to take a torsion-free limit at the level of the action. Nevertheless, at the level of the field equations, such a limit exists. The field equations of this sector can easily be found varying the action restricted to the linear terms of the torsion field with respect to these fields \cite{castillo-felisola18_einst_gravit_from_polyn_affin_model}, i.e. the terms with coefficients \(C_1\) and \(C_2\) in Eq. \eqref{eq:new-action}. A quick look at these terms of the action shows that the \emph{restricted} field equations for the torsion-free sector would be
\begin{equation}
  \nabla_{[\mu} \mathcal{R}_{\nu]\lambda} + C \, \nabla_{\lambda} \mathcal{R}_{\mu\nu}{}^{\sigma}{}_{\sigma} = 0.
  \label{eq:torsion-free-feqs}
\end{equation}
Note that the field equations for the symmetric affine connection are obtained varying with respect to the \(\mathcal{B}\)-field, which raises a mismatch between the number of equations (generically would be \(4 \frac{4 \cdot 3}{2} - 4 = 20\)) and unknowns (\(4 \frac{4 \cdot 5}{2} = 40\)). As mentioned in previous articles, this characteristic might arise from the non-uniqueness of the Lagrangian describing the system \cite{hojman_privat,castillo-felisola18_einst_gravit_from_polyn_affin_model}.

The quantity \(\mathcal{R}_{\mu\nu}{}^{\sigma}{}_{\sigma}\) in the second term of Eq. \eqref{eq:torsion-free-feqs}, called homothetic curvature or second Ricci curvature, vanishes in Riemannian geometries (and therefore in General Relativity) and also in (metric-)affine geometries with constant volume form \cite{schouten13_ricci,nomizu94_affin}. In either case, the above field equations would be simplified to
\begin{equation}
  \nabla_{[\mu} \mathcal{R}_{\nu]\lambda} = 0.
  \label{eq:codazzi-ricci}
\end{equation}
The above equation is the condition for the Ricci tensor to be a Codazzi tensor \cite{derdzinski83_codaz_tensor_field_curvat_pontr_forms,besse07_einst}, and it is a well-known generalisation of the Einstein's field equations. In addition, Eq. \eqref{eq:codazzi-ricci} is equivalent (via the differential Bianchi identity) to the condition of harmonic curvature,
\begin{equation}
  \nabla_{\sigma} \mathcal{R}_{\mu\nu}{}^{\sigma}{}_{\lambda} = 0,
  \label{eq:harmonic-curvature}
\end{equation}
which has been considered in the literature \cite{derdzinski80_class_certain_compac_rieman_manif,derdzinski82_compac_rieman_manif_with_harmon_curvat,derdzinski85_rieman}.

The field equations in Eq. \eqref{eq:codazzi-ricci} also are obtained from the variation with respect to the affine connection of the gravitational Yang--Mills action,
\begin{equation}
  S_{\text{gYM}} = \int_{\mathcal{M}}  \tr \left( \mathcal{R} \we \star \mathcal{R} \right),
  \label{eq:gravitational-yang-mills}
\end{equation}
where \(\mathcal{R} \in \Omega^2(\mathcal{M}, T^{*}\mathcal{M} \otimes T\mathcal{M})\) is the curvature two-form, the \(\star \colon \Omega^p(\mathcal{M}, \mathcal{E}) \to \Omega^{4-p}(\mathcal{M}, \mathcal{E})\) represents the Hodge operator and the trace is taken over the indices on the bundle \(\mathcal{E} = T^{*}\mathcal{M} \otimes T\mathcal{M}\). This model was considered by Stephenson, Kilmister, and Yang \cite{stephenson58_quadr_lagran_gener_relat,kilmister61_use_alg_struct_phys,yang74_integ_formal_gauge_field}, in the context of a metric model of gravity. Although the Stephenson--Kilmister--Yang model is known to possess nonphysical solutions to the field equations \cite{pavelle75_unphy_solut_yangs_gravit_field_equat,thompson75_geomet_degen_solut_kilmis_yang_equat}, the arguments to declare those solutions nonphysical come from the field equations for the metric tensor field \cite{zanelli_privat}. The absence of a metric in our model allows one to bypass the arguments.

The Codazzi condition of the Ricci tensor has very recently gained interest due to the novel formulation of gravity proposed by Harada \cite{harada21_emerg_cotton_tensor_descr_gravit,harada22_cotton_gravit_galax_rotat_curves,mantica23_note_harad_confor_killin_gravit,harada23_dark_energ_confor_killin_gravit}. In Harada's model the geometrical contribution to the field equations comes through the Cotton tensor \cite{cotton99_sur}, defined as
\begin{equation}
  C_{\mu\nu\lambda} = 2 \nabla_{[\mu} \mathcal{R}_{\nu]\lambda} - \frac{1}{3} g_{\lambda[\mu} \nabla_{\nu]} \mathcal{R}.
  \label{eq:cotton-tensor}
\end{equation}
Note that a \emph{projective} version\footnote{Operationally, a projective equivalent quantity is defined by a similar expression where the terms containing the metric tensor field are dropped.} of vanishing Cotton tensor would be equivalent to the field equations in Eq. \eqref{eq:codazzi-ricci}.

There are other projective quantities of interest, for example, the Weyl projective curvature tensor (in \(n\) dimensions) is defined by \cite{weyl18_reine_infin,eisenhart27_non_rieman,schouten13_ricci,nomizu94_affin}
\begin{dmath}
  W^{(p)}_{\mu\nu}{}^{\lambda}{}_{\rho} = \mathcal{R}_{\mu\nu}{}^{\lambda}_{\rho} - \frac{1}{n-1} \left( \mathcal{R}_{\nu\rho} \delta^{\lambda}_{\mu} - \mathcal{R}_{\mu\rho} \delta^{\lambda}_{\nu} \right) - \frac{1}{n+1} \mathcal{R}_{\mu\nu}{}^{\sigma}{}_{\sigma} \delta^{\lambda}_{\rho} - \frac{1}{n^2 - 1} \left( \mathcal{R}_{\nu\rho}{}^{\sigma}{}_{\sigma} \delta^{\lambda}_{\mu} - \mathcal{R}_{\mu\rho}{}^{\sigma}{}_{\sigma} \delta^{\lambda}_{\nu} \right),
  \label{eq:projective-weyl}
\end{dmath}
and it is a \emph{curvature} tensor invariant under projective transformations. Note that if the trace of the curvature vanishes, the above expression reduces to the well-known definition of the Weyl conformal curvature, where the terms containing the metric tensor have been removed.

Equation \eqref{eq:codazzi-ricci} has three \emph{levels} solutions: (1) vanishing Ricci tensor; (2) parallel Ricci tensor; and (3) the Ricci tensor is a Codazzi tensor. Solutions to the field equations at certain levels include the solutions at previous levels; however, there might exist \emph{proper} solutions at the level of interest. In addition, at the second and third levels of solutions, the Ricci could be either degenerated or non-degenerated. When the Ricci is nondegenerate, its symmetric part might be interpreted as a metric tensor. Note that such a metric tensor, or similar, is not \emph{fundamental} from the point of view of the model building, and therefore we call it \emph{emergent metric} (see Sec. \ref{sec:metrics}).

In those cases where the Ricci tensor, evaluated at the space of solutions, is symmetric and nondegenerate, the condition of being parallel with respect to the connection is equivalent to restricting to a Riemannian geometry. Similarly, nontrivial solutions to the Codazzi condition on the Ricci tensor is equivalent to focus on non-Riemannian manifolds with completely symmetric nonmetricity, i.e. \(\nabla_{\lambda} g_{\mu\nu} = \mathcal{Q}_{\lambda\mu\nu} \in C^{\infty}(S^3(T^{*}\mathcal{M}))\).
These types of manifolds are subjects of interest in Information Geometry, where they are known as \emph{statistical manifolds}, and provide a geometrical framework for understanding and analysing statistical models \cite{amari10_infor_geomet_optim_machin,suzuki14_infor_geomet_statis_manif,amari16_infor_geomet_its_applic}.

\section{Covariant field equations}
\label{sec:cov-feqs}

In order to obtain the field equations of the model, the action Eq. (\ref{eq:new-action}) has to be varied with repect to the fields $\Gamma$, $\mathcal{B}$ and $\mathcal{A}$, leading to the Euler--Lagrange equations,
\begin{align}
  \label{eq:euler-lagrange-G}
  \partial_{\mu}\left( \frac{\partial \mathcal{L}}{\partial(\partial_{\mu}\Gamma_{\nu}{}^{\lambda}{}_{\rho})} \right) -\frac{\partial \mathcal{L}}{\partial \Gamma_{\nu}{}^{\lambda}{}_{\rho}} &= 0, \\
  \label{eq:euler-lagrange-B}
  \partial_{\mu}\left( \frac{\partial \mathcal{L}}{\partial(\partial_{\mu}\mathcal{B}_{\nu}{}^{\lambda}{}_{\rho})} \right) -\frac{\partial \mathcal{L}}{\partial \mathcal{B}_{\nu}{}^{\lambda}{}_{\rho}} &= 0, \\
  \label{eq:euler-lagrange-A}
  \partial_{\mu}\left( \frac{\partial \mathcal{L}}{\partial(\partial_{\mu}\mathcal{A}_{\nu})}\right) - \frac{\partial \mathcal{L}}{\partial \mathcal{A}_{\nu}} &= 0. 
\end{align}
Although explicit equations in terms of the component of the fields can be obtained from here, it turns out to be more convenient to find a covariant version of them. To this end, from the equation for \(\Gamma\), the canonical conjugated momentum is defined as,
\begin{equation*}
  {\Pi_{\Gamma}}^{\mu\nu}{}_{\lambda}{}^{\rho} := \frac{\partial \mathcal{L}}{\partial(\partial_{\mu}\Gamma_{\nu}{}^{\lambda}{}_{\rho})} = \frac{\partial \mathcal{L}}{\partial \Gamma_{\mu\nu}{}^{\lambda}{}_{\rho}}.
\end{equation*}
Since the derivative of the field \(\Gamma\) only appears in the curvature terms, the chain rule can be used and write,
\begin{equation*}
  \frac{\partial \mathcal{L}}{\partial \Gamma_{\mu\nu}{}^{\lambda}{}_{\rho}} = \frac{\partial \mathcal{L}}{\partial \mathcal{R}_{\alpha \beta}{}^{\gamma}{}_{\delta}} \frac{\partial \mathcal{R}_{\alpha \beta}{}^{\gamma}{}_{\delta}}{\partial \Gamma_{\mu \nu}{}^{\lambda}{}_{\rho}},
\end{equation*}
where the second factor can be directly computed,
\begin{equation*}
  \frac{\partial \mathcal{R}_{\alpha\beta}{}^{\gamma}{}_{\delta}}{\partial \Gamma_{\mu\nu}{}^{\lambda}{}_{\rho}} = 4 \delta^{\gamma}_{\lambda} \delta^{\mu}_{[\alpha} \delta^{(\nu}_{\beta]} \delta^{\rho)}_{\delta}.
\end{equation*}
Defining the auxiliary variable \({z_{\Gamma}}^{\alpha \beta}{}_{\gamma}{}^{\delta} \equiv \frac{\partial\mathcal{L}}{\partial \mathcal{R}_{\alpha\beta}{}^{\gamma}{}_{\delta}}\), the canonical conjugated momentum can be written as,
\begin{equation*}
  {\Pi_{\Gamma}}^{\mu\nu}{}_{\lambda}{}^{\rho} = {z_{\Gamma}}^{\alpha\beta}{}_{\gamma}{}^{\delta} 4 \delta^{\gamma}_{\lambda} \delta^{\mu}_{[\alpha} \delta^{(\nu}_{\beta]} \delta^{\rho)}_{\delta} = 2 {z_{\Gamma}}^{[\mu \nu]}{}_{\lambda}{}^{\rho} + 2 {z_{\Gamma}}^{[\mu\rho]}{}_{\lambda}{}^{\nu}.
\end{equation*}

Similarly, the second term in the Euler--Lagrange equation can be expressed as,
\begin{equation*}
  \frac{\partial \mathcal{L}}{\partial \Gamma_{\nu}{}^{\lambda}{}_{\rho}} = \frac{\partial \mathcal{L}}{\partial \mathcal{R}_{\alpha\beta}{}^{\gamma}{}_{\delta}} \frac{\partial \mathcal{R}_{\alpha\beta}{}^{\gamma}{}_{\delta}}{\partial \Gamma_{\nu}{}^{\lambda}{}_{\rho}} = {z_{\Gamma}}^{\alpha\beta}{}_{\gamma}{}^{\delta} \frac{\partial \mathcal{R}_{\alpha\beta}{}^{\gamma}{}_{\delta}}{\partial \Gamma_{\nu}{}^{\lambda}{}_{\rho}}.
\end{equation*}
But this is only valid for the terms in the action that contain the curvature tensor.

Using the result,
\begin{equation*}
  \frac{\partial\mathcal{R}_{\alpha\beta}{}^{\gamma}{}_{\delta}}{\partial \Gamma_{\nu}{}^{\lambda}{}_{\rho}} = 4\left(\delta^{\gamma}_{\lambda} \delta^{(\nu}_{[\alpha} \Gamma_{\beta]}{}^{\rho)}{}_{\delta} + \delta^{\rho}_{\delta} \delta^{(\nu}_{[\beta}\Gamma_{\alpha]}{}^{\gamma)}{}_{\lambda}  \right),
\end{equation*}
it can be obtained that,
\begin{dmath}
  \frac{\partial \mathcal{L}}{\partial \Gamma_{\nu}{}^{\lambda}{}_{\rho}} = 2 \left( {z_{\Gamma}}^{[\beta \rho]}{}_{\gamma}{}^{\nu} \Gamma_{\beta}{}^{\gamma}{}_{\lambda} + {z_{\Gamma}}^{[\beta \nu]}{}_{\gamma}{}^{\rho} \Gamma_{\beta}{}^{\gamma}{}_{\lambda} + {z_{\Gamma}}^{[\nu \beta]}{}_{\lambda}{}^{\delta} \Gamma_{\beta}{}^{\rho}{}_{\delta} + {z_{\Gamma}}^{[\rho \beta]}{}_{\lambda}{}^{\delta} \Gamma_{\beta}{}^{\nu}{}_{\delta} \right )
  = {\Pi_{\Gamma}}^{\mu \nu}{}_{\gamma}{}^{\rho} \Gamma_{\mu}{}^{\gamma}{}_{\lambda} - {\Pi_{\Gamma}}^{\mu \nu}{}_{\lambda}{}^{\delta} \Gamma_{\mu}{}^{\rho}{}_{\delta} - {\Pi_{\Gamma}}^{\mu \rho}{}_{\lambda}{}^{\delta} \Gamma_{\mu}{}^{\nu}{}_{\delta}.
\end{dmath}
These two results can be replaced in the Euler--Lagrange equation for \(\Gamma\) and it can be expressed in a covariant form,
\begin{dmath}
  \nabla_{\mu}{{\Pi_{\Gamma}}^{\mu \nu}{}_{\lambda}{}^{\rho}} = \frac{\partial^{*} \mathcal{L}}{\partial \Gamma_{\nu}{}^{\lambda}{}_{\rho}},
  \label{eq:field_eq_g}
\end{dmath}
where the asterisk indicate that the partial derivative is only in the terms that do not contain the curvature. In obtaining this expression, it was used the fact that the conjugate momentum is a density, so its covariant derivative is
\begin{dmath}
  \nabla_{\sigma} {\Pi_{\Gamma}}^{\mu \nu}{}_{\lambda}{}^{\rho} = \partial_{\sigma} {\Pi_{\Gamma}}^{\mu \nu}{}_{\lambda}{}^{\rho} + \Gamma_{\sigma}{}^{\mu}{}_{\tau} {\Pi_{\Gamma}}^{\tau \nu}{}_{\lambda}{}^{\rho} + \Gamma_{\sigma}{}^{\nu}{}_{\tau} {\Pi_{\Gamma}}^{\mu \tau}{}_{\lambda}{}^{\rho} - \Gamma_{\sigma}{}^{\tau}{}_{\lambda} {\Pi_{\Gamma}}^{\mu \nu}{}_{\tau}{}^{\rho} + \Gamma_{\sigma}{}^{\rho}{}_{\tau} {\Pi_{\Gamma}}^{\mu \nu}{}_{\lambda}{}^{\tau} - \Gamma_{\sigma}{}^{\tau}{}_{\tau} {\Pi_{\Gamma}}^{\mu \nu}{}_{\lambda}{}^{\rho}.
\end{dmath}
Where contracting \(\mu\) with \(\sigma\) makes the second and last terms cancel each other.

Following the same procedure, the covariant version of the equations for \(\mathcal{B}\) and \(\mathcal{A}\) can be found,
\begin{equation}
  \nabla_{\mu} {\Pi_{\mathcal{B}}}^{\mu \nu}{}_{\lambda}{}^{\rho} = \frac{\partial \mathcal{L}}{\partial \mathcal{B}_{\nu}{}^{\lambda}{}_{\rho}},
  \label{eq:field_eq_b}
\end{equation}
\begin{equation}
  \nabla_{\mu} {\Pi_{\mathcal{A}}}^{\mu \nu} = \frac{\partial \mathcal{L}}{\partial \mathcal{A}_{\nu}}.
  \label{eq:field_eq_a}
\end{equation}
It is important to note that in this case, the variation with respect to \(\mathcal{B}\) is given by,
\begin{dmath}
  \frac{\partial \mathcal{B}_{\alpha}{}^{\beta}{}_{\gamma}}{\partial \mathcal{B}_{\nu}{}^{\lambda}{}_{\rho}} = 2 \delta^{\beta}_{\lambda} \delta^{\nu \rho}_{\alpha \gamma} + \frac{2}{3} \delta^{\beta}_{\alpha} \delta^{\nu \rho}_{\gamma \lambda} - \frac{2}{3} \delta^{\beta}_{\gamma} \delta^{\nu \rho}_{\alpha \lambda},
\end{dmath}
in order to count the traceless character of the field \(\mathcal{B}\). The explicit covariant field equations for the action of the model are shown in the appendix \ref{sec:appen-a}.

\subsection{Torsion-less sector}
\label{sec:tor-less-sect}

The field equations can be analysed in different sectors. One of such sectors is the torsion-free sector in which both torsional tensors vanish, i.e. $\mathcal{A} \to 0$ and $\mathcal{B} \to 0$. The field equations cannot be obtained by setting the torsion to zero in the action because it would vanish, but this can be done at the covariant equation level.

In this case, the only nontrivial equation comes from the variation of the action with respect to $\mathcal{B}$, which results in
\begin{dmath}
  \nabla_{\mu}\left(
    - \mathcal{R}_{\sigma\alpha}{}^{\sigma}{}_{\lambda} \mathfrak{E}^{\mu\nu\rho\alpha}
    + \frac{2}{3} \mathcal{R}_{\sigma \alpha}{}^{\sigma}{}_{\tau} \delta_{\lambda}^{[\nu}\mathfrak{E}^{\rho]\mu\tau\alpha}
    + C \mathcal{R}_{\alpha\beta}{}^{\sigma}{}_{\sigma} \delta_{\lambda}^{\mu} \mathfrak{E}^{\nu\rho\alpha\beta}
    + \frac{2}{3} C \mathcal{R}_{\alpha\beta}{}^{\sigma}{}_{\sigma}\delta_{\lambda}^{[\nu}\mathfrak{E}^{\rho]\mu\alpha\beta} \right) = 0,
  \label{eq:complete-feqs-torsionless}
\end{dmath}
where $C = \frac{C_1}{C_2}$. However, the contraction of three indices of the tensor density \(\mathfrak{E}\) with the curvature tensor and the covariant derivative ensures, via the Bianchi identities and their contractions, that the second and fourth terms in Eq. \eqref{eq:complete-feqs-torsionless} vanish identically. Therefore, the field equations of the polynomial affine model of gravity in its torsion-free sector are
\begin{dmath}
  \nabla_{\mu}
  \left(
    - \mathcal{R}_{\sigma\alpha}{}^{\sigma}{}_{\lambda} \mathfrak{E}^{\mu\nu\rho\alpha}
    + C \mathcal{R}_{\alpha\beta}{}^{\sigma}{}_{\sigma} \delta_{\lambda}^{\mu} \mathfrak{E}^{\nu\rho\alpha\beta}
  \right)
  = 0,
  \label{eq:feqs-torsionless}
\end{dmath}
which can be rewritten as in Eq. \eqref{eq:torsion-free-feqs}.

Moreover, if the affine connection preserves the volume, this type of geometries is called \emph{equi-affine}, the trace of the curvature (\({\mathcal{R}}_{\mu\nu}{}^{\sigma}{}_{\sigma}\)) vanishes and the Ricci tensor is symmetric. In this scenario, the field equations simplify further to the condition that the Ricci tensor is a Codazzi tensor.

\section{Building ansatz for the connection}
\label{sec:conn-ansatz}

In gravitational theories, the field equations are in general a very complicated system of nonlinear partial differential equations whose unknown functions are the components of the fundamental geometrical objects, e.g. in General Relativity the unknown functions of Einstein equations are the components of the metric tensor field. The unknown functions in our model are the components of the affine connection.\footnote{Note that there are sixty-four of these functions in four dimensions.}

The general strategy to tackle the problem of finding solutions to the field equations is to propose an ansatz for the geometrical object. For that end, one demands our object to be compatible with the symmetries of the system one wants to model, e.g. spherical symmetry for modelling a \emph{round} astronomical body, axial symmetry for rotating bodies, isotropy and homogeneity for cosmological evolution, plane-parallelism for domain walls, etc.

The set of transformations preserving the symmetry of the system forms a group, in particular, for continuous transformations, they form a Lie group, \(G\). The local structure of a Lie group is determined by a set of generators and their (Lie) algebra, \(\mathfrak{g} \in T_{e}G\). The set of generators forms a vector basis for the tangent space of the symmetry group based on the identity. Each of these vectors generate a flow on the manifold, \(\mathcal{M}\), in the sense that they form one-parametric subgroups of the symmetry group \(G\) which act on the manifold \(\mathcal{M}\).

A geometrical object \(\mathcal{O}\) is said to be compatible with the symmetry (Lie) group \(G\) if its variation along the integral curves generated by the set of generators of the Lie algebra \(\mathfrak{g}\) vanishes, in mathematical terms,
\begin{equation}
  \pounds_V \mathcal{O} = 0, \quad \forall \; V \in \mathfrak{g}.
  \label{eq:lie-der-object}
\end{equation}
Although the explicit formulas for the Lie derivative of tensor expressions can be found in almost any textbook on differential geometry or General Relativity, the expression of the Lie derivative of an affine connection is much less known, and it is given by (see for example Ref. \cite{schouten13_ricci})
\begin{dmath}
  \pounds_{\xi} \Gamma_{\mu}{}^{\lambda}{}_{\nu} = \xi^{\rho} \partial_{\rho} \Gamma_{\mu}{}^{\lambda}{}_{\nu} - \Gamma_{\mu}{}^{\rho}{}_{\nu} \partial_{\rho} \xi^{\lambda} + \Gamma_{\rho}{}^{\lambda}{}_{\nu} \partial_{\mu} \xi^{\rho} + \Gamma_{\mu}{}^{\lambda}{}_{\rho} \partial_{\nu} \xi^{\rho} + \frac{\partial^2 \xi^{\lambda}}{\partial x^{\mu} \partial x^{\nu}},
  \label{eq:lie-der-conn}
\end{dmath}
or in covariant form
\begin{equation}
  \pounds_{\xi} \Gamma_{\mu}{}^{\lambda}{}_{\nu} = \xi^{\rho} \mathcal{R}_{\rho\mu}{}^{\lambda}{}_{\nu} + \nabla_{\mu} \nabla_{\nu} \xi^{\lambda} - \nabla_{\mu} \left( \mathcal{T}_{\nu}{}^{\lambda}{}_{\rho} \xi^{\rho} \right).
  \label{eq:cov-lie-der-conn}
\end{equation}

In concordance with the above, it is possible to restrict even further the form of the connection by requiring discrete symmetries to the system, such as the time-reversal or parity (which are of utmost importance in quantum field theory). In the following, we shall denote by \(T\) and \(P\) the time and azimuthal angle \(\varphi\) reversal operators, whose action on the base vectors is
\begin{equation}
  \label{eq:P-T-action}
  P \colon \vec{e}_\varphi \to -\vec{e}_\varphi, \qquad T \colon \vec{e}_t \to -\vec{e}_t
\end{equation}
and by \(PT\) the simultaneous action of both operators. For example, the tensor \(\vec{e}_t \otimes \vec{e}_{\varphi}\) is odd under the action of \(P\) or \(T\), but it is invariant under the action of \(PT\).

In gravitational physics one usually analyses configurations with lots of symmetries, since they simplify the form of the geometrical objects and hence the field equations driving the dynamics of the system. A brief (and incomplete) list of customary symmetry conditions is presented in the Tab. \ref{tab:symmetries-connection}.

Notice that for vectors with constant components, such as \(\vec{e}_t\) or \(\vec{e}_{\varphi}\) in our coordinate system (explained below), the vanishing Lie derivative of the connection is equivalent to the independence of the component of the connection on that coordinate. For example, a stationary solution is invariant under translations along the time-like coordinate, i.e. the connection satisfy that
\begin{equation}
  \label{eq:stationary-condition}
  \pounds_{\vec{e}_t} \Gamma_{\mu}{}^{\lambda}{}_{\nu} = \partial_{t} \Gamma_{\mu}{}^{\lambda}{}_{\nu} = 0,
\end{equation}
and therefore none of the components of the connection depend on the time-like coordinate,
\begin{equation}
  \label{eq:stationary-condition-solution}
  \Gamma_{\mu}{}^{\lambda}{}_{\nu} = \Gamma_{\mu}{}^{\lambda}{}_{\nu}(r,\theta,\varphi). 
\end{equation}
The Eq. \eqref{eq:stationary-condition} restricts the dependence on the coordinates, but does not restrict the number of components of the connection. The same is true if one wants par-axisymmetry (the prefix ``par'' means ``partially'', in the sense that the symmetry group does not contain discrete transformations, such as \(P\)).

The effect of the discrete transformations cannot be seen at the level of the Lie derivative, but it can be seen using the definition of the connection coefficients. For example, let us consider the variation of the time-like base vector for a stationary connection,
\begin{equation}
  \nabla_{\vec{e}_t} \vec{e}_t = \Gamma_{t}{}^{t}{}_{t} \vec{e}_{t} + \Gamma_{t}{}^{r}{}_{t} \vec{e}_r.
\end{equation}
Applying the time-reversal transformation \(\vec{e}_t \mapsto -\vec{e}_t\), one gets
\begin{equation}
  \nabla_{-\vec{e}_t} (-\vec{e}_t) = - \Gamma_{t}{}^{t}{}_{t} \vec{e}_t + \Gamma_{t}{}^{r}{}_{t} \vec{e}_r.
\end{equation}
In order for the covariant derivative to remain invariant under time reversal, the \(\Gamma_{t}{}^{t}{}_{t}\) component of the connection must vanish,
\begin{equation}
  \Gamma_{t}{}^{t}{}_{t} = 0.
\end{equation}
A similar analysis of the behaviour of the variation of other vector basis under time-reversal reveals that the components of the connection with an odd number of time-like indices must vanish.

The same type of analysis can be applied if we require invariance under reversal of the azimuthal angle \(\varphi\), say transforming \(\vec{e}_{\varphi} \mapsto - \vec{e}_{\varphi}\). The conclusion is analogous; the components of the affine connection with an odd number of \(\varphi\)-indices must vanish.

\begin{table}[ht]
  \centering
  \caption{Non exhaustive list of symmetries and their constraints on the functions characterising the affine connection. In the column ``Symmetry'' we list the algebra of the symmetry group, and the operators \(T\) and \(P\) represent the time-reversal and \(\varphi\)-parity.}
  \label{tab:symmetries-connection}
  \begin{ruledtabular}
    \begin{tabular}{llcr}
      Condition & Symmetry & Functions & Coordinates\\[0pt]
      \hline
      \hline
      General &  & \(40\) & \(t,r,\theta,\varphi\)\\[0pt]
      \hline
      Stationary & \(\partial_{t}\) & \(40\) & \(r,\theta,\varphi\)\\[0pt]
      \hline
      Par-axisymmetry & \(\partial_{\varphi}\) & \(40\) & \(t,r,\theta\)\\[0pt]
      \hline
      Stationary & \([\partial_{t},\partial_{\varphi}] = 0\) & \(40\) & \(r,\theta\)\\[0pt]
      par-axisymmetric &  &  & \\[0pt]
      \hline
      Static & \(\partial_{t}\), \(T\) & \(24\) & \(r,\theta,\varphi\)\\[0pt]
      \hline
      Axisymmetric & \(\partial_{\varphi}\), \(P\) & \(24\) & \(t,r,\theta\)\\[0pt]
      \hline
      Circular & \([\partial_{t},\partial_{\varphi}] = 0\), \(PT\) & \(20\) & \(r,\theta\)\\[0pt]
      \hline
      Static & \([\partial_{t},\partial_{\varphi}] = 0\), & \(16\) & \(r,\theta\)\\[0pt]
      axisymmetric  & \(P\), \(T\) &  & \\[0pt]
      \hline
      Par-spherical & \(\mathfrak{o}(3)\) & \(12\) & \(t,r\)\\[0pt]
      \hline
      Stationary & \(\mathfrak{o}(3)\), \(\partial_{t}\) & \(12\) & \(r\)\\[0pt]
      par-spherical &  &  & \\[0pt]
      \hline
      Spherical & \(\mathfrak{o}(3)\), \(P\) & \(10\) & \(t,r\)\\[0pt]
      \hline
      Static & \(\mathfrak{o}(3)\), & \(6\) & \(r\)\\[0pt]
      par-spherical & \(\partial_{t}\), \(T\) &  & \\[0pt]
      \hline
      Static circular & \(\mathfrak{o}(3)\), \(\partial_{t}\), & \(6\) & \(r\)\\[0pt]
      spherical & \(PT\) &  & \\[0pt]
      \hline
      Static & \(\mathfrak{o}(3)\), \(P\) & \(5\) & \(r\)\\[0pt]
      spherical & \(\partial_{t}\), \(T\) &  & \\[0pt]
      \hline
    \end{tabular}
  \end{ruledtabular}
\end{table}

Interesting cases, usually considered in gravitational physics, are spherically symmetric and isotropic-homogeneous spaces. Note that isotropy and homogeneity refer to the \emph{spacial} section of \emph{spacetime}. We emphasise the names \emph{space} and \emph{spacetime}, because without an explicit metric on the manifold, we cannot distinguish between a coordinate similar to time and one similar to space. However, we treat $t$ as a time-like coordinate which is not equivalent to the remaining coordinates under the symmetry transformations.

In the remaining part of this section we shall review the general ansätze of the affine connection compatible with the isotropic (or spherical) and cosmological symmetries. The results below were found in Ref. \cite{castillo-felisola18_beyond_einstein}.\footnote{The Lie derivatives of the affine connection were computed using the free and open mathematical software \texttt{SAGE} \cite{stein18_sage_mathem_softw_version}, through its module \texttt{sagemanifolds} \cite{gourgoulhon18_sagem_version,gourgoulhon15_tensor_calcul_with_open_sourc_softw,gourgoulhon18_symbol_tensor_calcul_manif}.} Note, however, that one can treat the symmetric affine connection separately from the torsional fields \(\mathcal{A}\) and \(\mathcal{B}\), as suggested in Ref. \cite{toloza13_cosmol_with_scalar_euler_form_coupl}.

\subsection{Par-spherically symmetric connections}
\label{sec:spher-conn}

The isotropy group in three (real) dimensions is \(O(3,\mathbb{R})\), which has dimension \(\dim(O_3) = 3\). In spherical coordinates (\(t,r,\theta,\varphi\)) its generators are the following vectors,
\begin{equation}
  \begin{aligned}
    J_1 & =
          \begin{pmatrix}
            0 & 0 & - \cos(\varphi) & \cot(\theta) \sin(\varphi)
          \end{pmatrix},
    \\
    J_2 & =
          \begin{pmatrix}
            0 & 0 & \sin(\varphi) & \cot(\theta) \cos(\varphi)
          \end{pmatrix},
    \\
    J_3 & =
          \begin{pmatrix}
            0 & 0 & 0 & 1
          \end{pmatrix}.
  \end{aligned}
  \label{eq:spher-generators}
\end{equation}

\subsubsection*{Par-spherical \(\mathcal{A}\)-field}

A par-spherical vector has the following functional form,
\begin{equation}
  \begin{aligned}
    \mathcal{A}_{t} & = A_{0}(t,r), & \mathcal{A}_r & = A_1(t,r), \\
    \mathcal{A}_{\theta} & = A_2(t,r), & \mathcal{A}_\varphi & = A_2(t,r) \sin(\theta).
  \end{aligned}
  \label{eq:spherical-A}
\end{equation}

\subsubsection*{Par-spherical \(\mathcal{B}\)-field}

The non-trivial components of the \(\mathcal{B}\)-field can be found with some ease. The par-spherical ansatz is parameterised by six functions,
\begin{widetext}
  \begin{equation}
    \begin{aligned}
      \mathcal{B}_{\theta}{}^{t}{}_{\varphi} = - \mathcal{B}_{\varphi}{}^{t}{}_{\theta} & = B_{203}(t,r) \sin(\theta),
      &
        \mathcal{B}_{\theta}{}^{r}{}_{\varphi} = - \mathcal{B}_{\varphi}{}^{r}{}_{\theta} & = B_{213}(t,r) \sin(\theta), \\
      \mathcal{B}_{t}{}^{\theta}{}_{\varphi} = - \mathcal{B}_{\varphi}{}^{\theta}{}_{t} & = B_{023}(t,r) \sin(\theta),
      &
        \mathcal{B}_{r}{}^{\theta}{}_{\varphi} = - \mathcal{B}_{\varphi}{}^{\theta}{}_{r} & = B_{123}(t,r) \sin(\theta), \\
      \mathcal{B}_{t}{}^{\varphi}{}_{\theta} = - \mathcal{B}_{\theta}{}^{\varphi}{}_{t} & = \frac{B_{032}(t,r)}{\sin(\theta)},
      &
        \mathcal{B}_{r}{}^{\varphi}{}_{\theta} = - \mathcal{B}_{\theta}{}^{\varphi}{}_{r} & = \frac{B_{132}(t,r)}{\sin(\theta)}.
    \end{aligned}
    \label{eq:spherical-B}
  \end{equation}
\end{widetext}

\subsubsection*{Par-spherical symmetric connection}

The par-spherical (symmetric) connection can be obtained by solving the differential equations determined by the vanishing Lie derivative of the connection, Eq. \eqref{eq:lie-der-conn}.

After a straightforward but long manipulation, one finds that the non-trivial components of the par-spherical symmetric connection are
\begin{widetext}
  \begin{equation}
    \begin{aligned}
      \Gamma_{t}{}^{t}{}_{t} & = F_{000}(t,r), &
                                            \Gamma_{t}{}^{t}{}_{r} = \Gamma_{r}{}^{t}{}_{t} & = F_{001}(t,r),
      \\
      \Gamma_{r}{}^{t}{}_{r} & = F_{101}(t,r), &
                                            \Gamma_{\theta}{}^{t}{}_{\theta} & = F_{202}(t,r),
      \\
      \Gamma_{\varphi}{}^{t}{}_{\varphi} & = F_{202}(t,r) \sin^2(\theta), &
                                                      \Gamma_{t}{}^{r}{}_{t} & = F_{010}(t,r),
      \\
      \Gamma_{t}{}^{r}{}_{r} = \Gamma_{r}{}^{r}{}_{t} & = F_{011}(t,r), &
                                                                \Gamma_{r}{}^{r}{}_{r} & = F_{111}(t,r),
      \\
      \Gamma_{\theta}{}^{r}{}_{\theta} & = F_{212}(t,r), &
                                            \Gamma_{\varphi}{}^{r}{}_{\varphi} & = F_{212}(t,r)  \sin^2(\theta),
      \\
      \Gamma_{t}{}^{\theta}{}_{\theta} = \Gamma_{\theta}{}^{\theta}{}_{t} & = F_{022}(t,r), &
                                                                \Gamma_{t}{}^{\theta}{}_{\varphi} = \Gamma_{\varphi}{}^{\theta}{}_{t} & = F_{023}(t,r) \sin(\theta),
      \\
      \Gamma_{r}{}^{\theta}{}_{\theta} = \Gamma_{\theta}{}^{\theta}{}_{r} & = F_{122}(t,r), &
                                                                \Gamma_{r}{}^{\theta}{}_{\varphi} = \Gamma_{\varphi}{}^{\theta}{}_{r} & = F_{123}(t,r) \sin(\theta),
      \\
      \Gamma_{\varphi}{}^{\theta}{}_{\varphi} & = - \cos(\theta) \sin(\theta), &
                                                 \Gamma_{t}{}^{\varphi}{}_{\theta} = \Gamma_{\theta}{}^{\varphi}{}_{t} & = - \frac{F_{023}(t,r)}{\sin(\theta)},
      \\
      \Gamma_{t}{}^{\varphi}{}_{\varphi} = \Gamma_{\varphi}{}^{\varphi}{}_{t} & = F_{022}(t,r), &
                                                                \Gamma_{r}{}^{\varphi}{}_{\theta} = \Gamma_{\theta}{}^{\varphi}{}_{r} & = - \frac{F_{123}(t,r)}{\sin(\theta)},
      \\
      \Gamma_{r}{}^{\varphi}{}_{\varphi} = \Gamma_{\varphi}{}^{\varphi}{}_{r} & = F_{122}(t,r), &
                                                                \Gamma_{\theta}{}^{\varphi}{}_{\varphi} = \Gamma_{\varphi}{}^{\varphi}{}_{\theta} & = \frac{\cos(\theta)}{\sin(\theta)}.
    \end{aligned}
    \label{eq:spherical-gamma}
  \end{equation}
\end{widetext}
The symmetric connection is determined by twelve functions of the coordinates \(t\) and \(r\).

\subsection{Cosmological connections}
\label{sec:cosm-conn}

In order to obtain the cosmological connection, we could start from the results of the previous section and require, in addition, the invariance (in the sense of vanishing Lie derivative) under the generators of \emph{translations}. However, such an extension is not unique and depends on whether or not the generators of \emph{translations} commute.

The dimension of the symmetry group compatible with the cosmological principle is six, and therefore its algebra could be homomorphic to \(\mathfrak{o}(4)\), \(\mathfrak{io}(3)\) or \(\mathfrak{o}(3,1)\). The Lie algebras of these groups can be written in terms of the generators \(J_{AB} = \left\{ J_{ab}, J_{a\ast} \right\}\) as
\begin{equation}
  \begin{aligned}
    [J_{ab}, J_{cd}] & = \delta_{bc} J_{ad} - \delta_{ac} J_{bd} + \delta_{ad} J_{bc} - \delta_{bd} J_{ac}, \\
    [J_{ab}, J_{c\ast}] & = \delta_{bc} J_{a\ast} - \delta_{ac} J_{b\ast}, \\
    [J_{a\ast}, J_{b\ast}] & = - \kappa J_{ab},
  \end{aligned}
  \label{eq:cosmological-algebra}
\end{equation}
with \(\kappa = 1,0,-1\) for \(\mathfrak{o}(4)\), \(\mathfrak{io}(3)\) and \(\mathfrak{o}(3,1)\) respectively.

The generators \(P_a = J_{a\ast}\) can be expressed in spherical coordinates as
\begin{equation}
  \begin{aligned}
    P_1 & = \sqrt{1 - \kappa r^2}
          \begin{pmatrix}
            0 & \sin(\theta) \cos(\varphi) & \frac{\cos(\theta) \cos(\varphi)}{r} & - \frac{\sin(\varphi)}{r \sin(\theta)}
          \end{pmatrix}, \\
    P_2 & = \sqrt{1 - \kappa r^2}
          \begin{pmatrix}
            0 & \sin(\theta) \sin(\varphi) & \frac{\cos(\theta) \sin(\varphi)}{r} & \frac{\cos(\varphi)}{r \sin(\theta)}
          \end{pmatrix}, \\
    P_3 & = \sqrt{1 - \kappa r^2}
          \begin{pmatrix}
            0 & \cos(\theta) & - \frac{\sin(\theta)}{r} & 0
          \end{pmatrix}.
  \end{aligned}
  \label{eq:translational-generators}
\end{equation}
In practice, since we require isotropy and homogeneity simultaneously, if the ansatz is compatible with par-spherical symmetry and we add homogeneity along a single direction, the geometrical object would be symmetric along the other directions.

The homogeneity implies that none of the functions characterising the spherical connection defined in Sec. \ref{sec:spher-conn} would depend on the radial coordinate.

\subsubsection*{Cosmological \(\mathcal{A}\) field}

Requiring the vanishing Lie derivative of the isotropic \(\mathcal{A}\) along the vectors \(P_i\) has as consequence that
\begin{equation}
  \begin{aligned}
    \mathcal{A}_{t} & = A_0(t) \equiv \eta(t), & \mathcal{A}_r = \mathcal{A}_{\theta} = \mathcal{A}_{\varphi} & = 0.
  \end{aligned}
  \label{eq:cosmo-A}
\end{equation}
Hence, a vector compatible with the cosmological symmetries is determined by a single function depending on the time-like coordinate \(t\), which we have called \(\eta\).

\subsubsection*{Cosmological \(\mathcal{B}\) field}

Similarly to what happened for the \(\mathcal{A}\) field, the invariance of the \(\mathcal{B}\) field along the vectors \(P_{i}\) would restrict the functions that characterise the par-spherical field. In this particular case, the cosmological \(\mathcal{B}\)-field would be determined by
\begin{equation}
  \begin{aligned}
    \mathcal{B}_{\theta}{}^{r}{}_{\varphi} = - \mathcal{B}_{\varphi}{}^{r}{}_{\theta} & = B_{123}(t) \, \sqrt{1 - \kappa r^2} r^2 \sin(\theta), \\
    \mathcal{B}_{r}{}^{\theta}{}_{\varphi} = - \mathcal{B}_{\varphi}{}^{\theta}{}_{r} & = - B_{123}(t) \frac{\sin(\theta)}{\sqrt{1 - \kappa r^2}}, \\
    \mathcal{B}_{r}{}^{\varphi}{}_{\theta} = - \mathcal{B}_{\theta}{}^{\varphi}{}_{r} & = B_{123}(t) \frac{1}{\sqrt{1 - \kappa r^2}\sin(\theta)}.
  \end{aligned}
  \label{eq:cosmo-B}
\end{equation}

Interestingly, the cosmological field \(\mathcal{B}\) is defined by a single function of the time-like coordinate, \(B_{123}(t)\). In the following this function would be renamed,
\begin{equation*}
  B_{123}(t) \equiv \psi(t).
\end{equation*}

\subsubsection*{Cosmological symmetric connection}

Interestingly, since the group of cosmological symmetries is six-dimensional, acting on the three-dimensional (spacial) submanifold, the group determines a nondegenerated symmetric \(\binom{0}{2}\)-tensor on the submanifold, i.e., a spacial metric
\begin{equation}
  s_{ij} =
  \begin{pmatrix}
    \frac{1}{\sqrt{1 - \kappa r^2}} & 0 & 0 \\
    0 & r^2 & 0 \\
    0 & 0 & r^2 \sin^2(\theta)
  \end{pmatrix},
  \label{eq:cosmological-spacial-metric}
\end{equation}
which shall be used to characterise the nontrivial components of the symmetric connection \(\Gamma\).

Starting with the par-spherical ansatz and requiring the vanishing Lie derivative along the symmetry generators \(P_i\), the symmetric cosmological affine connection is characterised by the following components
\begin{equation}
  \begin{aligned}
    \Gamma_{t}{}^{t}{}_{t} & = G_{000}(t),
    &
      \Gamma_{i}{}^{t}{}_{j} & = G_{101}(t) \, s_{ij},
    \\
    \Gamma_{t}{}^{i}{}_{j} = \Gamma_{j}{}^{i}{}_{t} & = G_{011}(t) \delta^i_j,
    &
      \Gamma_{i}{}^{j}{}_{k} & = \gamma_i{}^j{}_k,
  \end{aligned}
  \label{eq:cosmo-gamma}
\end{equation}
with \(\gamma_i{}^j{}_k\) the Levi-Civita connection associated with the three-dimensional metric \(s_{ij}\), i.e.
\begin{equation}
  \begin{aligned}
    \gamma_{r}{}^{r}{}_{r} & = \frac{\kappa r}{1 - \kappa r^{2}},
    \\
      \gamma_{r}{}^{\theta}{}_{\theta} = \gamma_{\theta}{}^{\theta}{}_{r} & = \frac{1}{r},
    \\
    \gamma_{\theta}{}^{r}{}_{\theta} & = - r (1 - \kappa r^2),
    \\
      \gamma_{\varphi}{}^{\theta}{}_{\varphi} & = - \cos(\theta) \sin(\theta),
    \\
    \gamma_{\varphi}{}^{r}{}_{\varphi} & = - r (1 - \kappa r^2) \sin^2(\theta),
    \\
      \gamma_{r}{}^{\varphi}{}_{\varphi} = \gamma_{\varphi}{}^{\varphi}{}_{r} & = \frac{1}{r},
    \\
    \gamma_{\theta}{}^{\varphi}{}_{\varphi} = \gamma_{\varphi}{}^{\varphi}{}_{\theta} & = \frac{\cos(\theta)}{\sin(\theta)}.
  \end{aligned}
  \label{eq:cosmo-three-connection}
\end{equation}

In the following sections, we discuss the functions determining the cosmological connection, Eq. \eqref{eq:cosmo-gamma}, will be renamed,
\begin{equation}
  G_{000}(t) \equiv f(t), \qquad G_{101}(t) \equiv g(t), \qquad G_{011}(t) \equiv h(t).
  \label{eq:cosmo-renamed-functions}
\end{equation}
In addition, it is important to notice that one might reparameterise the coordinate \(t\) to require that \(f(t) = 0\) \cite{castillo-felisola21_aspec_polyn_affin_model_gravit_three}, so we shall use this parameterisation.

\section{Cosmological solutions in four dimensions}
\label{sec:cosm-sol}

The set of cosmological equations is obtained by replacing the cosmological ansatz with the covariant field equations. The set of field equations for the field $\mathcal{A}$ leads to a first-order differential equation.
\begin{dmath}
  \label{eq:field_a}
  \left(B_3\left(\dot{g} + gh + 2\kappa\right) - 2B_4\left(\dot{g} - gh\right) + 2D_6\eta g - 2F_3\psi^2\right)\psi = 0.
\end{dmath}
The field $\mathcal{B}$ leads to a second-order differential equation.
\begin{dmath}
  \label{eq:field_b}
  B_3(\dot{g} + gh + 2\kappa)\eta - 2B_4\left(\dot{g} - gh\right)\eta + C_1(2\kappa h + 4gh^2  + 2g\dot{h}- \ddot{g}) - 6h\psi^2\left(D_1 - 2D_2 + D_3\right) + D_6 \eta^2 g - 6F_3\eta\psi^2 = 0  
\end{dmath}
and finally, the $\Gamma$ field leads to three differential equations,
\begin{dmath}
  \label{eq:field_gamma_1}
  (B_3\eta\psi -2B_4\eta\psi + C_1(\dot{\psi} - 2h\psi))g = 0,
\end{dmath}
\begin{dmath}
  \label{eq:field_gamma_2}
  \left(B_3 + 2B_4\right)\eta g\psi + 2C_1(\kappa\psi + 4gh\psi - g\dot{\psi} - \psi\dot{g})- 2\psi^3\left(D_1 - 2D_2 + D_3\right) = 0,
\end{dmath}
\begin{dmath}
  \label{eq:field_gamma_3}
  B_3(\eta(h\psi -\dot{\psi}) - \psi\dot{\eta}) - 2B_4(\eta(-h\psi - \dot{\psi}) - \psi\dot{\eta}) + C_1(4h^2\psi + 2\psi\dot{h} -\ddot{\psi}) + D_6\eta^2\psi = 0.
\end{dmath}

Although the system is overdetermined, we will prove that it is indeed possible to find analytical solutions without any type of assumption on the functions. First, notice that Eqs.~\eqref{eq:field_a} and \eqref{eq:field_gamma_1} can be written in a more \emph{compact} manner as follows,
\begin{dmath}
  \label{Feq_F_a}
  \mathcal{F}(g, \dot{g}, h,\psi,\eta)\psi = 0,
\end{dmath}
\begin{dmath}
  \label{Feq_G_gamma_1}
  \mathcal{G}(h,\psi, \dot{\psi}, \eta)g = 0,
\end{dmath}
where the functions $\mathcal{F}$ and $\mathcal{G}$ are defined as
\begin{dmath}
  \mathcal{F}(g,\dot{g},h,\psi,\eta)  \equiv B_3\left(\dot{g} + gh + 2\kappa\right) - 2B_4\left(\dot{g} - gh\right) + 2D_6\eta g - 2F_3\psi^2.
\end{dmath}
\begin{dmath}
  \mathcal{G}(h,\psi, \dot{\psi}, \eta)  \equiv B_3\eta\psi -2B_4\eta\psi + C_1\left(\dot{\psi} - 2h\psi\right).
\end{dmath}
Thus, using Eqs. \eqref{Feq_F_a} and \eqref{Feq_G_gamma_1} it is possible to distinguish four different branches:
\begin{itemize}
\item {\bf First branch:}  { $ \ \ \ \mathcal{F}(g,h,\psi,\eta) \! = 0 \ \wedge \ \mathcal{G}(h,\psi,\eta) \!= \! 0$}.
\item {\bf Second branch:} $ \mathcal{F}(g,h,\psi,\eta)  = 0 \ \wedge \ g  = 0$.
\item {\bf Third branch:}   $\ \mathcal{G}(h,\psi,\eta)   = 0   \wedge  \psi  = 0$.
\item {\bf Fourth branch:} $\psi  = 0  \wedge  g  = 0$.
\end{itemize}
Clearly, the first branch has the least restrictions on the field equations, and therefore, it has
more information than the rest of the branches.

\subsection*{First branch}

The first branch is the most general case where functions $g(t) \neq 0$ and $\psi(t) \neq 0$, then the system to be solve
is given by
\begin{dmath}
  \label{eq:field_B1_a}
  B_3\left(\dot{g} + gh + 2\kappa\right) - 2B_4\left(\dot{g} - gh\right) + 2D_6\eta g - 2F_3\psi^2 = 0,
\end{dmath}
\begin{dmath}
  \label{eq:field_B1_b}
  B_3(\dot{g} + gh + 2\kappa)\eta - 2B_4\left(\dot{g} - gh\right)\eta + C_1(2\kappa h + 4gh^2  + 2g\dot{h}- \ddot{g}) - 6h\psi^2\left(D_1 - 2D_2 + D_3\right) + D_6 \eta^2 g - 6F_3\eta\psi^2 = 0  
\end{dmath}
\begin{dmath}
  \label{eq:field_gamma_B1_1}
  B_3\eta\psi -2B_4\eta\psi + C_1(\dot{\psi} - 2h\psi) = 0,
\end{dmath}
\begin{dmath}
  \label{eq:field_gamma_B2_2}
  \left(B_3 + 2B_4\right)\eta g\psi + 2C_1(\kappa\psi + 4gh\psi - g\dot{\psi} - \psi\dot{g})- 2\psi^3\left(D_1 - 2D_2 + D_3\right) = 0,
\end{dmath}
\begin{dmath}
  \label{eq:field_gamma_B3_3}
  B_3(\eta(h\psi -\dot{\psi}) - \psi\dot{\eta}) - 2B_4(\eta(-h\psi - \dot{\psi}) - \psi\dot{\eta}) + C_1(4h^2\psi + 2\psi\dot{h} -\ddot{\psi}) + D_6\eta^2\psi = 0.
\end{dmath}
In the following steps, we shall show how to solve the differential equation system \eqref{eq:field_B1_a}-\eqref{eq:field_gamma_B3_3} exactly without any assumption.
First, from Eq. \eqref{eq:field_gamma_B1_1} we found an expression for $\eta(t)$ as
\begin{equation}
  \label{B1_eta}
  \eta(t) = \left(2h\psi - \dot{\psi}\right)\left(\frac{C_1}{B_3 - 2B_4}\right),
\end{equation}
where we required that $B_3 \neq 2B_4$. Replacing the expression of $\eta(t)$ in the system leads to
\begin{equation}
  \label{B1_h}
  h(t) =
  \begin{cases}
    h_1 = \dfrac{\dot{\psi}}{2\psi}
    \\
    h_2 = \dfrac{\dot{\psi}}{\psi} \left(\dfrac{C_1D_6}{3B_3^2 - 8B_3B_4 + 4B_4^2 + 2C_1D_6}\right).
  \end{cases}
\end{equation}
Although there are two possible choices of $h(t)$, the choice of $h_2(t)$ leads to inconsistencies in the system of 
differential equations, therefore, we will take $h_1(t)$, and the system is reduced to
\begin{dmath}
  \label{Branch_1_1}
  4\kappa B_3 + 2\dot{g}\left(B_3 - 2B_4\right) + \frac{\dot{\psi}g}{\psi}\left(B_3 + 2B_4\right) - 4F_3\psi^2 = 0
\end{dmath}
\begin{dmath}
  \label{Branch_1_2}
  \left(D_1 - 2D_2 + D_3\right)\psi^3 - C_1\left(\psi\left(\kappa - \dot{g}\right) + g\dot{\psi}\right) = 0
\end{dmath}
\begin{dmath}
  \label{Branch_1_3}
  \left(\frac{\kappa \dot{\psi} - \psi\ddot{g} + g\ddot{\psi}}{\psi}\right)C_1 - 3\left(D_1 - 2D_2 + D_3\right)\psi\dot{\psi}  = 0
\end{dmath}
From Eq. \eqref{Branch_1_2} it is possible to find an expression $g(t)$ in terms of $\psi(t)$
\begin{equation}
  \label{B1_g}
  g(t) = \psi\left(g_0 + \int \left(\frac{\kappa}{\psi} - \left(\frac{D_1 - 2D_2 + D_3}{C_1}\right)\psi\right)\mathrm{d}\tau\right),
\end{equation}
where $g_0$ is an integration constant. Replacing the above expression of $g(t)$ automatically solves \eqref{Branch_1_3}, and
Eq. \eqref{Branch_1_1} leads into a first order integro-differential equation
\begin{widetext}
  \begin{equation}
    (3B_3 - 2 B_4) \left(2 \kappa + \dot{\psi}\left(g_0 + \int \left(\frac{\kappa}{\psi} - \left(\tfrac{D_1 - 2D_2 + D_3}{C_1}\right)\psi\right)\mathrm{d}\tau\right) \right) =  \psi^2\left(\tfrac{2\left(B_3 - 2B_4\right)\left(D_1 - 2D_2 + D_3\right)}{C_1} + 4F_3\right).
  \end{equation}
\end{widetext}
As a standard practice in cosmology, we shall take $\kappa = 0$ and therefore the above equation is reduced even further to
\begin{dmath}
  \left(3B_3 - 2B_4\right)\dot{\psi}\left(g_0 -  \left(\frac{D_1 - 2D_2 + D_3}{C_1}\right)\int \psi\mathrm{d}\tau\right) = 
  \psi^2\left(\frac{2\left(B_3 - 2B_4\right)\left(D_1 - 2D_2 + D_3\right) }{C_1} + 4F_3\right) 
\end{dmath}
Using the following definitions
\begin{equation}
  \begin{aligned}
    \alpha & = \frac{D_1 - 2D_2 + D_3}{C_1} \\ \beta & = \frac{3B_3 - 2B_4}{2} \\ \gamma & = \left(\beta - 2B_3\right)\alpha  + 2F_3
  \end{aligned}
\end{equation}
the dynamical equation for $\psi$ can be written in a more compact manner
\begin{equation}
  \beta \dot{\psi}\left(g_0 - \alpha\int \psi\mathrm{d}\tau\right) = \psi^2\gamma
\end{equation}
The above equation can be solve analytically, to prove this consider the variable change $\psi (t) \equiv \dot{\phi}$, which applied to the above equation leads to
\begin{equation}
  \label{eq:integro_differential_phi}
  \ddot{\phi}\left(g_0 - \phi\alpha\right)\beta - \dot{\phi}^2\gamma = 0,
\end{equation}
whose solution is
\begin{equation}
  \phi(t) = \frac{g_0}{\alpha} + \lambda\left(t - t_0\right)^{\frac{\alpha\beta}{\alpha\beta + \gamma}}
\end{equation}
where $\lambda$ and $t_0$ are integration constants. Using the solution $\phi(t)$, it is straightforward to recover to original function
\begin{equation}
  \psi(t) =\frac{ \lambda\alpha\beta}{\alpha\beta + \gamma}\left(t -t_0\right)^{-\frac{\gamma}{\alpha\beta + \gamma}}.
\end{equation}
Now, knowing the $\psi(t)$ function, and using the relations defined in Eqs. \eqref{B1_eta}, \eqref{B1_h} and \eqref{B1_g} a straightforward computation
allow us to obtain analytical expression for the affine functions
\begin{dmath}
  \eta(t)  = 0
\end{dmath}
\begin{dmath}
  \label{sol_B1_h}
  h(t) =  -\frac{\gamma}{2\left(\alpha\beta + \gamma\right)\left(t - t_0\right)}
\end{dmath}
\begin{dmath}
  \label{sol_B1_g}
  g(t) =g_1\left(t - t_0\right)^{-\frac{\gamma}{\alpha\beta + \gamma}}
  -\frac{\alpha^2\beta\lambda^2 }{\alpha\beta + \gamma}\left(t - t_0\right)^{\frac{\alpha\beta-\gamma}{\alpha\beta + \gamma}}\end{dmath}
where $g_1$ is an integration constant.

\subsection*{Second branch}

The second branch imposes the restrictions $\mathcal{F}(g,h,\psi,\eta)  = 0$ and $g(t) = 0$, leading to 
\begin{dmath}
  \label{Feq_B2_1}
  \kappa B_3 \psi - F_3\psi^3= 0,
\end{dmath}
\begin{dmath}
  \label{Feq_B2_3}
  \kappa C_1\psi - \psi^3 \left(D_1 - 2D_2 + D_3\right) = 0 ,
\end{dmath}
\begin{dmath}
  \label{Feq_B2_4}
  B_3(\eta(h\psi -\dot{\psi}) -\psi\dot{\eta}) - 2B_4(\eta(-h\psi - \dot{\psi}) - \psi\dot{\eta}) + C_1(4h^2\psi + 2\psi\dot{h} -\ddot{\psi}) + D_6\eta^2\psi = 0, 
\end{dmath}
\begin{dmath}
  \label{Feq_B2_5}
  B_3\kappa\eta + C_1\kappa h - 3h\psi^2 \left(D_1 - 2D_2 + D_3\right) - 3F_3\eta\psi^2 = 0.
\end{dmath}
From Eq.~\eqref{Feq_B2_1} it is possible to find an expression for $\psi(t)$ in the form
\begin{equation}
  \label{B2_Psi}
  \psi(t) = \pm \sqrt{\frac{\kappa B_3}{F_3}} .
\end{equation}
Using the compatibility condition from Eq. \eqref{Feq_B2_3} (as long as $\kappa \neq 0$), leads to a relation between
the coupling constant
\begin{equation}
  C_1 F_3 = \left(D_1 - 2D_2 + D_3\right) B_3.
\end{equation}
Solving the algebraic expression for $C_1$\footnote{Assuming that $D_1 - 2D_2 + D_3 \neq 0, F_3 \neq 0, B_3 \neq 0$} and replacing Eq.~\eqref{B2_Psi} in Eq.~\eqref{Feq_B2_5} leads to the equation
\begin{equation}
  h\left(D_1 - 2D_2 + D_3\right) + F_3\eta = 0,
\end{equation} 
which establish a relation between the functions $h(t)$ and $\eta(t)$ as follow
\begin{equation}
  h(t) = - \eta(t)\left(\frac{F_3}{D_1 - 2D_2 + D_3}\right)
\end{equation}
Combining the above result along with Eq. \eqref{B2_Psi} turns Eq. \eqref{Feq_B2_4} into a first order differential equation of the form
\begin{equation}
  \dot{\eta} - \eta^2\left(\frac{D_6}{3B_3 - 2B_4} + \frac{F_3}{D_1 - 2D_2 + D_3}\right) = 0,
\end{equation}
whose solution is
\begin{widetext}
  \begin{equation}
    \eta(t) = \frac{\left(3B_3 - 2B_4\right)\left(D_1 - 2D_2 + D_3\right)}{\left(D_1 - 2D_2 + D_3\right)\left(\eta_0 \left(3B_3 - 2B_4\right) + tD_6\right) + tF_3\left(3B_3 - 2B_4\right)}
  \end{equation}
  where $\eta_0$ is an integration constant. Then, $h(t)$ is given by
  \begin{equation}
    h(t) = \frac{F_3\left(3B_3 - 2B_4\right)}{\left(D_1 - 2D_2 + D_3\right)\left(\eta_0\left(3B_3 - 2B_4\right) + tD_6\right) + tF_3\left(3B_3 - 2B_4\right)}
  \end{equation}
  It is important to note that the solutions mentioned were derived for this particular case $\kappa \neq 0$. 

  If $\kappa = 0$, then Eq.~\eqref{Feq_B2_1} tells us that $\psi(t) = 0$ completely solves the other equations, and the remaining functions $h(t)$ and $\eta(t)$ cannot be determined.
\end{widetext}

\subsection*{Third branch}

The restriction $\mathcal{G}(h,\psi,\eta) = 0$ and $\psi(t)=0$ imposes a strong constraint on Eqs.~\eqref{eq:field_a}--\eqref{eq:field_gamma_3}, 
condensing the equations down to a single second-order differential equation.
\begin{dmath}
  g\eta^2D_6 + 2B_4\eta \left(gh - \dot{g}\right) + B_3\eta\left(2\kappa + gh + \dot{g}\right) + 
  C_1\left(2h\left(\kappa + 2gh\right) + 2g\dot{h} - \ddot{g}\right) = 0.
\end{dmath}
The above differential equation has three unknown functions of time $h(t)$, $g(t)$, and $\eta(t)$ that cannot be solved without further restriction or by providing an ansatz for two functions.

\subsection*{Fourth branch}

The restrictions for this branch required that $g(t) = \psi(t) = 0$, therefore the set of field Eqs.~\eqref{eq:field_a}--\eqref{eq:field_gamma_3} is reduced to one algebraic equation for two unknown functions
\begin{equation}
  \kappa\left(hC_1 + B_3\eta\right) = 0.
\end{equation}
The system is underdetermined and cannot be solved analytically.

\subsection*{Special cases}

Although the first branch leads to an analytical solution without any assumption, Eq.\eqref{eq:integro_differential_phi} has special cases
that are given when $\alpha = 0$ and $\alpha\beta + \gamma = 0$. The first comes directly from the structure of Eq.\eqref{eq:integro_differential_phi} and
setting $\alpha = 0$ changes its structures, whereas the second restriction comes from the solution space where the relation $\alpha\beta + \gamma = 0$ 
appears in the denominator of the function $\psi(t)$. We will address both cases now; first, the former restriction simplifies
Eq.\eqref{eq:integro_differential_phi} to
\begin{dmath}
  \label{diff_eq_phi_alpha_0}
  \ddot{\phi}g_0\beta - \dot{\phi}^2 \gamma  = 0,
\end{dmath}
which can be solved exactly by
\begin{equation}
  \phi(t) = \phi_0 + \frac{\beta g_0}{\gamma}\log\left(\gamma(t-t_0)\right),
\end{equation}
where $\phi_0$ and $t_0$ are integration constants. From this, it is straightforward to recover the original function
\begin{align}
  \label{special_first_solution}
  \psi(t) & = \frac{\beta g_0}{2\gamma(t-t_0)} & h(t) & = -\frac{\gamma}{2\gamma(t-t_0)} \\
  g(t) & = \frac{g_1}{\gamma(t-t_0)}
\end{align}
where $g_1$ is another integration constant.

The latter constraint leads to the following,
\begin{dmath}
  \label{diff_eq_phi_alpha_beta_gamma_0}
  \ddot{\phi}\left(g_0 - \alpha\phi\right) + \dot{\phi}^2 \alpha  = 0,
\end{dmath}
whose solution is given by
\begin{equation}   
  \phi(t) = \phi_0 e^{\alpha\phi_1\left(t -t_0\right)} + \frac{g_0}{\alpha},
\end{equation}
where the integration constants are $\phi_0$ and $t_0$. From simple algebra we can recover the rest of the affine functions
\begin{align}
  \label{special_second_solution}
  \psi(t) & = \psi_0 e^{\frac{\left(t -t_0\right)}{\tau_0}} & h(t) & = \frac{\alpha }{2} \\
  g(t) & =\psi(t)\left(g_1 - \frac{\psi(t)}{\phi_1}\right)  \nonumber
\end{align}
where we have defined the constant $\psi_0 = \alpha\phi_0$ and $\tau_0^{-1} = \alpha\phi_1$, and $g_1$ is an integration constant.

\section{Cosmological (affine) perturbations}
\label{sec:coms-pert}

In order to build the method to analyse cosmological perturbations affine models of gravity we shall follow the same steps as in metric models of gravitation. Hence, let us first review the perturbation technique in these theories (additional details can be found in Refs. \cite{mukhanov92_theor_cosmol_pertur,weinberg08_cosmol,poisson14_gravit,dodelson20_moder}).

The algorithm for cosmological perturbations in metric gravities can be summarised as follows: (i) Take an isotropic and homogeneous (background) metric \(\bar{g}_{\mu \nu}\), solution of the cosmological field equations; (ii) Assumed the \emph{physical} metric \(g\) is a deformation of the background one,
\begin{equation}
  g_{\mu \nu} = \bar{g}_{\mu \nu} + h_{\mu \nu},
\end{equation}
with \(h_{\mu \nu}(\ll \bar{g}_{\mu\nu})\) representing the \emph{perturbation};\footnote{It should be noted as the metric \(g\) has to be symmetric, the perturbation field \(h\) is symmetric too.} (iii) Splitting of the perturbation in a \((3+1)\)-decomposition, e.g., \(h_{\mu\nu} \to \left\{ h_{t t}, h_{t i}, h_{i j} \right\}\); (iv) Decomposition of the fields into longitudinal and transversal components, following the Helmholtz algorithm, see for example Ref. \cite{straumann08_proof_decom_theor_symmet}; (v) Define the composite fields that are invariant under coordinate transformations and express the field equations in \emph{gauge invariant} form. The dynamics of the perturbation fields can be analysed after this \emph{gauge analysis}.

The pertubation technique could be implemented in our affine theory with a similar treatment. The results reported in the following are a summary of the method described in Ref. \cite{castillo-felisola24_model_cosmol_pertur_affin_gravit}.\footnote{A cosmological perturbation technique for metric-affine theories has been proposed in Ref. \cite{aoki23_cosmol_pertur_theor_metric}.}

As in the metric theory, we consider the generic physical connection as the sum of a background cosmological connection, \(\bar{\Gamma}_\mu{}^\lambda{}_\nu\), like the ones found in Sec.~\ref{sec:cosm-sol}, and a small perturbation \(C_\mu{}^\lambda{}_\nu (\ll \Gamma_\mu{}^\lambda{}_\nu)\), 
\begin{equation}
  \Gamma_{\mu}{}^{\lambda}{}_{\nu} = \bar{\Gamma}_{\mu}{}^{\lambda}{}_{\nu} + C_{\mu}{}^{\lambda}{}_{\nu}.
  \label{eq:connection-perturbation}
\end{equation}
As the perturbation \(C\) results from the difference between two connections, it is a tensor field, i.e. \(C_{\mu}{}^{\lambda}{}_{\nu} \in C^{\infty}(T\mathcal{M} \otimes^2 T^{ * }\mathcal{M})\).

From the point of view of group theory, the affine perturbation $C$ behaves as a third-order tensor under the (local) group $GL(4,\mathbb{R})$, and in four dimensions it has \(64\) components. However, in the cosmological scenario, where three dimensions are \emph{equivalent}, the components of the perturbation obtained after the \((3+1)\)-decomposition are tensors of \(GL(3,\mathbb{R})\).

The \((3+1)\)-decomposition of the perturbation \(C\) yields the following fields: \(C_{t}{}^{t}{}_{t}\), \(C_{t}{}^{i}{}_{t}\), \(C_{i}{}^{t}{}_{t}\), \(C_{t}{}^{t}{}_{i}\), \(C_{i}{}^{t}{}_{j}\), \(C_{t}{}^{i}{}_{j}\), \(C_{j}{}^{i}{}_{t}\), and \(C_{i}{}^{j}{}_{k}\), comprehending a scalar field, three vector fields, three \(2\)-tensor fields, and one \(3\)-tensor field, all defined under (local) \(GL(3,\mathbb{R})\) transformations. 

It is worthwhile to introduce the notation
\begin{equation}
  \begin{aligned}
    \Sigma_{\mu\nu\lambda} & = \frac{1}{2} \left( C_{\mu\nu\lambda} + C_{\lambda\nu\mu} \right), \\
    \Lambda_{\mu\nu\lambda} & = \frac{1}{2} \left( C_{\mu\nu\lambda} - C_{\lambda\nu\mu} \right),
  \end{aligned}
  \label{eq:perturbation-tensor-symmetry}
\end{equation}
where, as the isotropy and homogeneity symmetries can induce (spatial) metric structures \(s_{i j}\) emerging from the background connection fields, we can relate the perturbation \((3+1)\)-decomposition fields with lower indices as,\footnote{In Eq. \eqref{eq:relation-c-indices} the \emph{lowering} of the \(t\)-index should be considered an identification to simplify the presentation of the forthcoming expression, and not as the action of a metric.}
\begin{equation}
  C_{\mu t \nu} \equiv C_{\mu}{}^{t}{}_{\nu}, \quad C_{\mu i \nu} \equiv s_{i j} \,C_{\mu}{}^{j}{}_{\nu},
  \label{eq:relation-c-indices}
\end{equation}
considering that the components can be parameterised by unrelated objects in the former equation and they could be related by the \(s_{i j}\) object in the other equation. The contributions to the components of the symmetric part of the affine connection or to the torsion (antisymmetric part) are shown in the Tab. \ref{tab:number-components}.

\begin{table}[htbp]
  \caption{\label{tab:number-components} Number of contributions of each term in the scalar-vector-tensor decomposition, to the symmetric and anti-symmetric components of the affine perturbation.}
  \centering
  \begin{ruledtabular}
    \begin{tabular}{l|c|c}
      Terms & Symm. (\(\Sigma\)) & Anti-symm. (\(\Lambda\))\\
      \hline
      \(C_{ttt}\) & \(1\) & \(0\)\\
      \(C_{tit}\) & \(3\) & \(0\)\\
      \(C_{itt}, \, C_{tti}\) & \(3\) & \(3\)\\
      \(C_{itj}\) & \(6\) & \(3\)\\
      \(C_{ijt}, \, C_{tji}\) & \(9\) & \(9\)\\
      \(C_{ijk}\) & \(18\) & \(9\)\\
      \hline
      Total components: & \(40\) & \(24\)
    \end{tabular}
  \end{ruledtabular}
\end{table}

The existence of a metric tensor (induced by the cosmological symmetries imposed on the \emph{spatial} sector of our spacetime) allows an additional decomposition of the irreducible representations of $GL(3,\mathbb{R})$ into those of $SO(3,\mathbb{R})$ \cite{landsberg11_tensor,itin21_decom_third_order_const_tensor}.

For example, the generic \(GL(3,\mathbb{R})\) \(3\)-tensor \(C_{ijk}\) splits into irreducible representations as
\begin{equation}
  \begin{aligned}
    \ydiagram{1} \otimes \ydiagram{1} \otimes \ydiagram{1}
    & =
      \left( \ydiagram{2} \oplus \ydiagram{1,1} \right) \otimes \ydiagram{1}
    \\
    & =
      \underbrace{\ydiagram{3} \oplus \ydiagram{2,1}}_{\stackrel{18}{\mathrm{symmetric}}} \oplus \underbrace{\ydiagram{2,1} \oplus \ydiagram{1,1,1}}_{\stackrel{9}{\mathrm{anti-symmetric}}}
    \\
    & =
      10_{GL_{3}} \oplus 8_{GL_{3}} \oplus 8_{GL_{3}} \oplus 1_{GL_{3}},
  \end{aligned}
  \label{eq:young-decomposition-3-tensor}
\end{equation}
which decomposes onto \(SO(3,\mathbb{R})\) as follows
\begin{equation}
  \begin{aligned}
    10_{GL_{3}} & \to 7_{SO_{3}} \oplus 3_{SO_{3}}, \\
    8_{GL_{3}} & \to 5_{SO_{3}} \oplus 3_{SO_{3}}, \\
    1_{GL_{3}} & \to 1_{SO_{3}}.
  \end{aligned}
  \label{eq:decomposition-gl3-to-so3}
\end{equation}
A similar analysis decomposition can be made to the other components of the \(C\)-field in Tab. \ref{tab:number-components}. Tab. \ref{tab:so3-decomposition-summary} summarises the results of that decomposition.

\begin{table}[htbp]
  \caption{\label{tab:so3-decomposition-summary}Summary of irreducible representations of \(SO(3,\mathbb{R})\) obtained from the irreducible components of the affine perturbation tensor \(C\), in terms of its symmetric (\(\Sigma\)) and anti-symmetric (\(\Lambda\)) parts.}
  \centering
  \begin{ruledtabular}
    \begin{tabular}{c|c|c|c}
      Term & Components & \(GL(3,\mathbb{R})\) & \(SO(3,\mathbb{R})\)\\[0pt]
      \hline
      \(\Sigma_{ttt}\) & \(1_s\) & \(1_{GL_3}\) & \(1_{SO_3}\)\\[0pt]
      \(\Sigma_{tit}\) & \(3_s\) & \(3_{GL_3}\) & \(3_{SO_3}\)\\[0pt]
      \(\Sigma_{tti}\) & \(3_s\) & \(3_{GL_3}\) & \(3_{SO_3}\)\\[0pt]
      \(\Lambda_{tti}\) & \(3_a\) & \(3_{GL_3}\) & \(3_{SO_3}\)\\[0pt]
      \(\Sigma_{itj}\) & \(6_s\) & \(6_{GL_3}\) & \(5_{SO_3} \oplus 1_{SO_3}\)\\[0pt]
      \(\Lambda_{itj}\) & \(3_a\) & \(3_{GL_3}\) & \(3_{SO_3}\)\\[0pt]
      \(\Sigma_{tij}\) & \(9_s\) & \(6_{GL_3} \oplus 3_{GL_3}\) & \(5_{SO_3} \oplus 1_{SO_3} \oplus 3_{SO_3}\)\\[0pt]
      \(\Lambda_{tij}\) & \(9_a\) & \(6_{GL_3} \oplus 3_{GL_3}\) & \(5_{SO_3} \oplus 1_{SO_3} \oplus 3_{SO_3}\)\\[0pt]
      \(\Sigma_{ijk}\) & \(18_s\) & \(10_{GL_3} \oplus 8_{GL_3}\) & \(7_{SO_3} \oplus 3_{SO_3} \oplus 5_{SO_3} \oplus 3_{SO_3}\)\\[0pt]
      \(\Lambda_{ijk}\) & \(9_a\) & \(6_{GL_3} \oplus 3_{GL_3}\) & \(5_{SO_3} \oplus 1_{SO_3} \oplus 3_{SO_3}\)\\[0pt]
    \end{tabular}
  \end{ruledtabular}
\end{table}

In order to obtain the Helmholtz decomposition of the affine perturbation, it is convenient to see this process as the decomposition of representations of \(SO(3,\mathbb{R})\) into irreducible representations of \(SO(2,\mathbb{R})\), given that fixing a longitudinal direction still leaves a transverse plane of symmetry \cite{castillo-felisola24_model_cosmol_pertur_affin_gravit}. Therefore, all the objects in the Tab. \ref{tab:so3-decomposition-summary} can be decomposed into a trivial \emph{longitudinal} one-dimensional representation and a family of non-equivalent two-dimensional (irreducible) representations of \(SO(2,\mathbb{R})\) labelled by \emph{winding number}, as it is summarised in the Tab. \ref{tab:count-helmholtz-fields}.

\begin{table}[htbp]
  \caption{\label{tab:count-helmholtz-fields}Number of scalars (\(T_0\)), vectors (\(T_1\)), \(2\)-tensors (\(T_2\)) and \(3\)-tensors (\(T_3\)), obtained from the Helmholtz decomposition of the irreducible components of the affine connection.}
  \centering
  \begin{ruledtabular}
    \begin{tabular}{c|c|c|c|c}
      Component & \(T_0\) & \(T_1\) & \(T_2\) & \(T_3\)\\[0pt]
      \hline
      \(1_s\) & \(1\) &  &  & \\[0pt]
      \(3_s\) & \(1\) & \(1\) &  & \\[0pt]
      \(3_a\) & \(1\) & \(1\) &  & \\[0pt]
      \(6_s\) & \(2\) & \(1\) & \(1\) & \\[0pt]
      \(9_a\) & \(3\) & \(2\) & \(1\) & \\[0pt]
      \(18_s\) & \(4\) & \(4\) & \(2\) & \(1\)\\[0pt]
    \end{tabular}
  \end{ruledtabular}
\end{table}

With these considerations and taking account of the indexes symmetries in each object, we have the Helmholtz decomposition of the fields in the Tab. \ref{tab:so3-decomposition-summary} can be written as
\begin{dgroup}
  \begin{dmath}
    \label{eq:helmholtz-sttt}
    \Sigma_{ttt} = A,
  \end{dmath}
  \begin{dmath}
    \label{eq:helmholtz-stit}
    \Sigma_{tit} = D_i B + C_i,
  \end{dmath}
  \begin{dmath}
    \label{eq:helmholtz-stti}
    \Sigma_{tti} = D_i D + E_i,
  \end{dmath}
  \begin{dmath}
  \Lambda_{tti} = D_i \tilde{B} + \tilde{C}_i,
  \label{eq:helmholtz-ltti}
  \end{dmath}
  \begin{dmath}
    \label{eq:helmholtz-litj}
    \Lambda_{itj} = \sqrt{s} \, \mathfrak{e}_{ijk} \, s^{kl} (D_l \tilde{D} + \tilde{E}_l) ,
  \end{dmath}
  \begin{dmath}
    \label{eq:helmholtz-sitj}
    \Sigma_{itj} = \frac{s_{ij}}{3} F + \left( D_i D_j - \frac{s_{ij}}{3} D^2 \right) G + 2 D_{(i} H_{j)} + I_{ij},
  \end{dmath}
  \begin{dmath}
    \label{eq:helmholtz-stij}
    \Sigma_{tij} = \sqrt{s} \mathfrak{e}_{ijk} s^{kl} \left( D_l J + K_l \right) + \frac{s_{ij}}{3} L + \left( D_i D_j - \frac{s_{ij}}{3} D^2 \right) M + 2 D_{(i} N_{j)} + O_{ij},
  \end{dmath}
  \begin{dmath}
    \label{eq:helmholtz-ltij}
    \Lambda_{tij} = \sqrt{s} \mathfrak{e}_{ijk} s^{kl} \left( D_l \tilde{J} + \tilde{K}_l \right) + \frac{s_{ij}}{3} \tilde{L} + \left( D_i D_j - \frac{s_{ij}}{3} D^2 \right) \tilde{M} + 2 D_{(i} \tilde{N}_{j)} + \tilde{O}_{ij},
  \end{dmath}
  \begin{dmath}
    \Sigma_{ijk}
    =
    \frac{3}{5} \left( s_{(ij} D_{k)} P + s_{(ij} Q_{k)} \right) +
    \left( D_{(i} D_j D_{k)}
      - \frac{2}{5} D^2  s_{(ij} D_{k)}
      - \frac{1}{5} s_{(ij} D_{k)} D^2 \right) R
    + D_{(i} D_j S_{k)}
    - \frac{1}{5} D^2 s_{(ij} S_{k)}
    - \frac{1}{5} s_{(ij} D^m D_{k)} S_m
    + D_{(i} T_{jk)} + U_{ijk}
    + \frac{1}{2} \sqrt{s} s^{pq} \left( \mathfrak{e}_{ijp} \delta_k^r + \mathfrak{e}_{kjp} \delta_i^r \right)
    \left[ \left(D_q D_r - \frac{1}{3} s_{qr} D^2 \right) V + 2 D_{(q} W_{r)} + X_{qr} + \sqrt{s} \mathfrak{e}_{qrm} s^{mn} (D_n Y + Z_n) \right],
    \label{eq:helmholtz-sijk}
  \end{dmath}
  \begin{dmath}
    \Lambda_{ijk}
    = \sqrt{s} \mathfrak{e}_{ijk} \tilde{A} + \frac{1}{2} \sqrt{s} s^{pq} \left( 2 \mathfrak{e}_{ikp} \delta_j^r + \mathfrak{e}_{ijp} \delta_k^r - \mathfrak{e}_{kjp} \delta_i^r \right) 
    \left[ \left( D_q D_r - \frac{1}{3} s_{qr} D^2 \right) \tilde{V} + 2 D_{(q} \tilde{W}_{r)} + \tilde{X}_{qr} + \sqrt{s} \mathfrak{e}_{qrm} s^{mn} (D_n \tilde{Y} + \tilde{Z}_n) \right], 
    \label{eq:helmholtz-lijk}
  \end{dmath}
\end{dgroup}
where the tensor objects are symmetric, traceless and transverse. A summary of the fields obtained in the Helmholtz decomposition of $C$ can be found in the Tab. \ref{tab:list-fields}

\begin{table}[htbp]
  \centering
  \caption{Classification of the modes obtained after the Helmholtz decomposition of the perturbation \(C\)-field.}
  \label{tab:list-fields}
  \begin{ruledtabular}
    \begin{tabular}{l|l}
      Scalars & $A,B,D,F,G,L,M,P,R,Y,\tilde B,\tilde L,\tilde M,\tilde Y$\\[0.05cm]
      Pseudoscalars & $J,V,\tilde A, \tilde D,\tilde J, \tilde V$\\[0.05cm]
      Vectors & $C_i,E_i,H_i,N_i,Q_i,S_i,Z_i,\tilde C_i,\tilde N_i,\tilde Z_i$\\[0.05cm]
      Pseudovectors & $K_i,W_i,\tilde E_i,\tilde K_i,\tilde W_i$\\[0.05cm]
      2-tensor & $I_{ij},O_{ij},T_{ij},\tilde O_{ij}$ \\[0.05cm]
      Pseudo 2-tensor & $X_{ij},\tilde X_{ij}$ \\[0.05cm]
      3-tensor & $U_{ijk}$\\[0.05cm]
    \end{tabular}
  \end{ruledtabular}
\end{table}

Now, the infinitesimal \emph{gauge} transformation of the perturbation field \(C\) is given by the relation,
\begin{equation}
  \begin{aligned}
    \delta C_{\mu}{}^{\lambda}{}_{\nu}
    & = \pounds_{\xi} \Gamma_{\mu}{}^{\lambda}{}_{\nu} \\
    & = \xi^\sigma \mathcal{R}_{\sigma \mu}{}^{\lambda}{}_{\nu} + \nabla_\mu \nabla_\nu \xi^\lambda 
      - \nabla_\mu (\mathcal{T}_\nu{}^\lambda{}_\sigma \xi^\sigma ).
  \end{aligned}
  \label{eq:transformation-perturbation}
\end{equation}
Next, we (Helmholtz) decompose the spacial component of the generator of the transformation, \(\xi^i\), as
\begin{equation}
  \xi^i \to D^i \psi + \zeta^i \text{ where } D_i \zeta^i = 0,
  \label{eq:split-generator}
\end{equation}
which allows us to obtain the transformation rules of the fields under coordinates infinitesimal transformations:
\begin{align}
  \label{eq:dA}
  \delta A & = \ddot{\xi}^t,\\
  \label{eq:dB}
  \delta B & = \ddot{\psi} + 2 h \dot{\psi}, \\
  \label{eq:dC}
  \delta C^i & = \ddot{\zeta}^{i} + 2 h \dot{\zeta}^{i},\\
  \label{eq:dJK}
  \delta \left( D_l J + K_l \right) & = - \frac{1}{2 \sqrt{s}} s_{l k} \epsilon^{k i j} D_i \zeta_j, \quad \text{and} \\
  \label{eq:dW}
  \delta W_i & = \frac{1}{3} \sqrt{s} \epsilon_{ijk} D^j \zeta^k.
\end{align}

From this analysis, we can see that \(24\) of the \(64\) components of $C$ are invariant under infinitesimal coordinate transformations, and the following conditions are satisfied:
\begin{itemize}
\item As $\xi^t$ affects the fields $(G,P,Y,\tilde{A},\tilde{B},\tilde{D},\tilde{Y})$,
\item  As $\psi$ affects the fields $(F,G,R,\tilde{A},\tilde{V})$, and
\item  As $\zeta^i$ affects the fields $(H_i,S_i,W_i,\tilde{W}_i,D_i J + K_i)$.
\end{itemize}
Four out of the \(64\) components of the perturbation can be disregarded with a particular choice of \(\xi^t\), \(\psi\) and \(\zeta^i\) (or \emph{gauge}) since they can be associated to a particular coordinate frame. Therefore, only the remaining \(60\) components can be associated with gravitational interactions. Under these circumstances, a combinatorial factor allows us to count \(165\) different possible gauge choices.    

\section{Perspectives of the models\label{sec:perspectives}}

The content in the preceding sections has been extensively explored by our research group, and the presented results today have a solid ground and well-understood interpretation.

The purpose of this section is to overview some additional edges, which we have explored in a yet non-exhaustive way. The content can be seen as a compendium of preliminary results of our ongoing investigations.

\subsection{Metric independence of the model\label{sec:background-indep}}

While it is not strictly necessary to use a metric for achieving diffeomorphism invariance when discussing gravitational phenomena, it frequently offers a more intuitive framework. In the polynomial affine model of gravity, the symmetric connection can be described using a metric, although this approach is subject to gauge symmetries that relate different metric choices through nonmetricity transformations.\footnote{Generically the independence of the affine connection under a change of metric would require also a transformation of the torsion tensor.}

To make this clearer, let us break down the connection $\Gamma_{\mu}{}^{\lambda}{}_{\nu}$ into two components: the Riemannian part corresponding to a reference metric \(g_{\mu\nu}\) and the non-Riemannian part related to nonmetricity
\begin{equation}
  \label{eq:conn-decomp-lc-y-s}
  \Gamma_{\mu}{}^{\lambda}{}_{\nu} = \frac{1}{2} g^{\lambda\kappa} (\partial_{\mu} g_{\nu\kappa} + \partial_{\nu} g_{\mu\kappa} - \partial_{\kappa} g_{\mu\nu})
  +\hat{Y}^\lambda{}_{\mu\nu} 
  + \hat{S}^\lambda{}_{\mu\nu},
\end{equation}
where \(\hat{Y}_{\lambda\mu\nu} = \frac{1}{2} (\hat{Y}_{[\lambda\mu]\nu} + \hat{Y}_{[\lambda\nu]\mu})\)
and \(\hat{S}_{\lambda\mu\nu} = \hat{S}_{(\lambda\mu\nu)}\).

In terms of the connection, we have:
\begin{equation}
  \label{eq:cov-der-g}
  \nabla^{\Gamma}_\lambda g_{\mu\nu} = 2\hat{Y}_{\lambda\mu\nu} + 2\hat{S}_{\lambda\mu\nu}.
\end{equation}
For the connection \(\Gamma_{\mu}{}^{\lambda}{}_{\nu}\) to remain invariant under infinitesimal transformations of the metric:
\begin{equation}
  \label{eq:metric-transformation}
  g_{\mu\nu} \to g'_{\mu\nu} = g_{\mu\nu} + s_{\mu\nu},
\end{equation}
where \(s_{\mu\nu}\) is symmetric, the nonmetricity components must transform as follows:
\begin{equation}
  \hat{Y}_{\lambda\mu\nu} \to \hat{Y}'_{\lambda\mu\nu} = \hat{Y}_{\lambda\mu\nu} + \frac{2}{3} (\nabla^\Gamma_{[\lambda}s^{\ }_{\mu]\nu} + \nabla^\Gamma_{[\lambda}s^{\ }_{\nu]\mu})
\end{equation}
and
\begin{equation}
  \hat{S}_{\lambda\mu\nu} \to \hat{S}'_{\lambda\mu\nu} = \hat{S}_{\lambda\mu\nu} - \frac{1}{2} \nabla^\Gamma_{(\lambda}s^{\ }_{\mu\nu)}.
\end{equation}

Although the polynomial affine model of gravity is generally invariant under changes in the metric, using a metric simplifies the comparison with Einstein's gravity and helps in understanding the solutions of the polynomial affine model. In this sense, there is an absolute sense in which we can affirm the total background independence of the model, where by background we choose to refer to the metric on which we perform an expansion of the field equations.\footnote{In fact, making a functional derivative of the action with respect to the metric is going to return a trivial output.}

On many symmetric subspaces of solutions, the description of the connection can be more familiarly described by a metric instead of using non-metricity. For instance, in cosmological and spherically symmetric spaces. To show this, we are going to decompose the non-metricity into its traceless and trace parts
\begin{dmath}
  \label{nonmetricityConnection}
  \Gamma_\mu{}^\lambda{}_\nu = \Gamma_\mu{}^\lambda{}_\nu(g) + S{}^\lambda{}_{\mu\nu} + Y{}^\lambda{}_{\mu\nu} + V^\lambda g_{\mu\nu} + 2 W_{(\mu} \delta^\lambda_{\nu)}, 
\end{dmath} 

The introduction of a metric relates the representation of the connection to a specific choice of the full set of geodesics of the spacetime regardless of the full set of autoparallels associated to the specific connection. Moreover, it can be shown that in the cosmological scenario there is a metric whose geodesics are the autoparallels of the connection, according to the criteria in Refs. \cite{ehlers12_repub_of,matveev13_criter_compat_confor_projec_struc}.

\subsubsection*{Autoparallels and geodesics in cosmology}

In order to represent the split between the dimensions of homogeneous space and time, \emph{in this section we use greek letters for the full space and latin letters from the beginning of the alphabet for spacial coordinates}, such that \(x^\mu \to (t,x^a)\).

We propose the cosmological metric  \(g_{\mu\nu} = \diag(-N^2,a^2 s_{ab})\), where \(s_{ab} = \diag((1-\kappa r^2)^{-1},r^2, r^2 \sin^2(\theta))\), with \(\kappa=-1,0, \text{ or }1\). 
The split of the connection reveals that the only symmetric components of the connection are
\begin{align}
  \Gamma_{0}{}^{0}{}_{0} & = J = \frac{\dot N}{N}-N^2V^0+2W_0,\\
  \Gamma_{a}{}^{0}{}_{b} & = g s_{ab} = \Big(\frac{a\dot a}{N^2}+a^2V^0\Big)s_{ab},\\
  \Gamma_{0}{}^{a}{}_{b} & = h \delta^a_b = \Big(\frac{\dot a}{a}+W_0\Big)\delta^a_b,\\
  \Gamma_{a}{}^{c}{}_{b} & = \gamma_{a}{}^{c}{}_{b}(s),
\end{align}
where $\gamma_{a}{}^{c}{}_{b}(s_{ab})$ is the  connection $\nabla^\gamma_c s_{ab}=0$. As the reader may have noticed, there are 3 equations to relate $(J,g,h)$ to $(N,a,W_0,V^0)$, but there is some ambiguity left to the reader's choice to find unique solutions to these equations. An additional condition may impose that geodesics are also autoparallels, thus we set $V^0=0$.

The system of equations for $(N,a,W_0)$ 
\begin{equation}
  \label{eq:3}
  g=\frac{a\dot a}{N^2},\
  h=\frac{\dot a}{a}+W_0 \text{, and }
  J=\frac{\dot N}{N}+2W_0,
\end{equation}
can be used to obtain \(2h-J=2\frac{\dot a}{a}-\frac{\dot N}{N}\), whose solution is
\begin{equation}
  \frac{a^2}{N}=\frac{a^2_0}{N_0} \exp\Big(\int_0^t \textrm{d}t' \, (2h-J)\Big).
\end{equation}
We also get \(g\Big(\frac{a^2}{N}\Big)^{-2}=\frac{\dot a}{a^3}\), from which we obtain
\begin{equation}
  a^2 = \frac{a_0^2}{1 - \dfrac{N^2_0}{2a^2_0} \int_0^t \textrm{d}t' \, g \exp\Big(-2\int_0^{t'} \textrm{d}t'' \, (2h-J)\Big) }
\end{equation}
that can be used, together with the previous solution, to obtain $N$. Finally, we obtain $W_0$ using any of the equations where we find it. 

One concludes that any set of specific solutions in polynomial affine model of gravity cosmology can be expressed in terms of a metric whose geodesics are the autoparallels and a vector which is a combination of projective transformations of the connection and the Weyl connection.

\subsubsection*{Autoparallels and geodesics in spherically symmetric spacetimes}

In this section, we propose a splitting of the indices, corresponding to the coordinates \(x^{\mu} = (t,r,\theta,\varphi)\), as \(\mu=(a,i)\), where the letters of the initial part of the alphabet take values \(a = (t,r)\) while mid-alphabetic letters correspond to the angular coordinates \(i = (\theta,\varphi)\). This will allow us to establish a naming convention that is useful when we further restrict the model.

From a general decomposition of the connection such that it is parity invariant and spherically symmetric, using the metric
\begin{equation}
  \label{eq:spheical-metric}
  g_{\mu\nu} = \delta_\mu^b \delta_\nu^b q_{ab} + \delta_\mu^i \delta_\nu^j r^2 s_{ij},
\end{equation}
where
\begin{equation}
  \label{eq:two-dimensional-metric}
  s_{ij}=\begin{pmatrix}
    1&0\\
    0&\sin^2\theta
  \end{pmatrix}.
\end{equation}
As in two dimensions, some of the nonmetricity components in Eq. \eqref{nonmetricityConnection} can be expressed in terms of the metric connection, thus, we can set the traceless parts of the nonmetricity to
$$S_{\lambda\mu\nu}=\Bigg(\delta_{(\lambda}^a\delta_{\mu}^b\delta_{\nu)}^c  q_{bc}-4\delta_{(\lambda}^a\delta_{\mu}^i\delta_{\nu)}^j r^2 s_{ij}\Bigg)S_a$$ and $$Y_{\lambda\mu\nu}=\Bigg(2\delta_{\lambda}^{[a}\delta_{(\mu}^{b]}\delta_{\nu)}^c  q_{bc}-\delta_{\lambda}^{[a}\delta_{(\mu}^{i]}\delta_{\nu)}^j r^2 s_{ij}\Bigg)Y_a,$$
while $W_\mu=\delta_\mu^a W_a$ and $V^\mu=\delta^ \mu_a V^a$.

\paragraph{Static black hole-like connections:} Stationary black holes can be studied by imposing time independent variables, static solutions additionally have time reversal symmetry, thus
\begin{equation}
  \label{eq:static-bh-qab}
  q_{ab}=
  \begin{pmatrix}
    - F(r) G(r) & 0\\
    0 & \frac{1}{F(r)}
  \end{pmatrix},
\end{equation}
and
\begin{equation}
  \begin{aligned}
    S_{a} & = S(r) \delta_{a}^r, & Y_{a} & = Y(r) \delta_{a}^r, \\ 
    V^{\lambda} & = V(r) \delta_r^\lambda, & W_{\lambda} & = W(r) \delta^r_\lambda.
  \end{aligned}
\end{equation}

From the general autoparallel equation, we get
\begin{dmath}
  \frac{DU^\mu}{D\tau} + U^\mu W_\lambda U^\lambda + U^2V^\mu + S^\mu{}_{\lambda\kappa} U^\lambda U^\kappa + Y^\mu{}_{\lambda\kappa} U^\lambda U^\kappa = 0.
\end{dmath} 
Here, the presence of $S^\mu{}_{\lambda\kappa}U^\lambda U^\kappa+Y^\mu{}_{\lambda\kappa}U^\lambda U^\kappa$ makes it improbable for us to be able to choose a metric whose geodesics coincide with autoparallels. We can concentrate our efforts on studying radial geodesics. In such a case, transformations of the metric make it possible to rewrite these equations into geodesic equations with a different choice of the affine parameter. Thus, using radial geodesics, black holes can be defined as regions of spacetime where radial null geodesics (the paths followed by massless particles such as photons) that enter the region cannot escape back to infinity.

\subsection{Emerging metrics in the space of solutions}
\label{sec:metrics}

In order to provide a physical interpretation of the solutions of the field equations, we explore the descendent metric structures that emerge from the fundamental fields of the connection. Hence, it is convenient to revisit the definition of a metric tensor.

\begin{definition}
  Let \(\mathcal{M}\) be an \(n\)-dimensional smooth manifold. A section \(g \in C^{\infty}(T^{*}\mathcal{M} \otimes T^{*}\mathcal{M})\) is said to be \emph{metric tensor field} in \(\mathcal{M}\) if its action on vector fields \(X,Y,Z \in C^{\infty}(T\mathcal{M})\) satisfies: (i) it is symmetric, \(g(X,Y) = g(Y,X)\); (ii) it is \(C^{\infty}(\mathcal{M})\)-bilinear, \(g(f_1 X + f_2 Y, Z) = f_1 \, g(X,Z) + f_2 \, g(Y,Z)\); (iii) it is nondegenerate, i.e., if at a point \(g(X,Y)_p = 0\) for all \(Y_{p}\) implies \(X_p = 0\).
\end{definition}

All three points must always be satisfied simultaneously in order to have a proper metric tensor, however, the last point plays a crucial role, because it allows us to ensure the existence of the inverse tensor of \(g_{\mu\nu}\) denoted by \(g^{\mu\nu}\). The metric structure allows us to provide a notion of distance.

In gravitational physics the signature of the metric tensor is mainly required to be Lorentzian (\(\operatorname{sig}(g) = \pm(n-2)\)) or Euclidean (\(\operatorname{sig}(g) = \pm n\)). In the former there is a notion of light-cone, and causal structure, while the latter is useful to analyse soliton configurations.

In the literature there are examples of \emph{derived} metric tensors in affinely connected manifolds \cite{eisenhart27_non_rieman,schouten13_ricci,poplawski14_affin_theor_gravit,castillo-felisola20_emerg_metric_geodes_analy_cosmol,castillo-felisola23_corrig}, but these emergent metrics are defined on the space of solutions of the field equations of the gravitational model.

A first example of emergent metric is the symmetrised Ricci tensor \(\mathcal{R}_{(\mu\nu)}\), defined by the contraction of the Riemann curvature tensor
\begin{equation}
  \label{chain}
  \mathcal{R}_{\beta\delta} = \mathcal{R}_{\alpha\beta}{}^{\alpha}{}_{\delta} .
\end{equation}
A second metric structure, comes from the contraction of the product of two torsion tensors. This idea was first introduced by Poplawski, and the metric structure is defined as follows,\footnote{Note that one could restrict to the traceless part of the torsion, and obtain another metric as the trace of the squared \(\mathcal{B}\)-tensor.}
\begin{equation}
  \mathcal{P}_{\alpha\delta} = \left(\mathcal{B}_{\alpha}{}^{\beta}{}_{\gamma} + \delta^{\beta}_{[\gamma}\mathcal{A}_{\alpha]}\right)\left(\mathcal{B}_{\beta}{}^{\gamma}{}_{\delta} + \delta^{\gamma}_{[\delta}\mathcal{A}_{\beta]}\right).
\end{equation}
Finally, the third candidate of metric tensor, comes from the covariant derivative (symmetrised) of the vectorial part of the torsion tensor, defined by the \(\mathcal{A}\) field as follow
\begin{equation}
  \mathrm{A}_{\mu\nu} = \nabla_{(\mu}\mathcal{A}_{\nu)}.
\end{equation}

Using the cosmological ansatz for the symmetric part of the connection, defined in Sec. \ref{sec:cosm-conn}, the non-trivial
components of the symmetrised Ricci tensor are
\begin{align}
  \label{Ricci_tensor}
  \mathcal{R}_{tt} & = - 3 ( \dot{h} + h^2 ), & \mathcal{R}_{ij} & = {\dot{g} + gh + 2\kappa} s_{ij}.
\end{align}
The Poplawski metric is computed using the ansatz for the antisymmetric part of the affine connection (see Sec. \ref{sec:cosm-conn})
\begin{align}
  \label{Pop_tensor}
  \mathcal{P}_{tt} & = \eta^2, & \mathcal{P}_{ij} & = - {2\psi^2} s_{ij}.
\end{align}
The final emergent metric coming from $\mathrm{A}_{\mu\nu}$ is
\begin{align}
  \mathrm{A}_{tt} & = \dot{\eta}, & \mathrm{A}_{ij} & = {\eta g} s_{ij}.
\end{align}

Notice that, from the definition of a metric tensor, we have provided three different metrics candidates that do not match to each other, and that is, because they are built from different/combined parts of the affine connection. Moreover, from the definition, since the tensor must be invertible, and in the space of solutions we have found that $\eta (t) = 0$, then we can discard $\mathcal{P}_{\mu\nu}$ and $\mathrm{A}_{\mu\nu}$ as suitable candidates (due to degeneracy). For that reason, the only viable candidate is the Ricci tensor.

If this Ricci tensor can be identified with a homogeneous and isotropic metric, we should have $g_{tt} = \mathcal{R}_{tt}/\mathcal{R}_0 $ and \(a^2(t) s_{ij} = \mathcal{R}_{ij}/\mathcal{R}_0\), and $\mathcal{R}_0$ is some constant with curvature dimensions used to get a dimensionless metric. From the above analysis it can be deduced that the signature of the metric would depend on the explicit form of the functions \(g\) and \(h\), and the values of the coupling constants of the model.

\subsection{Coupling scalar matter}
\label{sec:scalar}

The simplest (fundamental) type of matter to couple with gravity is a scalar field, \(\phi\). An essential term for the matter field is \emph{kinetic energy}. In the absence of a metric, one can only write the term \(\nabla_{\mu}\phi \nabla_{\nu}\phi\),\footnote{In the literature on Horndeski gravity (see for example Ref. \cite{horndeski24_years_hornd_gravit}), this tensor is usually denoted by \(X_{\mu\nu}\).} so we need a symmetric \(\binom{2}{0}\)-tensor whose transvection with the former yields a scalar.

Due to the nature of the fields, we have to include the skew-symmetric tensor density, \(\mathfrak{E}\), to obtain the aforementioned tensor. Hence, the resulting quantity would be a symmetric \(\binom{2}{0}\)-tensor density. Using the analysis of the indices structure explained in Sec. \ref{sec:model}, it is easily demonstrable that the expected tensor has the form
\begin{dmath}
  \mathfrak{g}^{\mu\nu} = \alpha \, \nabla_{\lambda} \mathcal{B}_{\rho}{}^{(\mu}{}_{\sigma} \mathfrak{E}^{\nu)\lambda\rho\sigma}
  + \beta \, \mathcal{A}_{\lambda} \mathcal{B}_{\rho}{}^{(\mu}{}_{\sigma} \mathfrak{E}^{\nu)\lambda\rho\sigma}
  + \gamma \, \mathcal{B}_{\kappa}{}^{\mu}{}_{\lambda} \mathcal{B}_{\rho}{}^{\nu}{}_{\sigma} \mathfrak{E}^{\kappa\lambda\rho\sigma},
  \label{eq:inv_metr_dens}
\end{dmath}
where the parameters \(\alpha\), \(\beta\) and \(\gamma\) are arbitrary constants.

The action of the scalar field would be a \emph{kinetic term},
\begin{equation}
  S_{\phi} = - \int \textrm{d}^4 x \, \mathfrak{g}^{\mu\nu} \nabla_{\mu} \phi \nabla_{\nu} \phi.
  \label{eq:scalar-action}
\end{equation}

Although one could add the term in Eq. \eqref{eq:scalar-action} with the complete action of polynomial affine gravity, Eq. \eqref{eq:new-action}, it is interesting that the restriction to the torsion-less sector is well defined, giving us the opportunity to focus our attention to a simplified model.

The only two terms which would contribute to the field equations in the torsion-free sector are those linear in the \(\mathcal{B}\)-field,
\begin{equation}
  S = \int \textrm{d}^4 x \, \mathfrak{E}^{\alpha\beta\gamma\delta} \left( {\mathcal{R}}_{\mu\alpha}{}^{\mu}{}_{\nu} - C \nabla_{\alpha} \phi \nabla_{\nu} \phi \right) \nabla_{\beta} \mathcal{B}_{\gamma}{}^{\nu}{}_{\delta}.
  \label{eq:simplified-scalar-pag-action}
\end{equation}
The variation of the action in Eq. \eqref{eq:simplified-scalar-pag-action} with respect to the \(\mathcal{B}\)-field yields the field equations
\begin{equation}
  \nabla_{\mu} \left( \mathcal{R}_{\alpha\lambda} - C \nabla_{\alpha} \phi \nabla_{\lambda} \phi \right) \mathfrak{E}^{\mu\nu\rho\alpha} = 0,
  \label{eq:simplified-scalar-pag-feqs}
\end{equation}
while the variation with respect to either the symmetric connection (\(\Gamma\)) or the scalar field (\(\phi\)) turn into identities in the torsion-free sector, \(\mathcal{B} \to 0\).

The coupling of the scalar field via its kinetic term is not enough to introduce non-trivial effects on the simplified, torsion-free, sector of the polynomial affine model of gravity, and does not allow self-interaction of the scalar field. Since the scalar field does not have an index structure, it would be possible to include \emph{non-minimal} couplings, e.g., multiplying the terms of the action by functions of the scalar field.

An interesting proposal considered by Kijowski in Ref. \cite{kijowski07_univer_affin_formul_gener_relat}, and implemented further by Azri and collaborators in Refs. \cite{azri17_affin_inflat,azri18_induc_affin_inflat,azri18_cosmol_implic_affin_gravit}, is the \emph{scaling} of the action terms by the function \(\mathcal{V}(\phi)\). This scaling---which might be think as an analogous to the substitution, \(g_{\mu\nu}(x) \to g_{\mu\nu}(\phi)\), used to obtain non-linear \(\sigma\)-models from the standard kinetic term of the scalar field action---, is implemented through the substitution
\begin{equation}
  \mathfrak{E}^{\alpha\beta\gamma\delta} \mapsto \frac{\mathfrak{E}^{\alpha\beta\gamma\delta}}{\mathcal{V}(\phi)}.
  \label{eq:volume-scaling}
\end{equation}
The consequence of the \emph{scaling} in Eq. \eqref{eq:volume-scaling}, is the inclusion of a scalar field self-interaction potential \(\mathcal{V}(\phi)\) in the model field equations.\footnote{Notice that in general one could \emph{scale} each term of the action by different functions of the scalar field.}

In our simplified coupled model, Eq. \eqref{eq:simplified-scalar-pag-action} with the scaling included, the field equations are just a modification of the ones derived in the torsion-free sector,
\begin{equation}
  \nabla_{[\lambda} \mathcal{S}_{\mu]\nu} = 0,
  \label{eq:scalar-pag-feq}
\end{equation}
with
\begin{equation}
  \mathcal{S}_{\mu\nu} = \frac{\mathcal{R}_{\mu\nu} - C \, \nabla_{\mu} \phi \nabla_{\nu} \phi}{\mathcal{V}(\phi)}.
  \label{eq:definition-S-tensor}
\end{equation}

In particular, if we restrict our focus to cosmological scenarios, the compatibility of the scalar field with the cosmological principle requires that \(\phi = \phi(t)\), and therefore the \(\mathcal{S}\)-tensor (up to the scaling) differs from the Ricci tensor by a \emph{deformation} of its \((t,t)\)-component.

The solutions to the field equations in Eq. \eqref{eq:scalar-pag-feq} can be classified into three types: (i) vanishing \(\mathcal{S}\)-tensor, \(\mathcal{S}_{\mu\nu} = 0\); (ii) covariantly constant \(\mathcal{S}\)-tensor, \(\nabla_{\lambda}\mathcal{S}_{\mu\nu} = 0\), and; (iii) the tensor \(\mathcal{S}\) is a Codazzi tensor, \(\nabla_{[\lambda}\mathcal{S}_{\mu]\nu} = 0\). 

The field equations in the \(\mathcal{S}\)-flat cosmological scenario are
\begin{equation}
  \begin{aligned}
    \dot{h} + h^2 + \frac{C}{3} (\dot{\phi})^2 & = 0, \\
    \dot{g} + g h + 2 \kappa & = 0,
  \end{aligned}
  \label{eq:S-flat-equations}
\end{equation}
whose solutions are parameterised by the \(h\)-function,
\begin{equation}
  \begin{aligned}
    \phi(t) & = \phi_0 \pm \sqrt{- \frac{3}{C}} \int \textrm{d}t \, \sqrt{\dot{h} + h^2}, \\
    g(t) & = e^{- \int \textrm{d}t \, h} \left( g_0 - 2 \kappa \int \textrm{d}t \, e^{\int \textrm{d}t \, h} \right).
  \end{aligned}
  \label{eq:S-flat-solution}
\end{equation}
Accordingly with the interpretation of the component of the affine connection \(\Gamma_{t}{}^{i}{}_{j} = h \delta^i_j\), from to the autoparallel equation, the \(h\)-function takes the role of the Hubble function.\footnote{In metric Friedmann--Robertson--Walker scenario, the Hubble function is expressed in terms of the scale factor as \(H(t) = \dot{a}/a\).}

Although the Eqs. \eqref{eq:S-flat-solution} lake of predictability due to the arbitrariness of the \(h\)-function, if one could determine the Hubble function from the observations (e.g. from the latest observations of the DESI Collaboration \cite{collaboration24_desi_vi,roy24_dynam_dark_energ_light}), the cosmological model would be completely determined.

In Ref. \cite{castillo-felisola23_inflat_scenar_effec_polyn}, we shown that the case of parallel \(\mathcal{S}\)-tensor is (somehow) equivalent to the minimally coupled Einstein--Klein--Gordon system. The equivalence is ensured by the existence of a symmetric, nondegenerate, and parallel \(\binom{0}{2}\)-tensor, say \(g_{\mu\nu}\). Hence, the field equations of the polynomial affine model of gravity might be written as follows
\begin{dgroup}
  \begin{dmath}
    {\mathcal{R}}_{\mu\nu} - \frac{1}{2} \mathcal{R} g_{\mu\nu} = C \left( \partial_{\mu} \phi \partial_{\nu} \phi - \frac{1}{2} g_{\mu\nu} (\partial \phi)^2 \right) - \Sigma \mathcal{V}(\phi) g_{\mu\nu},
  \end{dmath}
  \begin{dmath}
    C \nabla^{\mu} \nabla_{\mu} \phi = \Sigma \mathcal{V}^{\prime}(\phi),
  \end{dmath}
  \label{eq:parallel-S-equations}
\end{dgroup}
with \(\Sigma \in \mathbb{R}\) an arbitrary constant and \(\nabla^{\mu} = g^{\mu\nu} \nabla_{\nu}\). A key point to be highlighted is that the field equation for the scalar field is a consequence of the symmetries of the system (not of a least action principle), following the method proposed in Ref. \cite{bekenstein15_is_princ_least_action}.

In the case of \(\mathcal{S}\) being a Codazzi tensor, there is a single field equation, say
\begin{dmath}
  C \mathcal{V}(\phi) g \dot{\phi}^2
  + \mathcal{V}(\phi) (4 g h^2 + 2 \kappa h + 2 g \dot{h} - \ddot{g})
  + \mathcal{V}'(\phi) \dot{\phi} (g h + 2 \kappa + \dot{g}) = 0.
  \label{eq:codazzi-S}
\end{dmath}
Therefore, one has to complement the Eq. \eqref{eq:codazzi-S} with additional equations to search for solutions.\footnote{This strategy is similar to the utilised in thermodynamics, where one supplements the thermodynamic equations with the equation of states.}

\subsection{Toward the spherically symmetric solutions}
\label{sec:black-holes}

In this section we shall inquire the space of solutions of the field equations of the polynomial affine model of gravity, using the static spherical connection starting from the ansatz found in Sec.~\ref{sec:spher-conn}. The requirement of invariance under the action of the time and \(\varphi\) reversal operators (\(T\) and \(P\) respectively) eliminate seven of the functions characterising the stationary par-spherical connection, see Eq. \eqref{eq:spherical-gamma}.

With all this considerations we rename the nonzero components of the static spherical symmetric connection in Eq.~\eqref{eq:spherical-gamma}, as follows
\begin{equation}
  \begin{aligned}
    F_{001}(t,r) & = a(r)  \\
    F_{010}(t,r) & = b(r)  \\
    F_{111}(t,r) & = c(r) \\
    F_{212}(t,r) & = f(r) \\
    F_{122}(t,r) & = g(r)
  \end{aligned}
  \label{spheric_simplified_conection}
\end{equation}

The system of equations will categorise as before in three groups: Ricci flat, parallel Ricci, and Ricci as a Codazzi tensor.

In the Ricci flat case, \(\mathcal{R}_{\mu\nu} = 0\), the corresponding set of equations is
\begin{equation}
  \begin{aligned}
    -a b + b' + b \left( c + 2 g \right) & = 0, \\
    -a' + a c - a^2 + 2 c g - 2 g' - 2 g^2 & = 0, \\
    f \left( a + c \right) + f' + 1 & = 0.
  \end{aligned}
  \label{eq:sph-ricci-flat}
\end{equation}

For the case where the Ricci tensor is parallel, \(\nabla_{\lambda} \mathcal{R}_{\mu\nu} = 0\), the field equations are as follows
\begin{dmath}
  b \left( -a' + c' + 2 g' \right) - a \left( 3 b' + 2 b \left( c + 2 g \right) \right) + 2 a^2 b + b'' + b' \left( c + 2 g \right) = 0,
  \label{prs1}
\end{dmath}
\begin{dmath}
  b \left( a' + 2 g \left( g - c \right) + 2 g' \right) - a \left( b' + 2 b \left( c + g \right) \right) + 2 a^2 b = 0,
  \label{prs2}
\end{dmath}
\begin{dmath}
  -a'' + a \left( c' - 2 \left( a' + c^2 \right) \right) + c \left( 3 a' + 6 g' + 4 g^2 \right) + 2 a^2 c + 2 g \left( c' - 2 g' \right) - 4 c^2 g - 2 g'' = 0,
  \label{prs3}
\end{dmath}
\begin{dmath}
  f \left( a' + 2 g' \right) - a f \left( c + g \right) + a^2 f - g \left( 3 c f + f' + 1 \right) + 2 f g^2 = 0,
  \label{prs4}
\end{dmath}
\begin{dmath}
  -f \left( a' + c' \right) + 2 g \left( f \left( a + c \right) + f' + 1 \right) - (a + c) f' - f'' = 0.
  \label{prs5}
\end{dmath}

Finally, in the case where the Ricci is a Codazzi tensor, \(\nabla_{[\lambda} \mathcal{R}_{\mu]\nu} = 0\), the system of equations reduces to
\begin{dmath}
  b \left( -2 \left( a' + g^2 \right) + c' + 2 c g \right) - 2 a \left( b' + b g \right) + b'' + b' \left( c + 2 g \right) = 0,
  \label{cod_1}
\end{dmath}
\begin{dmath}
  -f' \left( a + c - g \right) + f g \left( a - c \right) + a f \left( a - c \right) - f c' - f'' + 2 f g' + 2 f g^2 + g = 0.
  \label{cod_2}
\end{dmath}

From the above, we observe that, in general, the number of field equations is not enough to solve the whole set of unknown functions. Interestingly, the case of parallel Ricci is the case which could allow us to solve all the unknowns, but an analysis like the one presented in Ref. \cite{castillo-felisola23_inflat_scenar_effec_polyn}, shows that this would be a spherically symmetric metric tensor (if nondegenerated).

Hence, it is interesting to analyse the case where the connection is the Levi-Civita connection associated to a symmetrically spherical metric.

Let us then consider the spherically symmetric line element,
\begin{equation}
  ds^2 = -p(r) dt^2 + \frac{dr^2}{p(r)} + d\Omega^2.
  \label{esfericmetric}
\end{equation}
Therefore, the five functions in Eq. \eqref{spheric_simplified_conection} reduces to solve only for the \(p\)-function,
\begin{dmath}
  \label{60}
  \begin{aligned}
    a & = \frac{1}{2} \frac{p^{\prime}}{p}, &
                                              b & = \frac{1}{2} p p^{\prime}, \\
    c & = - a, &
                 f & =  - r p, \\
    g & = \frac{1}{r}.
  \end{aligned}
\end{dmath}
Inserting the functions in Eq.~\eqref{60} into the Codazzi field equations, Eqs.~\eqref{cod_1}-\eqref{cod_2}, yields
\begin{equation}
  \label{pcods}
  \begin{aligned}
    p \left(r p^{(3)} + 2 p'' - \frac{2 p'}{r}\right) & = 0,
    \\
    r p'' + \frac{2-2 p}{r} & = 0,
  \end{aligned}
\end{equation}
whose solution is
\begin{dmath}
  p(r) = 1 + \frac{c_1}{r} + c_2 r^2.
\end{dmath}
Hence, the geometry is a Schwarzschild--(Anti-)de-Sitter Riemannian spacetime.

For the parallel Ricci case we also obtain Schwarzschild--(Anti-)de-Sitter geometry, while for Ricci flat case there is a restriction on the parameters and consequently the solution restricts to a Schwarzschild spacetime.

\subsection{Affine foliations and dimensional reduction}
\label{sec:foliations}

A formalism that allows to define foliations in an affine theory is crucial for advancing in the incorporation of matter and the study of symmetries and degrees of freedom in the model.

In general relativity, the Arnowitt--Deser--Misner (ADM) formalism, based on the foliation of spacetime into three-dimensional hypersurfaces, facilitates the development of the Hamiltonian formalism that enables detailed analysis of these aspects. Similarly, Kaluza--Klein theory provides a way to incorporate bosonic matter, such as electromagnetic fields, through the process of dimensional reduction, which should be understood as a  local projection of the higher-dimensional space onto a reduced space.

In the case of an affine theories of gravity, which relies solely on an affine connection and lacks a metric, incorporating matter and analysing symmetries becomes significantly more challenging. In particular, the absence of a metric prevents the use of canonical orthogonal projections, which are a common tool in metric spaces.

A possible solution to these challenges is to decompose the space locally using the direct product of subspaces. This approach could provide a basis for defining projections and foliations in the context of affine gravities.

In this section we shall use the modern language of differential geometry, incorporating the formalism of fibre bundles to our discussion \cite{percacci83_kaluz_klein_theor_bundl_with_homog_fiber,choquet-bruhat89_analy,ivan93_natur_operat_differ_geomet,nakahara05_geomet_topol_physic,dudek18_ehres_theor_connec_princ_bundl_compen_physic}. Our aim would be to resolve the dimensional reduction a la Kaluza--Klein in geometrical terms using the notion of projection.

The standard setup of the Kaluza--Klein model consists of a higher--dimensional space \(\hat{\mathcal{M}}\) and a lower-dimensional space \(\mathcal{M}\), which could be thought of as embedded in the higher-dimensional one. However, as a bundle the space \(\hat{\mathcal{M}}\) would be the total space based in \(\mathcal{M}\), the bundle projection \(\pi : \hat{\mathcal{M}} \to \mathcal{M}\) defines the fibre as the preimage of a point \(m \in \mathcal{M}\), i.e., \(F = \operatorname{preim}_{\pi}(m)\).

The original Kaluza--Klein model has a \(U(1)\)--fiber (\(F \cong G = U(1)\)) so that \(\dim(\hat{\mathcal{M}}) = \dim(\mathcal{M}) + \dim(G) = \dim(\mathcal{M}) + 1\). Hence, the set up is a \(U(1)\)-principal bundle based on \(\mathcal{M}\).

On each manifold of the Kaluza--Klein-bundle one can define their tangent bundles, and the projection \(\pi\) induces the derived projection on the tangent bundles denoted by \(T\pi\),\footnote{The notation \(T\pi\) refers to the induced map in any \emph{tangent} bundle (including tensor bundles) \cite{ivancevic07_applied_differ_geomet}. If we would like to refer to an specific induced map, e.g over the tangent or cotangent bundles, we would use the more standard notation \(\pi_{*}\) and \(\pi^{*}\) respectively.}
\begin{equation*}
  \begin{tikzcd}
    T\hat{\mathcal{M}} \ar[r, "T\pi"] \ar[d] & T\mathcal{M} \ar[d] \\
    \hat{\mathcal{M}} \ar[r, "\pi"] & \mathcal{M}
  \end{tikzcd}.
\end{equation*}
A \(1\)-form field \(\hat{\theta} \in C^{\infty}(T^{*}\hat{\mathcal{M}})\) induces a natural splitting of the tangent bundle \(T\hat{\mathcal{M}}\) into two sub-bundles, vertical (\(V\hat{\mathcal{M}}\)) and horizontal (\(H\hat{\mathcal{M}}\)), defined as vectors along the directions of the fibre and the base manifold respectively. The \(1\)-form \(\hat{\theta}\) is known as the \emph{Ehresmann connection}.

A vector field \(\tilde{v} \in C^{\infty}(T\hat{\mathcal{M}})\) is said to be a \emph{horizontal lift} of a vector field \(v \in C^{\infty}(T\mathcal{M})\) if at each point \(p \in T\hat{\mathcal{M}}\) the map \(T\pi\) projects \(\tilde{v}\) into \(v\), and in addition \(\tilde{v}\) belongs to the horizontal sub-bundle, i.e.
\begin{equation}
  \label{eq:definition-lift}
  \hat{\theta}(\tilde{v}) = 0.
\end{equation}
Let \(\eta \in C^{\infty}(V\hat{\mathcal{M}})\) be the unique vertical vector field (\(T\pi(\eta) = 0\)) such that,
\begin{equation}
  \label{eq:nomalization-eta}
  \hat{\theta}(\eta) = 1.
\end{equation}

For illustrative purposes, we shall introduce the downward arrow operation (\(\downarrow\)) which represents the induced map of the \(\pi\)-projection, \(\downarrow{} \cong T\pi\). The downward arrow operator acts removing the hat to the tensor fields, i.e.
\begin{equation}
  \label{eq:down-arrow-operator}
  \downarrow(\hat{v}) = v,
\end{equation}
for \(\hat{v} \in C^{\infty}(\otimes^p T\hat{\mathcal{M}} \otimes \otimes^q T^{*}\hat{\mathcal{M}})\) and \(v \in C^{\infty}(\otimes^p T{\mathcal{M}} \otimes \otimes^q T^{*}{\mathcal{M}})\).

Similarly, we shall define the \emph{lifting operator}, denoted with the upward arrow, \(\uparrow\), as the unique linear map (up to the action of an element of the Lie group \(G\)) that satisfies
\begin{equation}
  \label{eq:uparrow-properties}
  \downarrow(\uparrow(v)) = v \text{ and } \hat{\theta}(\uparrow(v)) = 0.
\end{equation}
Its action should be understood as the addition of the tilde, i.e.
\begin{equation}
  \label{eq:up-arrow-operator}
  \uparrow({v}) = \tilde{v},
\end{equation}
for \({v} \in C^{\infty}(\otimes^p T{\mathcal{M}} \otimes \otimes^q T^{*}{\mathcal{M}})\) and \(\tilde{v} \in C^{\infty}(\otimes^p T\hat{\mathcal{M}} \otimes \otimes^q T^{*}\hat{\mathcal{M}})\). Moreover, this map generates a point-wise isomorphism \(T_{\pi(p)}M \cong T_p\tilde{\mathcal{M}} \subset T_p\hat{\mathcal{M}}\), so we can locally identify the reduced space as part of the total space.

The above isomorphism allows us to identify the action of operators \(\hat{L}\) over the sections of the lifted sub-bundle, with that of projected operators \(L = \downarrow(L)\) on the un-lifted bundle, as follows,
\begin{equation}
  \label{eq:identification-operation-bundles}
  \hat{L}(\tilde{v}) = L(v),
\end{equation}
with \(\tilde{v} = \uparrow(v)\).

In the remainder of this section, we shall use the described setup to reproduce the metric Kaluza--Klein decomposition of the metric, as a warming up of the affine Kaluza--Klein decomposition of the connection, which would be detailed in a future article.

Consider a bundle \(\hat{\mathcal{M}} \to \mathcal{M}\) with projection \(\pi: \hat{\mathcal{M}} \to \mathcal{M}\), and fibre \(G\). Let \(\hat{g}\) be a metric on $\hat{\mathcal{M}}$, i.e. a section \(g \in C^{\infty}(S^2(T^{*}\hat{\mathcal{M}}))\). The metric \(\hat{g}\) naturally induces a metric \(g \in C^{\infty}(S^2(T^{*}{\mathcal{M}}))\) on the reduced space \(\mathcal{M}\), defined as
\begin{equation}
  \label{eq:induced-metric}
  g = \downarrow(\hat{g}).
\end{equation}
Additionally, we define the module of a vector lying on the fibre \(G\) as \(\hat{\phi} = \sqrt{\hat{g}(\eta, \eta)}\), and the one-form \(\hat{\alpha}\), $\hat{\alpha}: T\hat{\mathcal{M}} \rightarrow \mathbb{R}$, such that $\hat{\alpha}(\hat{x}) = \hat{g}(\hat{x}, \eta)$.

Assuming that at each point the vertical sub-bundle $V\hat{\mathcal{M}}$ is orthogonal to the horizontal lifting $T\tilde{\mathcal{M}}$, then $T\tilde{\mathcal{M}} = \ker(\hat{\alpha})$. This implies that $\hat{\alpha} = \hat{\phi}^2 \hat{\theta}$ and $\hat{g}(\hat{x}, \eta) = 0$.

Since every vector field \(\hat{X} \in C^{\infty}T\hat{\mathcal{M}}\) admits a decomposition of the form
\begin{equation}
  \label{eq:kk-vector-decomposition}
  \hat{X} = \tilde{X} + \hat{\theta}(\hat{X}) \eta,
\end{equation}
it follows that the metric \(\hat{g}\) acts as follows,
\begin{equation}
  \label{eq:kk-metric-decomposition}
  \begin{aligned}
    \hat{g}(\hat{X}, \hat{Y})
    & = \hat{g} \left( \tilde{X} + \hat{\theta}(\hat{X}) \eta, \tilde{Y} + \hat{\theta}(\hat{Y}) \eta \right) \\
    & = \hat{g}(\tilde{X}, \tilde{Y}) + \hat{\phi}^2 \hat{\theta}(\hat{X}) \hat{\theta}(\hat{Y}) \\
    & = \tilde{g}(\tilde{X}, \tilde{Y}) + \hat{\phi}^2 \hat{\theta}(\hat{X}) \hat{\theta}(\hat{Y}).
  \end{aligned}
\end{equation}
In free index notation, $\hat{g} = \tilde{g} + \hat{\phi}^2 \hat{\theta} \otimes \hat{\theta}$.\footnote{It worth to clarify that the quantity \(\tilde{g}(\tilde{X},\tilde{Y}) = \hat{g}(\tilde{X},\tilde{Y})\) would be, according with Eq. \eqref{eq:identification-operation-bundles}, the effective metric in the manifold \(\mathcal{M}\).}

Finally, projecting in a basis $B = \left\{ \hat{e}_{0}, \hat{e}_{1}, \dots, \hat{e}_{D} = \eta \right\}$, we recover:
\begin{equation}
  \label{eq:kk-metric-matrices}
  \hat{g} = \left(
    \begin{array}{c|c}
      \hat{g}_{ij} & \hat{g}_{iD} \\
      \hline
      \hat{g}_{Dj} & \hat{g}_{DD}
    \end{array}
  \right) = \left(
    \begin{array}{c|c}
      \tilde{g}_{ij} + \hat{\phi}^2 \hat{\theta}_i\hat{\theta}_j & \hat{\phi}^2 \hat{\theta}_i \\
      \hline
      \hat{\phi}^2 \hat{\theta}_j & \hat{\phi}^2
    \end{array}
  \right).
\end{equation}

The above is the ansatz usually employed in Kaluza--Klein theory, which has been obtained using our geometric formalism. Although we initially assumed that $\hat{\theta}$ was independent of the metric, we found a relationship between them.

We can highlight that this formalism does not require a metric a priori, so it can be applied to decompose the connection in purely affine models of gravity.

\section{Conclusions}
\label{sec:concl}

This article examines the progress and advancements in the polynomial affine model of gravity that have taken place over a decade since its introduction.

The model presents an alternative, or a generalization, to the affine model introduced by Einstein and Eddington, distinguished by its polynomial action. The action, as shown in Eq. \eqref{eq:new-action}, encompasses numerous terms, especially when compared to other alternative models. The characteristics discussed in detail in Sec. \ref{sec:model} motivate a thorough investigation into the model's dynamic properties.

We showed that the space of solutions of polynomial affine gravity contains the space of Einstein manifolds, and in general the affine generalisations of Einstein manifolds are parametric families that contain the solutions of pure General Relativity as \emph{points} of those families. Interestingly, the restriction to the torsion-free sector is still well defined, and for equiaffine connections, the space of solutions is equivalent to the space of statistical manifolds. This indicates that such a space of solutions can be seen as a projective manifold.

We have found diverse ansatz for the affine connection, specially for the par-spherical and cosmological symmetries, and determine the role of the discrete symmetries \(P\) and \(T\) in constraining their components. We used those ansatz to analyse the explicit cosmological models, even when the polynomial affine model of gravity is coupled with a scalar field, and note that even if completely determined exact solutions can be obtained, the most interesting solution (proper solutions of the Ricci as a Codazzi tensor) are parameterised by an undetermined function, e.g. \(h\). Even if this type of solution is not suitable as a physical model, we believe that this scenario provided a unique possibility to test our model. For example, the latest observations reported by the DESI collaboration favour a modification of the \(\Lambda\)CDM  over the standard model of cosmology \cite{collaboration24_desi_vi,roy24_dynam_dark_energ_light}, we could use those results to determine the function \(h\) and then compare other cosmological observables with the predictions derived from our fitted model. We also have all the ingredients to consider scenarios with affine inflation.

Even though our model is based on the lack of a fundamental metric structure, in the space of solutions metric structures might emerge. The emergence of metric structures allows us to define distances on the manifold and provide a tool to discriminate between \emph{time-like} and \emph{space-like} geodesics (or autoparallel curves), or even analyse the causal structure of the model (on-shell).

We built up the method of cosmological affine perturbations with the idea of analysing the phenomena of structure formation, and also the \emph{stability} of the cosmological models. We are pointing toward the use of the tools from dynamical systems to obtain qualitative information about the model. These ideas are currently under development.

Although in affine models there is no necessity of considering a fundamental metric structure, one could choose to use a metric in the model, inducing a splitting of the affine information into a Levi-Civita connection, nonmetricity, and torsion. Based on the criteria analysed in Refs. \cite{ehlers12_repub_of,matveev13_criter_compat_confor_projec_struc}, we enquire about the condition of metric independence of an affine model of gravity. In this context, we found that cosmological models in polynomial affine gravity differ from metric cosmological formulations by a vector that encodes the projective transformation of the connection. A similar analysis was made for connections with spherical symmetry, but the equation of autoparallel curves contains terms that might turn the task of identifying autoparallels with geodesics impossible.

Despite the discussion above, in the space of solutions of affine models metrics could emerge. We show that in our model there are three possible emergent metric tensors, the symmetric component of the Ricci tensor, the Poplawski tensor (there is a variation if the contribution of the \(\mathcal{A}\)-field is ignored) or the symmetrisation of the covariant derivative of the \(\mathcal{A}\)-field.

Even in those cases where the space of solutions admits emergent metrics, since they are defined in terms of the components of the connection, using those metrics (or their inverse) to couple matter to the model would spoil the polynomial property. However, it is possible to build up a sort of \emph{inverse metric density} using the strategy of index structure analysis (called dimensional analysis in our earlier articles), which allows us to couple scalar fields to polynomial affine gravity. The study of couplings to other matter fields is a subject of great interest but still under development by our research group.

It is worth to highlight that the field equations of polynomial affine gravity coupled with a scalar field are an affine generalisation of the Einstein--Klein--Gordon equations, and we could use this coupled system to enquire inflationary scenarios within the context of affine gravity.

Another interesting subject is the study of the space of solutions with spherical symmetry. In Ref. \cite{castillo-felisola18_einst_gravit_from_polyn_affin_model}, we used a metric ansatz to try to say something about the spherical solutions in the polynomial affine model of gravity, concluding that starting from a static Schwarzschild-like metric (with a single undetermined function), the sole solution was the Schwarzschild(-Anti)-de~Sitter solution.\footnote{When we considered the Schwarzschild-like metric with two unknown functions, we were unable to solve the field equations exactly. However, a power series expansion of the unknown functions point to the necessity of three conditions to determine the solution, we hypothesise that those conditions are related to three different scales in gravity, which might be thought as celestial scale (Newtonian gravity plus corrections), galactic scale (which might explain the rotation curves of galaxies, i.e. dark matter), and cosmological scale (related to the cosmological constant, i.e. dark energy).}

In this opportunity, we started to enquire the most basic affine static spherical solutions to the field equations of polynomial affine gravity (without torsion), and found that the affine solutions were parameterised by five functions, which determine the connection. Interestingly, the number of field equations coming from the condition of parallel Ricci, \(\nabla_\lambda \mathcal{R}_{\mu\nu} = 0\), is five, and the connection could be integrated exactly. We conjecture that the solution to this interesting case is given by an affine generalisation of Schwarzschild geometry with cosmological constant. In the Ricci flat case, the number of field equations is three, and therefore generically the solutions would be parameterised by two arbitrary functions. Although those functions might be fixed by observations or restricted by boundary conditions, in a future article we shall consider that this pair of functions coincides with two of the functions determining the components of the Levi-Civita connection. However, finding proper solutions to the field equations coming from the Codazzi condition for the Ricci tensor is very difficult since the system has a third arbitrary function parameterising the connection.

In order to extend the richness of our model, we would like to be able to define conserved charges in or (as mentioned) coupling matter to the polynomial affine model of gravity. The foliation of affinely connected manifolds becomes an interesting tool to solve both problems. On the one hand, the foliation of the manifold is the starting point of the Arnowitt--Deser--Misner formalism, which allows us to define conserved charges in General Relativity; so an affine analogous would be the initial place to developing a similar programme. On the other hand, the dimensional reduction (a la Kaluza--Klein) could shed light on the sort of couplings between polynomial affine gravity and matter fields, in the same way the standard Kaluza--Klein model yields General Relativity coupled with gauge fields and scalars.

Clearly, some of the formal and phenomenological aspects of the model are still under development, but during these ten years we have been able to carry the idea of a polynomial affine gravity onto a viable model of gravitational interactions which encloses the successes of General Relativity but allows the flexibility of accommodating additional geometrical effects that might be helpful to unveil the current mystic of the dark sector of the Universe, and possibly hinting toward a (consistent) quantum theory of gravity.

\begin{acknowledgments}
  The authors are grateful to N. Zambra, J. Vaca, M. Morocho, A. Zerwekh and Ivan Schmidt\footnote{OCF wants to dedicate this article to the memory of Ivan, who has passed away recently.} for their fruitful discussions and invaluable support during the development of this research. OCF acknowledge the financial support received by ANID PIA/APOYO AFB230003 (Chile) and FONDECYT Regular No. 1230110 (Chile). B.G. acknowledges the support of ANID Becas/Doctorado Nacional 21242572.

  We are indebted to the authors and countless contributors of the free and open software community, for maintaining the projects \texttt{SAGEmath}, \texttt{sagemanifolds} and \texttt{cadabra} \cite{stein18_sage_mathem_softw_version,gourgoulhon18_sagem_version,peeters07_symbol_field_theor_with_cadab,peeters07_introd_cadab,peeters07_cadab,brewin19_using_cadab_tensor_comput_gener_relat,kulyabov19_new_featur_secon_version_cadab}, which has been used extensively to complete this work.
\end{acknowledgments}

\appendix

\section{Detailed contributions to the covariant field equations}
\label{sec:appen-a}
This appendix provides the field equations for \(\Gamma\), \(\mathcal{B}\), and \(\mathcal{A}\). Observe that because of the extensive nature of the equations, they were broken down based on the contribution of each term in the action \eqref{eq:new-action}. The complete set of field equations is derived by summing all the given relations, each multiplied by its respective coupling constant.

\subsection{Field Equations for \(\Gamma\)}
\label{sec:field-eq-gamma}

\begin{dgroup}
  \begin{dmath}
    B_{1}\colon \phantom{=} \quad  \nabla_{\mu}\left(2\delta^{[\mu}_{\lambda}\mathcal{B}_{\alpha}{}^{\nu]}{}_{\beta} \mathcal{B}_{\gamma}{}^{\rho}{}_{\tau} + 2\delta^{[\mu}_{\lambda} \mathcal{B}_{\alpha}{}^{\rho]}{}_{\beta} \mathcal{B}_{\gamma}{}^{\nu}{}_{\tau}\right) \mathfrak{E}^{\alpha \beta \gamma \tau}  
  \end{dmath}
  \begin{dmath}
    B_{2}\colon \phantom{=} \quad  \nabla_{\mu}\left(4 \mathcal{B}_{\gamma}{}^{\sigma}{}_{\delta} \mathcal{B}_{\sigma}{}^{(\rho}{}_{\lambda} \right) \mathfrak{E}^{\nu)\mu \gamma\delta}
  \end{dmath}
  \begin{dmath}
    B_{3}\colon \phantom{=} \quad  \nabla_{\mu}\left( 2\delta^{[\mu}_{\lambda} \mathcal{B}_{\beta}{}^{\nu]}{}_{\gamma} \mathcal{A}_{\delta} \mathfrak{E}^{\rho\beta\gamma\delta} + 2 \delta^{[\mu}_{\lambda} \mathcal{B}_{\beta}{}^{\rho]}{}_{\gamma} \mathcal{A}_{\delta} \mathfrak{E}^{\nu\beta\gamma\delta} \right)
  \end{dmath}
  \begin{dmath}
    B_{4}\colon \phantom{=} \quad  \nabla_{\mu}\left( -4 \mathcal{B}_{\gamma}{}^{(\rho}{}_{\delta} \mathcal{A}_{\lambda} \right)\mathfrak{E}^{\nu)\mu\gamma\delta}
  \end{dmath}
  \begin{dmath}
    B_{5}\colon \phantom{=} \quad  \nabla_{\mu}\left( -4 \mathcal{B}_{\gamma}{}^{\sigma}{}_{\delta} \mathcal{A}_{\sigma} \delta^{(\rho}_{\lambda} \right) \mathfrak{E}^{\nu)\mu\gamma\delta}
  \end{dmath}
  \begin{dmath}
    C_{1}\colon \phantom{=} \quad  \nabla_{\mu} \left( 2\nabla_{\beta} \mathcal{B}_{\gamma}{}^{\rho}{}_{\delta} \delta^{[\mu}_{\lambda} \mathfrak{E}^{\nu]\beta\gamma\delta} +2 \nabla_{\beta} \mathcal{B}_{\gamma}{}^{\nu}{}_{\delta} \delta^{[\mu}_{\lambda} \mathfrak{E}^{\rho]\beta\gamma\delta} \right) + 2 \mathcal{R}_{\mu\alpha}{}^{\mu}{}_{\lambda} \mathcal{B}_{\gamma}{}^{(\rho}{}_{\delta} \mathfrak{E}^{\nu)\alpha\gamma\delta}
  \end{dmath}
  \begin{dmath}
    C_{2}\colon \phantom{=} \quad  \nabla_{\mu} \left( -4 \nabla_{\sigma} \mathcal{B}_{\gamma}{}^{\sigma}{}_{\delta} \delta^{(\rho}_{\lambda} \mathfrak{E}^{\nu)\mu\gamma\delta} \right) + 2\mathcal{R}_{\alpha\beta}{}^{\sigma}{}_{\sigma} \left( 2 \mathcal{B}_{\lambda}{}^{(\nu}{}_{\delta} \mathfrak{E}^{\rho)\alpha\beta\delta} - \delta^{(\nu}_{\lambda} \mathcal{B}_{\gamma}{}^{\rho)}{}_{\delta} \mathfrak{E}^{\alpha\beta\gamma\delta} \right)
  \end{dmath}
  \begin{dmath}
    D_{1}\colon \phantom{=} \quad  2 \mathcal{B}_{\tau}{}^{\sigma}{}_{\lambda} \mathcal{B}_{\sigma}{}^{\tau}{}_{\alpha} \mathcal{B}_{\gamma}{}^{(\rho}{}_{\delta} \mathfrak{E}^{\nu)\alpha\gamma\delta}
  \end{dmath}
  \begin{dmath}
    D_{2}\colon \phantom{=} \quad  2 \mathcal{B}_{\alpha}{}^{\sigma}{}_{\beta} \mathcal{B}_{\sigma}{}^{(\nu\lvert}{}_{\tau} \left( 2\mathcal{B}_{\lambda}{}^{\tau}{}_{\delta} \mathfrak{E}^{\lvert\rho)\alpha\beta\delta} - \delta^{\tau}_{\lambda} \mathcal{B}_{\gamma}{}^{\lvert\rho)}{}_{\delta} \mathfrak{E}^{\alpha\beta\gamma\delta} \right)
  \end{dmath}
  \begin{dmath}
    D_{3}\colon \phantom{=} \quad  2\mathcal{B}_{\alpha}{}^{\sigma}{}_{\tau} \mathcal{B}_{\beta}{}^{(\nu}{}_{\gamma} \left( \delta^{\rho)}_{\sigma} \mathcal{B}_{\lambda}{}^{\tau}{}_{\delta} + \delta^{\rho)}_{\delta} \mathcal{B}_{\sigma}{}^{\tau}{}_{\lambda} - \delta^{\tau}_{\lambda} \mathcal{B}_{\sigma}{}^{\lvert \rho)}{}_{\delta} \right) \mathfrak{E}^{\alpha\beta\gamma\delta}
  \end{dmath}
  \begin{dmath}
    D_{4}\colon \phantom{=} \quad  2\mathcal{B}_{\alpha}{}^{(\nu}{}_{\beta} \mathcal{B}_{\gamma}{}^{\rho)}{}_{\delta} \mathcal{A}_{\lambda} \mathfrak{E}^{\alpha\beta\gamma\delta}
  \end{dmath}
  \begin{dmath}
    D_{5}\colon \phantom{=} \quad  2 \mathcal{B}_{\alpha}{}^{(\nu\lvert}{}_{\beta} \mathcal{A}_{\sigma} \left( 2 \mathcal{B}_{\lambda}{}^{\sigma}{}_{\delta} \mathfrak{E}^{\lvert\rho)\alpha\beta\delta} - \delta^{\sigma}_{\lambda} \mathcal{B}_{\gamma}{}^{\lvert\rho)}{}_{\delta} \mathfrak{E}^{\alpha\beta\gamma\delta} \right)
  \end{dmath}
  \begin{dmath}
    D_{6}\colon \phantom{=} \quad  -2 \mathcal{B}_{\alpha}{}^{(\nu}{}_{\beta}\mathcal{A}_{\gamma}\mathcal{A}_{\lambda} \mathfrak{E}^{\rho)\alpha\beta\gamma}
  \end{dmath}
  \begin{dmath}
    E_{1}\colon \phantom{=} \quad  4\nabla_{\sigma}\mathcal{B}_{\alpha}{}^{\sigma}{}_{\beta} \left( 2 \mathcal{B}_{\lambda}{}^{(\nu}{}_{\delta}\mathfrak{E}^{\rho)\alpha\beta\delta} - \delta^{(\nu}_{\lambda} \mathcal{B}_{\gamma}{}^{\rho)}{}_{\delta}\mathfrak{E}^{\alpha\beta\gamma\delta} \right)
  \end{dmath}
  \begin{dmath}
    E_{2}\colon \phantom{=} \quad  2\mathcal{F}_{\alpha\beta} \left( \mathcal{B}_{\lambda}{}^{(\nu}{}_{\delta} \mathfrak{E}^{\rho)\alpha\beta\delta} - \delta^{(\nu}_{\lambda} \mathcal{B}_{\gamma}{}^{\rho)}{}_{\delta} \mathfrak{E}^{\alpha\beta\gamma\delta} \right)
  \end{dmath}
\end{dgroup}

\subsection{Field Equations for \(\mathcal{B}\)}
\label{sec:field-eq-b}

\begin{dgroup}
  \begin{dmath}
    B_{1}\colon \phantom{=} \quad  -4 \mathcal{R}_{\mu(\sigma}{}^{\mu}{}_{\lambda)} \mathcal{B}_{\gamma}{}^{\sigma}{}_{\delta}\mathfrak{E}^{\nu\rho\gamma\delta} - \frac{4}{3} \mathcal{R}_{\mu\tau}{}^{\mu}{}_{\sigma} \mathcal{B}_{\gamma}{}^{\sigma}{}_{\delta} \delta^{[\nu}_{\lambda} \mathfrak{E}^{\rho]\tau\gamma\delta} - \frac{4}{3} \mathcal{R}_{\mu\sigma}{}^{\mu}{}_{\tau} \mathcal{B}_{\gamma}{}^{\sigma}{}_{\delta} \delta^{[\nu}_{\lambda} \mathfrak{E}^{\rho]\tau\gamma\delta}
  \end{dmath}
  \begin{dmath}
    B_{2}\colon \phantom{=} \quad  -2 \mathcal{R}_{\alpha\beta}{}^{\mu}{}_{\sigma} \mathcal{B}_{\mu}{}^{\sigma}{}_{\lambda} \mathfrak{E}^{\nu\rho\alpha\beta} - 2\mathcal{R}_{\alpha\beta}{}^{[\nu}{}_{\lambda} \mathcal{B}_{\gamma}{}^{\rho]}{}_{\delta} \mathfrak{E}^{\alpha\beta\gamma\delta} - \frac{4}{3} \mathcal{R}_{\alpha\beta}{}^{\mu}{}_{\sigma} \mathcal{B}_{\mu}{}^{\sigma}{}_{\tau} \delta^{[\nu}_{\lambda} \mathfrak{E}^{\rho]\tau\alpha\beta} - \frac{2}{3} \mathcal{R}_{\alpha\beta}{}^{\tau}{}_{\tau} \mathcal{B}_{\gamma}{}^{[\nu}{}_{\delta} \delta^{\rho]}_{\lambda} \mathfrak{E}^{\alpha\beta\gamma\delta} + \frac{2}{3}\mathcal{R}_{\alpha\beta}{}^{[\nu}{}_{\tau} \delta^{\rho]}_{\lambda} \mathcal{B}_{\gamma}{}^{\tau}{}_{\delta} \mathfrak{E}^{\alpha\beta\gamma\delta} 
  \end{dmath}
  \begin{dmath}
    B_{3}\colon \phantom{=} \quad  -2 \mathcal{R}_{\mu\lambda}{}^{\mu}{}_{\alpha} \mathcal{A}_{\beta} \mathfrak{E}^{\nu\rho\alpha\beta} - \frac{4}{3} \mathcal{R}_{\mu\tau}{}^{\mu}{}_{\alpha} \mathcal{A}_{\beta} \delta^{[\nu}_{\lambda} \mathfrak{E}^{\rho]\tau\alpha\beta}
  \end{dmath}
  \begin{dmath}
    B_{4}\colon \phantom{=} \quad  -2 \mathcal{R}_{\alpha\beta}{}^{\sigma}{}_{\lambda} \mathcal{A}_{\sigma} \mathfrak{E}^{\nu\rho\alpha\beta} - \frac{4}{3} \mathcal{R}_{\alpha\beta}{}^{\sigma}{}_{\tau} \mathcal{A}_{\sigma} \delta^{[\nu}_{\lambda} \mathfrak{E}^{\rho]\tau\alpha\beta}
  \end{dmath}
  \begin{dmath}
    B_{5}\colon \phantom{=} \quad  -2 \mathcal{R}_{\alpha\beta}{}^{\tau}{}_{\tau} \mathcal{A}_{\lambda} \mathfrak{E}^{\nu\rho\alpha\beta} - \frac{4}{3} \mathcal{R}_{\alpha\beta}{}^{\tau}{}_{\tau} \mathcal{A}_{\delta} \delta^{[\nu}_{\lambda} \mathfrak{E}^{\rho]\delta\alpha\beta}
  \end{dmath}
  \begin{dmath}
    C_{1}\colon \phantom{=} \quad  \nabla_{\mu} \left( -2\mathcal{R}_{\sigma\alpha}{}^{\sigma}{}_{\lambda}\mathfrak{E}^{\mu\nu\rho\alpha} + \frac{4}{3} \mathcal{R}_{\sigma\alpha}{}^{\sigma}{}_{\tau} \delta^{[\nu}_{\lambda} \mathfrak{E}^{\rho]\mu\tau\alpha} \right)
  \end{dmath}
  \begin{dmath}
    C_{2}\colon \phantom{=} \quad  \nabla_{\mu}\left( 2\mathcal{R}_{\alpha\beta}{}^{\sigma}{}_{\sigma} \delta^{\mu}_{\lambda} \mathfrak{E}^{\nu\rho\alpha\beta} + \frac{4}{3}\mathcal{R}_{\alpha\beta}{}^{\sigma}{}_{\sigma} \delta^{[\nu}_{\lambda} \mathfrak{E}^{\rho]\mu\alpha\beta} \right)
  \end{dmath}
  \begin{dmath}
    D_{1}\colon \phantom{=} \quad  \nabla_{\mu}\left( -2 \mathcal{B}_{\sigma}{}^{\theta}{}_{\lambda} \mathcal{B}_{\theta}{}^{\sigma}{}_{\alpha} \right)\mathfrak{E}^{\mu\nu\rho\alpha} - 2\mathcal{B}_{\lambda}{}^{[\nu}{}_{\alpha} \nabla_{\beta} \mathcal{B}_{\gamma}{}^{\rho]}{}_{\delta} \mathfrak{E}^{\alpha\beta\gamma\delta} - 2 \mathcal{B}_{\lambda}{}^{[\nu\lvert}{}_{\sigma} \nabla_{\beta} \mathcal{B}_{\gamma}{}^{\sigma}{}_{\delta} \mathfrak{E}^{\lvert\rho]\beta\gamma\delta} - \frac{2}{3} \delta^{[\nu}_{\lambda} \mathcal{B}_{\tau}{}^{\rho]}{}_{\alpha} \nabla_{\beta} \mathcal{B}_{\gamma}{}^{\tau}{}_{\delta} \mathfrak{E}^{\alpha\beta\gamma\delta} + \frac{2}{3}\mathcal{B}_{\tau}{}^{[\nu}{}_{\sigma}\delta^{\rho]}_{\lambda} \nabla_{\beta} \mathcal{B}_{\gamma}{}^{\sigma}{}_{\delta} \mathfrak{E}^{\tau\beta\gamma\delta}  
  \end{dmath}
  \begin{dmath}
    D_{2}\colon \phantom{=} \quad  \nabla_{\mu} \left( 2\mathcal{B}_{\alpha}{}^{\sigma}{}_{\beta} \mathcal{B}_{\sigma}{}^{\mu}{}_{\lambda} \mathfrak{E}^{\nu\rho\alpha\beta} +\frac{4}{3} \mathcal{B}_{\alpha}{}^{\sigma}{}_{\beta} \mathcal{B}_{\sigma}{}^{\mu}{}_{\tau} \delta^{[\nu}_{\lambda} \mathfrak{E}^{\rho]\tau\alpha\beta} \right) - 2\mathcal{B}_{\lambda}{}^{\mu}{}_{\sigma}\nabla_{\mu}\mathcal{B}_{\alpha}{}^{\sigma}{}_{\beta} \mathfrak{E}^{\nu\rho\alpha\beta} - 2\mathcal{B}_{\alpha}{}^{[\nu}{}_{\beta} \nabla_{\lambda} \mathcal{B}_{\gamma}{}^{\rho]}{}_{\delta} \mathfrak{E}^{\alpha\beta\gamma\delta} - \frac{4}{3} \mathcal{B}_{\tau}{}^{\mu}{}_{\sigma} \nabla_{\mu} \mathcal{B}_{\alpha}{}^{\sigma}{}_{\beta} \delta^{[\nu}_{\lambda} \mathfrak{E}^{\rho]\tau\alpha\beta} - \frac{2}{3}\mathcal{B}_{\alpha}{}^{\tau}{}_{\beta} \nabla_{\tau} \mathcal{B}_{\gamma}{}^{[\nu}{}_{\delta} \delta^{\rho]}_{\lambda} \mathfrak{E}^{\alpha\beta\gamma\delta} + \frac{2}{3} \mathcal{B}_{\alpha}{}^{[\nu}{}_{\beta} \delta^{\rho]}_{\lambda} \nabla_{\tau} \mathcal{B}_{\gamma}{}^{\tau}{}_{\delta}\mathfrak{E}^{\alpha\beta\gamma\delta}  
  \end{dmath}
  \begin{dmath}
    D_{3}\colon \phantom{=} \quad  \nabla_{\mu} \left( -2 \mathcal{B}_{\beta}{}^{\mu}{}_{\gamma} \mathcal{B}_{\alpha}{}^{[\nu}{}_{\lambda} \mathfrak{E}^{\rho]\alpha\beta\gamma} -\frac{2}{3} \mathcal{B}_{\beta}{}^{\mu}{}_{\gamma} \mathcal{B}_{\alpha}{}^{[\nu}{}_{\tau} \delta^{\rho]}_{\lambda}\mathfrak{E}^{\alpha\beta\gamma\tau} \right) - 2\mathcal{B}_{\beta}{}^{\mu}{}_{\gamma}\nabla_{\mu}\mathcal{B}_{\lambda}{}^{[\rho}{}_{\delta}\mathfrak{E}^{\nu]\beta\gamma\delta} - 2 \mathcal{B}_{\gamma}{}^{\mu}{}_{\sigma} \nabla_{\lambda} \mathcal{B}_{\mu}{}^{\sigma}{}_{\delta}\mathfrak{E}^{\nu\rho\gamma\delta} - \frac{2}{3} \mathcal{B}_{\beta}{}^{\mu}{}_{\gamma} \nabla_{\mu} \mathcal{B}_{\tau}{}^{[\nu}{}_{\delta} \delta^{\rho]}_{\lambda} \mathfrak{E}^{\tau\beta\gamma\delta} - \frac{4}{3} \mathcal{B}_{\gamma}{}^{\mu}{}_{\sigma} \nabla_{\tau} \mathcal{B}_{\mu}{}^{\sigma}{}_{\delta} \delta^{[\nu}_{\lambda} \mathfrak{E}^{\rho]\tau\gamma\delta}  
  \end{dmath}
  \begin{dmath}
    D_{4}\colon \phantom{=} \quad  -4\mathcal{B}_{\alpha}{}^{\sigma}{}_{\beta} \nabla_{(\lambda} \mathcal{A}_{\sigma)} \mathfrak{E}^{\nu\rho\alpha\beta} - \frac{4}{3} \mathcal{B}_{\alpha}{}^{\sigma}{}_{\beta}\nabla_{\tau}\mathcal{A}_{\sigma} \delta^{[\nu}_{\lambda} \mathfrak{E}^{\rho]\tau\alpha\beta} - \frac{4}{3}\mathcal{B}_{\alpha}{}^{\sigma}{}_{\beta} \nabla_{\sigma}\mathcal{A}_{\tau} \delta^{[\nu}_{\lambda}\mathfrak{E}^{\rho]\tau\alpha\beta}
  \end{dmath}
  \begin{dmath}
    D_{5}\colon \phantom{=} \quad  \nabla_{\mu} \left( 2\mathcal{B}_{\alpha}{}^{\mu}{}_{\beta} \mathcal{A}_{\lambda} \mathfrak{E}^{\nu\rho\alpha\beta} + \frac{4}{3}\mathcal{B}_{\alpha}{}^{\mu}{}_{\beta} \mathcal{A}_{\tau} \delta^{[\nu}_{\lambda} \mathfrak{E}^{\rho]\tau\alpha\beta} \right) - 2 \nabla_{\lambda} \mathcal{B}_{\alpha}{}^{\sigma}{}_{\beta}\mathcal{A}_{\sigma}\mathfrak{E}^{\nu\rho\alpha\beta} - \frac{4}{3}\nabla_{\tau} \mathcal{B}_{\alpha}{}^{\sigma}{}_{\beta} \mathcal{A}_{\sigma} \delta^{[\nu}_{\lambda}\mathfrak{E}^{\rho]\tau\alpha\beta}  
  \end{dmath}
  \begin{dmath}
    D_{6}\colon \phantom{=} \quad  -2 \mathcal{A}_{\gamma} \nabla_{\lambda} \mathcal{A}_{\delta}\mathfrak{E}^{\nu\rho\gamma\delta} - \frac{4}{3}\mathcal{A}_{\gamma}\nabla_{\tau} \mathcal{A}_{\delta} \delta^{[\nu}_{\lambda} \mathfrak{E}^{\rho]\tau\gamma\delta}
  \end{dmath}
  \begin{dmath}
    D_{7}\colon \phantom{=} \quad  -\mathcal{A}_{\lambda} \mathcal{F}_{\gamma\delta}\mathfrak{E}^{\nu\rho\gamma\delta} - \frac{4}{3}\mathcal{A}_{\tau} \mathcal{F}_{\gamma\delta}\delta^{[\nu}_{\lambda}\mathfrak{E}^{\rho]\tau\gamma\delta}
  \end{dmath}
  \begin{dmath}
    E_{1}\colon \phantom{=} \quad  \nabla_{\mu}\left( 4\delta^{\mu}_{\lambda} \nabla_{\sigma}\mathcal{B}_{\alpha}{}^{\sigma}{}_{\beta} \mathfrak{E}^{\nu\rho\alpha\beta} + \frac{8}{3} \nabla_{\sigma}\mathcal{B}_{\alpha}{}^{\sigma}{}_{\beta} \delta^{[\nu}_{\lambda} \mathfrak{E}^{\rho]\mu\alpha\beta} \right)
  \end{dmath}
  \begin{dmath}
    E_{2}\colon \phantom{=} \quad  \nabla_{\mu}\left( 2\delta^{\mu}_{\lambda} \mathcal{F}_{\alpha\beta} \mathfrak{E}^{\nu\rho\alpha\beta} + \frac{4}{3} \mathcal{F}_{\alpha\beta} \delta^{[\nu}_{\lambda} \mathfrak{E}^{\rho]\mu\alpha\beta} \right)
  \end{dmath}
  \begin{dmath}
    F_{1}\colon \phantom{=} \quad   -4 \mathcal{B}_{\alpha}{}^{\mu}{}_{\beta} \mathcal{B}_{\mu}{}^{\sigma}{}_{\tau} \mathcal{B}_{\lambda}{}^{\tau}{}_{\sigma} \mathfrak{E}^{\nu\rho\alpha\beta} - 4\mathcal{B}_{\alpha}{}^{\mu}{}_{\beta} \mathcal{B}_{\gamma}{}^{[\nu}{}_{\delta} \mathcal{B}_{\mu}{}^{\rho]}{}_{\lambda} \mathfrak{E}^{\alpha\beta\gamma\delta} - \frac{8}{3} \mathcal{B}_{\alpha}{}^{\mu}{}_{\beta} \mathcal{B}_{\mu}{}^{\sigma}{}_{\tau} \mathcal{B}_{\kappa}{}^{\tau}{}_{\sigma} \delta^{[\nu}_{\lambda} \mathfrak{E}^{\rho]\kappa\alpha\beta}   
  \end{dmath}
  \begin{dmath}
    F_{2}\colon \phantom{=} \quad  -2\mathcal{B}_{\alpha}{}^{\mu}{}_{\sigma} \mathcal{B}_{\beta}{}^{\sigma}{}_{\tau} \mathcal{B}_{\lambda}{}^{\tau}{}_{\mu} \mathfrak{E}^{\nu\rho\alpha\beta} + 2\mathcal{B}_{\alpha}{}^{\mu}{}_{\beta} \mathcal{B}_{\mu}{}^{\sigma}{}_{\lambda} \mathcal{B}_{\gamma}{}^{[\nu}{}_{\sigma} \mathfrak{E}^{\rho]\alpha\beta\gamma} - 2\mathcal{B}_{\alpha}{}^{\mu}{}_{\beta} \mathcal{B}_{\gamma}{}^{\sigma}{}_{\lambda} \mathcal{B}_{\mu}{}^{[\nu}{}_{\sigma} \mathfrak{E}^{\rho]\alpha\beta\gamma} -2 \mathcal{B}_{\alpha}{}^{[\nu}{}_{\beta} \mathcal{B}_{\gamma}{}^{\rho]}{}_{\sigma} \mathcal{B}_{\delta}{}^{\sigma}{}_{\lambda}\mathfrak{E}^{\alpha\beta\gamma\delta} - \frac{4}{3} \mathcal{B}_{\alpha}{}^{\mu}{}_{\sigma} \mathcal{B}_{\beta}{}^{\sigma}{}_{\tau} \mathcal{B}_{\kappa}{}^{\tau}{}_{\mu} \delta^{[\nu}_{\lambda} \mathfrak{E}^{\rho]\kappa\alpha\beta} - \frac{2}{3} \mathcal{B}_{\alpha}{}^{\mu}{}_{\beta} \mathcal{B}_{\mu}{}^{\sigma}{}_{\tau} \mathcal{B}_{\gamma}{}^{[\nu}{}_{\sigma} \delta^{\rho]}_{\lambda}\mathfrak{E}^{\alpha\beta\tau\gamma} +\frac{2}{3} \mathcal{B}_{\alpha}{}^{\mu}{}_{\beta} \mathcal{B}_{\gamma}{}^{\sigma}{}_{\tau} \mathcal{B}_{\mu}{}^{\tau}{}_{\sigma} \delta^{[\nu}_{\lambda} \mathfrak{E}^{\rho]\alpha\beta\gamma} - \frac{2}{3} \mathcal{B}_{\alpha}{}^{\mu}{}_{\beta} \mathcal{B}_{\gamma}{}^{\sigma}{}_{\tau} \mathcal{B}_{\mu}{}^{\tau}{}_{\sigma} \delta^{[\nu}_{\lambda} \mathfrak{E}^{\rho]\alpha\beta\gamma} - \frac{2}{3} \mathcal{B}_{\alpha}{}^{\tau}{}_{\beta} \mathcal{B}_{\delta}{}^{\sigma}{}_{\tau} \mathcal{B}_{\gamma}{}^{[\nu}{}_{\sigma}\delta^{\rho]}_{\lambda} \mathfrak{E}^{\alpha\beta\gamma\delta}
  \end{dmath}
  \begin{dmath}
    F_{3}\colon \phantom{=} \quad  -2 \mathcal{B}_{\lambda}{}^{[\nu}{}_{\alpha} \mathcal{B}_{\beta}{}^{\rho]}{}_{\gamma} \mathcal{A}_{\delta} \mathfrak{E}^{\alpha\beta\gamma\delta} - 2\mathcal{B}_{\alpha}{}^{\sigma}{}_{\beta} \mathcal{A}_{\gamma} \mathcal{B}_{\lambda}{}^{[\nu}{}_{\sigma} \mathfrak{E}^{\rho]\alpha\beta\gamma} -2 \mathcal{B}_{\sigma}{}^{\mu}{}_{\lambda} \mathcal{B}_{\mu}{}^{\sigma}{}_{\alpha} \mathcal{A}_{\beta} \mathfrak{E}^{\nu\rho\alpha\beta} + \frac{2}{3}\mathcal{B}_{\beta}{}^{\tau}{}_{\gamma} \mathcal{A}_{\delta} \mathcal{B}_{\tau}{}^{[\nu}{}_{\alpha} \delta^{\rho]}_{\lambda} \mathfrak{E}^{\alpha\beta\gamma\delta} + \frac{2}{3} \mathcal{B}_{\tau}{}^{[\nu}{}_{\sigma} \delta^{\rho]}_{\lambda} \mathcal{B}_{\alpha}{}^{\sigma}{}_{\beta}\mathcal{A}_{\gamma}\mathfrak{E}^{\tau\alpha\beta\gamma}
  \end{dmath}
  \begin{dmath}
    F_{4}\colon \phantom{=} \quad  -4 \mathcal{B}_{\alpha}{}^{\mu}{}_{\beta} \mathcal{A}_{\mu}\mathcal{A}_{\lambda} \mathfrak{E}^{\nu\rho\alpha\beta} - \frac{8}{3} \mathcal{B}_{\alpha}{}^{\mu}{}_{\beta} \mathcal{A}_{\mu} \mathcal{A}_{\tau} \delta^{[\nu}_{\lambda} \mathfrak{E}^{\rho]\tau\alpha\beta}
  \end{dmath}
\end{dgroup}

\subsection{Field Equations for \(\mathcal{A}\)}
\label{sec:field-eq-a}

\begin{dgroup}
  \begin{dmath}
    B_{3}\colon \phantom{=} \quad  - \mathcal{R}_{\sigma \tau}{}^{\sigma}{}_{\alpha} \mathcal{B}_{\beta}{}^{\tau}{}_{\gamma} \mathfrak{E}^{\alpha \beta \gamma \nu}  ,
  \end{dmath}
  \begin{dmath}
    B_{4}\colon \phantom{=} \quad  - \mathcal{R}_{\alpha \beta}{}^{\nu}{}_{\sigma} \mathcal{B}_{\gamma}{}^{\sigma}{}_{\tau} \mathfrak{E}^{\alpha \beta \gamma \tau}  ,
  \end{dmath}
  \begin{dmath}
    B_{5}\colon \phantom{=} \quad  - \mathcal{R}_{\alpha \beta}{}^{\rho}{}_{\rho} \mathcal{B}_{\gamma}{}^{\nu}{}_{\tau} \mathfrak{E}^{\alpha \beta \gamma \tau}  ,
  \end{dmath}
  \begin{dmath}
    D_{4}\colon \phantom{=} \quad  \nabla_{\mu}\left[\mathcal{B}_{\alpha}{}^{\mu}{}_{\beta} \mathcal{B}_{\gamma}{}^{\nu}{}_{\tau}\right] \mathfrak{E}^{\alpha \beta \gamma \tau}  ,
  \end{dmath}
  \begin{dmath}
    D_{5}\colon \phantom{=} \quad  -\mathcal{B}_{\alpha}{}^{\sigma}{}_{\beta} \nabla_{\sigma}{\mathcal{B}_{\gamma}{}^{\nu}{}_{\tau}} \mathfrak{E}^{\alpha \beta \gamma \tau}  ,
  \end{dmath}
  \begin{dmath}
    D_{6}\colon \phantom{=} \quad  \nabla_{\mu}\left[\mathcal{B}_{\alpha}{}^{\mu}{}_{\beta} \mathcal{A}_{\gamma}\right] \mathfrak{E}^{\alpha \beta \gamma \nu}+\mathcal{B}_{\alpha}{}^{\mu}{}_{\beta} \nabla_{\mu}{\mathcal{A}_{\gamma}} \mathfrak{E}^{\alpha \beta \gamma \nu}  ,
  \end{dmath}
  \begin{dmath}
    D_{7}\colon \phantom{=} \quad  \nabla_{\mu}\left[\mathcal{B}_{\alpha}{}^{\sigma}{}_{\beta} \mathcal{A}_{\sigma}\right] \mathfrak{E}^{\alpha \beta \mu \nu}-\mathcal{B}_{\alpha}{}^{\nu}{}_{\beta} \mathcal{F}_{\gamma \tau} \mathfrak{E}^{\alpha \beta \gamma \tau}  ,
  \end{dmath}
  \begin{dmath}
    E_{2}\colon \phantom{=} \quad  \nabla_{\mu}\left[\nabla_{\sigma}{\mathcal{B}_{\alpha}{}^{\sigma}{}_{\beta}}\right] \mathfrak{E}^{\alpha \beta \mu \nu}  ,
  \end{dmath}
  \begin{dmath}
    F_{3}\colon \phantom{=} \quad  -\mathcal{B}_{\sigma}{}^{\tau}{}_{\lambda} \mathcal{B}_{\tau}{}^{\sigma}{}_{\alpha} \mathcal{B}_{\beta}{}^{\lambda}{}_{\gamma} \mathfrak{E}^{\alpha \beta \gamma \nu}  ,
  \end{dmath}
  \begin{dmath}
    F_{4}\colon \phantom{=} \quad  -2 \mathcal{B}_{\alpha}{}^{\sigma}{}_{\beta} \mathcal{B}_{\gamma}{}^{\nu}{}_{\tau} \mathcal{A}_{\sigma} \mathfrak{E}^{\alpha \beta \gamma \tau}  .
  \end{dmath}
\end{dgroup}


\begin{thebibliography}{171}%
\makeatletter
\providecommand \@ifxundefined [1]{%
 \@ifx{#1\undefined}
}%
\providecommand \@ifnum [1]{%
 \ifnum #1\expandafter \@firstoftwo
 \else \expandafter \@secondoftwo
 \fi
}%
\providecommand \@ifx [1]{%
 \ifx #1\expandafter \@firstoftwo
 \else \expandafter \@secondoftwo
 \fi
}%
\providecommand \natexlab [1]{#1}%
\providecommand \enquote  [1]{``#1''}%
\providecommand \bibnamefont  [1]{#1}%
\providecommand \bibfnamefont [1]{#1}%
\providecommand \citenamefont [1]{#1}%
\providecommand \href@noop [0]{\@secondoftwo}%
\providecommand \href [0]{\begingroup \@sanitize@url \@href}%
\providecommand \@href[1]{\@@startlink{#1}\@@href}%
\providecommand \@@href[1]{\endgroup#1\@@endlink}%
\providecommand \@sanitize@url [0]{\catcode `\\12\catcode `\$12\catcode
  `\&12\catcode `\#12\catcode `\^12\catcode `\_12\catcode `\%12\relax}%
\providecommand \@@startlink[1]{}%
\providecommand \@@endlink[0]{}%
\providecommand \url  [0]{\begingroup\@sanitize@url \@url }%
\providecommand \@url [1]{\endgroup\@href {#1}{\urlprefix }}%
\providecommand \urlprefix  [0]{URL }%
\providecommand \Eprint [0]{\href }%
\providecommand \doibase [0]{https://doi.org/}%
\providecommand \selectlanguage [0]{\@gobble}%
\providecommand \bibinfo  [0]{\@secondoftwo}%
\providecommand \bibfield  [0]{\@secondoftwo}%
\providecommand \translation [1]{[#1]}%
\providecommand \BibitemOpen [0]{}%
\providecommand \bibitemStop [0]{}%
\providecommand \bibitemNoStop [0]{.\EOS\space}%
\providecommand \EOS [0]{\spacefactor3000\relax}%
\providecommand \BibitemShut  [1]{\csname bibitem#1\endcsname}%
\let\auto@bib@innerbib\@empty
\bibitem [{\citenamefont {Einstein}(1915)}]{einstein15_zur_allgem_relat}%
  \BibitemOpen
  \bibfield  {author} {\bibinfo {author} {\bibfnamefont {A.}~\bibnamefont
  {Einstein}},\ }\bibfield  {title} {\bibinfo {title} {Zur allgemeinen
  relativit{\"a}tstheorie},\ }\href@noop {} {\bibfield  {journal} {\bibinfo
  {journal} {Sitzungsber. Preuss. Akad. Wiss.}\ }\textbf {\bibinfo {volume}
  {1}},\ \bibinfo {pages} {778} (\bibinfo {year} {1915})}\BibitemShut {NoStop}%
\bibitem [{\citenamefont {Hilbert}(1915)}]{hilbert15_grund_physik}%
  \BibitemOpen
  \bibfield  {author} {\bibinfo {author} {\bibfnamefont {D.}~\bibnamefont
  {Hilbert}},\ }\bibfield  {title} {\bibinfo {title} {Die grundlagen der
  physik.(erste mitteilung.)},\ }\href@noop {} {\bibfield  {journal} {\bibinfo
  {journal} {Nachrichten von der Gesellschaft der Wissenschaften zu
  G{\"o}ttingen, Mathematisch-Physikalische Klasse}\ }\textbf {\bibinfo
  {volume} {1915}},\ \bibinfo {pages} {395} (\bibinfo {year}
  {1915})}\BibitemShut {NoStop}%
\bibitem [{\citenamefont
  {Schwarzschild}(1916)}]{schwarzschild16_on_the_gravit_fiel}%
  \BibitemOpen
  \bibfield  {author} {\bibinfo {author} {\bibfnamefont {K.}~\bibnamefont
  {Schwarzschild}},\ }\bibfield  {title} {\bibinfo {title} {{On the
  gravitational field of a mass point according to Einstein's theory}},\
  }\href@noop {} {\bibfield  {journal} {\bibinfo  {journal} {Sitzungsber.
  Preuss. Akad. Wiss. Berlin (Math. Phys.)}\ }\textbf {\bibinfo {volume}
  {1916}},\ \bibinfo {pages} {189} (\bibinfo {year} {1916})},\ \Eprint
  {https://arxiv.org/abs/physics/9905030} {arXiv:physics/9905030 [physics]}
  \BibitemShut {NoStop}%
\bibitem [{\citenamefont {Will}(2014)}]{will14_confr_between_gener_relat}%
  \BibitemOpen
  \bibfield  {author} {\bibinfo {author} {\bibfnamefont {C.~M.}\ \bibnamefont
  {Will}},\ }\bibfield  {title} {\bibinfo {title} {{The Confrontation Between
  General Relativity and experiment}},\ }\href
  {https://doi.org/10.12942/lrr-2014-4} {\bibfield  {journal} {\bibinfo
  {journal} {Living Rev. Relativ.}\ }\textbf {\bibinfo {volume} {17}},\
  \bibinfo {pages} {4} (\bibinfo {year} {2014})},\ \Eprint
  {https://arxiv.org/abs/1403.7377} {arXiv:1403.7377 [gr-qc]} \BibitemShut
  {NoStop}%
\bibitem [{\citenamefont {Will}(2018)}]{will18_theor_exper_gravit_physic}%
  \BibitemOpen
  \bibfield  {author} {\bibinfo {author} {\bibfnamefont {C.~M.}\ \bibnamefont
  {Will}},\ }\href@noop {} {\emph {\bibinfo {title} {Theory and Experiment in
  Gravitational Physics}}},\ \bibinfo {edition} {2nd}\ ed.\ (\bibinfo
  {publisher} {Cambridge University Press},\ \bibinfo {year}
  {2018})\BibitemShut {NoStop}%
\bibitem [{\citenamefont {Arnowitt}\ \emph {et~al.}(1959)\citenamefont
  {Arnowitt}, \citenamefont {Deser},\ and\ \citenamefont
  {Misner}}]{arnowitt59_dynam_struc_defin_energ_gener_relat}%
  \BibitemOpen
  \bibfield  {author} {\bibinfo {author} {\bibfnamefont {R.~L.}\ \bibnamefont
  {Arnowitt}}, \bibinfo {author} {\bibfnamefont {S.}~\bibnamefont {Deser}},\
  and\ \bibinfo {author} {\bibfnamefont {C.~W.}\ \bibnamefont {Misner}},\
  }\bibfield  {title} {\bibinfo {title} {{Dynamical Structure and Definition of
  Energy in General Relativity}},\ }\href
  {https://doi.org/10.1103/PhysRev.116.1322} {\bibfield  {journal} {\bibinfo
  {journal} {Phys. Rev.}\ }\textbf {\bibinfo {volume} {116}},\ \bibinfo {pages}
  {1322} (\bibinfo {year} {1959})}\BibitemShut {NoStop}%
\bibitem [{\citenamefont {Arnowitt}\ \emph {et~al.}(1960)\citenamefont
  {Arnowitt}, \citenamefont {Deser},\ and\ \citenamefont
  {Misner}}]{arnowitt60_canon_variab_gener}%
  \BibitemOpen
  \bibfield  {author} {\bibinfo {author} {\bibfnamefont {R.~L.}\ \bibnamefont
  {Arnowitt}}, \bibinfo {author} {\bibfnamefont {S.}~\bibnamefont {Deser}},\
  and\ \bibinfo {author} {\bibfnamefont {C.~W.}\ \bibnamefont {Misner}},\
  }\bibfield  {title} {\bibinfo {title} {{Canonical Variables for General
  relativity}},\ }\href {https://doi.org/10.1103/PhysRev.117.1595} {\bibfield
  {journal} {\bibinfo  {journal} {Phys. Rev.}\ }\textbf {\bibinfo {volume}
  {117}},\ \bibinfo {pages} {1595} (\bibinfo {year} {1960})}\BibitemShut
  {NoStop}%
\bibitem [{\citenamefont {Wheeler}(1964)}]{wheeler64_relat}%
  \BibitemOpen
  \bibfield  {author} {\bibinfo {author} {\bibfnamefont {J.~A.}\ \bibnamefont
  {Wheeler}},\ }\bibinfo {title} {{R}elativity, groups and topology}\ (\bibinfo
   {publisher} {Gordon and Breach},\ \bibinfo {year} {1964})\ Chap.\ \bibinfo
  {chapter} {{G}eometrodynamics and the issue of the final state}, p.\ \bibinfo
  {pages} {317}\BibitemShut {NoStop}%
\bibitem [{\citenamefont
  {DeWitt}(1967{\natexlab{a}})}]{dewitt67_quant_theor_gravit_i}%
  \BibitemOpen
  \bibfield  {author} {\bibinfo {author} {\bibfnamefont {B.~S.}\ \bibnamefont
  {DeWitt}},\ }\bibfield  {title} {\bibinfo {title} {{Quantum Theory of
  Gravity. 1. The Canonical Theory}},\ }\href
  {https://doi.org/10.1103/PhysRev.160.1113} {\bibfield  {journal} {\bibinfo
  {journal} {Phys. Rev.}\ }\textbf {\bibinfo {volume} {160}},\ \bibinfo {pages}
  {1113} (\bibinfo {year} {1967}{\natexlab{a}})}\BibitemShut {NoStop}%
\bibitem [{\citenamefont
  {DeWitt}(1967{\natexlab{b}})}]{dewitt67_quant_theor_gravit_ii}%
  \BibitemOpen
  \bibfield  {author} {\bibinfo {author} {\bibfnamefont {B.~S.}\ \bibnamefont
  {DeWitt}},\ }\bibfield  {title} {\bibinfo {title} {{Quantum Theory of
  Gravity. 2. The Manifestly Covariant Theory}},\ }\href
  {https://doi.org/10.1103/PhysRev.162.1195} {\bibfield  {journal} {\bibinfo
  {journal} {Phys. Rev.}\ }\textbf {\bibinfo {volume} {162}},\ \bibinfo {pages}
  {1195} (\bibinfo {year} {1967}{\natexlab{b}})}\BibitemShut {NoStop}%
\bibitem [{\citenamefont
  {DeWitt}(1967{\natexlab{c}})}]{dewitt67_quant_theor_gravit_iii}%
  \BibitemOpen
  \bibfield  {author} {\bibinfo {author} {\bibfnamefont {B.~S.}\ \bibnamefont
  {DeWitt}},\ }\bibfield  {title} {\bibinfo {title} {{Quantum Theory of
  Gravity. 3. Applications of the Covariant Theory}},\ }\href
  {https://doi.org/10.1103/PhysRev.162.1239} {\bibfield  {journal} {\bibinfo
  {journal} {Phys. Rev.}\ }\textbf {\bibinfo {volume} {162}},\ \bibinfo {pages}
  {1239} (\bibinfo {year} {1967}{\natexlab{c}})}\BibitemShut {NoStop}%
\bibitem [{\citenamefont {'t~Hooft}(1973)}]{hooft73_algor_poles_at_dimen_four}%
  \BibitemOpen
  \bibfield  {author} {\bibinfo {author} {\bibfnamefont {G.}~\bibnamefont
  {'t~Hooft}},\ }\bibfield  {title} {\bibinfo {title} {{An Algorithm for the
  Poles At Dimension Four in the Dimensional Regularization procedure}},\
  }\href {https://doi.org/10.1016/0550-3213(73)90263-0} {\bibfield  {journal}
  {\bibinfo  {journal} {Nucl. Phys. B}\ }\textbf {\bibinfo {volume} {62}},\
  \bibinfo {pages} {444} (\bibinfo {year} {1973})}\BibitemShut {NoStop}%
\bibitem [{\citenamefont {'t~Hooft}\ and\ \citenamefont
  {Veltman}(1974)}]{hooft74_one_loop_diver_theor}%
  \BibitemOpen
  \bibfield  {author} {\bibinfo {author} {\bibfnamefont {G.}~\bibnamefont
  {'t~Hooft}}\ and\ \bibinfo {author} {\bibfnamefont {M.~J.~G.}\ \bibnamefont
  {Veltman}},\ }\bibfield  {title} {\bibinfo {title} {{One Loop Divergencies in
  the Theory of gravitation}},\ }\href@noop {} {\bibfield  {journal} {\bibinfo
  {journal} {Annales Poincare Phys. Theor. A}\ }\textbf {\bibinfo {volume}
  {20}},\ \bibinfo {pages} {69} (\bibinfo {year} {1974})}\BibitemShut {NoStop}%
\bibitem [{\citenamefont
  {'t~Hooft}(1976{\natexlab{a}})}]{hooft76_comput_quant_effec_due_to}%
  \BibitemOpen
  \bibfield  {author} {\bibinfo {author} {\bibfnamefont {G.}~\bibnamefont
  {'t~Hooft}},\ }\bibfield  {title} {\bibinfo {title} {Computation of the
  quantum effects due to a four-dimensional pseudoparticle},\ }\href
  {https://doi.org/10.1103/physrevd.14.3432} {\bibfield  {journal} {\bibinfo
  {journal} {Phys. Rev. D}\ }\textbf {\bibinfo {volume} {14}},\ \bibinfo
  {pages} {3432} (\bibinfo {year} {1976}{\natexlab{a}})}\BibitemShut {NoStop}%
\bibitem [{\citenamefont
  {'t~Hooft}(1976{\natexlab{b}})}]{hooft76_symmet_break_throug_bell_jackiw_anomal}%
  \BibitemOpen
  \bibfield  {author} {\bibinfo {author} {\bibfnamefont {G.}~\bibnamefont
  {'t~Hooft}},\ }\bibfield  {title} {\bibinfo {title} {{Symmetry Breaking
  Through Bell-Jackiw Anomalies}},\ }\href
  {https://doi.org/10.1103/PhysRevLett.37.8} {\bibfield  {journal} {\bibinfo
  {journal} {Phys. Rev. Lett.}\ }\textbf {\bibinfo {volume} {37}},\ \bibinfo
  {pages} {8} (\bibinfo {year} {1976}{\natexlab{b}})}\BibitemShut {NoStop}%
\bibitem [{\citenamefont {'t~Hooft}(1978)}]{hooft78_errat}%
  \BibitemOpen
  \bibfield  {author} {\bibinfo {author} {\bibfnamefont {G.}~\bibnamefont
  {'t~Hooft}},\ }\bibfield  {title} {\bibinfo {title} {{Erratum: Computation of
  the Quantum Effects Due To a Four-Dimensional Pseudoparticle}},\ }\href
  {https://doi.org/10.1103/physrevd.18.2199.3} {\bibfield  {journal} {\bibinfo
  {journal} {Phys. Rev. D}\ }\textbf {\bibinfo {volume} {18}},\ \bibinfo
  {pages} {2199} (\bibinfo {year} {1978})}\BibitemShut {NoStop}%
\bibitem [{\citenamefont
  {Papantonopoulos}(2007)}]{papantonopoulos07_invis_univer}%
  \BibitemOpen
  \bibfield  {author} {\bibinfo {author} {\bibfnamefont {L.}~\bibnamefont
  {Papantonopoulos}},\ }\href@noop {} {\emph {\bibinfo {title} {The Invisible
  Universe: Dark Matter and Dark Energy}}},\ \bibinfo {edition} {1st}\ ed.,\
  \bibinfo {series} {Lecture Notes in Physics}, Vol.\ \bibinfo {volume} {720}\
  (\bibinfo  {publisher} {Springer-Verlag},\ \bibinfo {address} {Berlin},\
  \bibinfo {year} {2007})\BibitemShut {NoStop}%
\bibitem [{\citenamefont {Weinberg}(2008)}]{weinberg08_cosmol}%
  \BibitemOpen
  \bibfield  {author} {\bibinfo {author} {\bibfnamefont {S.}~\bibnamefont
  {Weinberg}},\ }\href@noop {} {\emph {\bibinfo {title} {Cosmology}}}\
  (\bibinfo  {publisher} {Oxford},\ \bibinfo {address} {London},\ \bibinfo
  {year} {2008})\BibitemShut {NoStop}%
\bibitem [{\citenamefont {Dodelson}\ and\ \citenamefont
  {Schmidt}(2020)}]{dodelson20_moder}%
  \BibitemOpen
  \bibfield  {author} {\bibinfo {author} {\bibfnamefont {S.}~\bibnamefont
  {Dodelson}}\ and\ \bibinfo {author} {\bibfnamefont {F.}~\bibnamefont
  {Schmidt}},\ }\href@noop {} {\emph {\bibinfo {title} {Modern cosmology}}}\
  (\bibinfo  {publisher} {Academic press},\ \bibinfo {year} {2020})\BibitemShut
  {NoStop}%
\bibitem [{\citenamefont
  {Lovelock}(1969)}]{lovelock69_uniquen_einst_field_equat_four_dimen_space}%
  \BibitemOpen
  \bibfield  {author} {\bibinfo {author} {\bibfnamefont {D.}~\bibnamefont
  {Lovelock}},\ }\bibfield  {title} {\bibinfo {title} {The uniqueness of the
  einstein field equations in a four-dimensional space},\ }\href
  {https://doi.org/10.1007/BF00248156} {\bibfield  {journal} {\bibinfo
  {journal} {Archive for Rational Mechanics and Analysis}\ }\textbf {\bibinfo
  {volume} {33}},\ \bibinfo {pages} {54} (\bibinfo {year} {1969})}\BibitemShut
  {NoStop}%
\bibitem [{\citenamefont {Lovelock}(1971)}]{lovelock71_einst_tensor_its_gener}%
  \BibitemOpen
  \bibfield  {author} {\bibinfo {author} {\bibfnamefont {D.}~\bibnamefont
  {Lovelock}},\ }\bibfield  {title} {\bibinfo {title} {The einstein tensor and
  its generalizations},\ }\href@noop {} {\bibfield  {journal} {\bibinfo
  {journal} {J. Math. Phys.}\ }\textbf {\bibinfo {volume} {12}},\ \bibinfo
  {pages} {498} (\bibinfo {year} {1971})}\BibitemShut {NoStop}%
\bibitem [{Note1()}]{Note1}%
  \BibitemOpen
  \bibinfo {note} {Lanczos--Lovelock models of gravity might be inspired by
  Sakharov's proposal that General Relativity might be an effective model that
  receives higher curvature corrections \cite
  {sakharov00_vacuum_quant_fluct_curved}.}\BibitemShut {Stop}%
\bibitem [{\citenamefont {Hehl}\ \emph {et~al.}(1995)\citenamefont {Hehl},
  \citenamefont {McCrea}, \citenamefont {Mielke},\ and\ \citenamefont
  {Ne'eman}}]{hehl95_metric_affin_gauge_theor_gravit}%
  \BibitemOpen
  \bibfield  {author} {\bibinfo {author} {\bibfnamefont {F.~W.}\ \bibnamefont
  {Hehl}}, \bibinfo {author} {\bibfnamefont {J.~D.}\ \bibnamefont {McCrea}},
  \bibinfo {author} {\bibfnamefont {E.~W.}\ \bibnamefont {Mielke}},\ and\
  \bibinfo {author} {\bibfnamefont {Y.}~\bibnamefont {Ne'eman}},\ }\bibfield
  {title} {\bibinfo {title} {{Metric Affine Gauge Theory of Gravity: Field
  Equations, Noether Identities, World Spinors, and Breaking of Dilation
  invariance}},\ }\href {https://doi.org/10.1016/0370-1573(94)00111-F}
  {\bibfield  {journal} {\bibinfo  {journal} {Phys. Rep.}\ }\textbf {\bibinfo
  {volume} {258}},\ \bibinfo {pages} {1} (\bibinfo {year} {1995})},\ \Eprint
  {https://arxiv.org/abs/gr-qc/9402012} {arXiv:gr-qc/9402012 [gr-qc]}
  \BibitemShut {NoStop}%
\bibitem [{\citenamefont {Weyl}(1921)}]{weyl21_zur_infin}%
  \BibitemOpen
  \bibfield  {author} {\bibinfo {author} {\bibfnamefont {H.}~\bibnamefont
  {Weyl}},\ }\bibinfo {title} {Zur infinitesimalgeometrie: Einordnung der
  projektiven und konformen auffassung}\ (\bibinfo  {publisher} {G\"ottingen
  Nachrichten},\ \bibinfo {year} {1921})\ Chap.~\bibinfo {chapter} {7}, pp.\
  \bibinfo {pages} {99--112}\BibitemShut {NoStop}%
\bibitem [{\citenamefont {Weyl}(1922)}]{weyl22_space}%
  \BibitemOpen
  \bibfield  {author} {\bibinfo {author} {\bibfnamefont {H.}~\bibnamefont
  {Weyl}},\ }\href@noop {} {\emph {\bibinfo {title} {Space--time--matter}}}\
  (\bibinfo  {publisher} {Dutton},\ \bibinfo {year} {1922})\BibitemShut
  {NoStop}%
\bibitem [{\citenamefont {Cartan}(1922)}]{cartan22_sur_une_de_la_notion}%
  \BibitemOpen
  \bibfield  {author} {\bibinfo {author} {\bibfnamefont {E.}~\bibnamefont
  {Cartan}},\ }\bibfield  {title} {\bibinfo {title} {Sur une g\'en\'eralisation
  de la notion de courbure de riemann et les espaces \`a torsion},\ }\href
  {http://gallica.bnf.fr/ark:/12148/bpt6k3127j.image.langFR} {\bibfield
  {journal} {\bibinfo  {journal} {C. R. Acad. Sci. Paris}\ }\textbf {\bibinfo
  {volume} {174}},\ \bibinfo {pages} {593} (\bibinfo {year}
  {1922})}\BibitemShut {NoStop}%
\bibitem [{\citenamefont {Cartan}(1923)}]{cartan23_sur_les_connex_affin_et}%
  \BibitemOpen
  \bibfield  {author} {\bibinfo {author} {\bibfnamefont {E.}~\bibnamefont
  {Cartan}},\ }\bibfield  {title} {\bibinfo {title} {Sur les vari{\'e}t{\'e}s
  {\`a} connexion affine et la th{\'e}orie de la relativit{\'e}
  g{\'e}n{\'e}ralis{\'e}e (premi{\`e}re partie)},\ }\href
  {http://archive.numdam.org/article/ASENS_1923_3_40__325_0.pdf} {\bibfield
  {journal} {\bibinfo  {journal} {Ann. Ec. Norm. Super.}\ }\textbf {\bibinfo
  {volume} {40}},\ \bibinfo {pages} {325} (\bibinfo {year} {1923})}\BibitemShut
  {NoStop}%
\bibitem [{\citenamefont {Cartan}(1924)}]{cartan24_sur_les_connex_affin_et}%
  \BibitemOpen
  \bibfield  {author} {\bibinfo {author} {\bibfnamefont {E.}~\bibnamefont
  {Cartan}},\ }\bibfield  {title} {\bibinfo {title} {Sur les vari\'et\'es \`a
  connexion affine, et la th\'eorie de la relativit\'e g\'en\'eralis\'ee
  (premi\`ere partie) (suite)},\ }\href
  {http://www.numdam.org/numdam-bin/item?id=ASENS_1924_3_41__1_0} {\bibfield
  {journal} {\bibinfo  {journal} {Ann. Ec. Norm. Super.}\ }\textbf {\bibinfo
  {volume} {41}},\ \bibinfo {pages} {1} (\bibinfo {year} {1924})}\BibitemShut
  {NoStop}%
\bibitem [{\citenamefont {Cartan}(1925)}]{cartan25_sur_les_connex_affin_et}%
  \BibitemOpen
  \bibfield  {author} {\bibinfo {author} {\bibfnamefont {E.}~\bibnamefont
  {Cartan}},\ }\bibfield  {title} {\bibinfo {title} {Sur les vari\'et\'es \`a
  connexion affine et la th\'eorie de la relativit\'e g\'en\'eralis\'ee,
  (deuxi\`eme partie)},\ }\href
  {http://www.numdam.org/numdam-bin/item?id=ASENS_1925_3_42__17_0} {\bibfield
  {journal} {\bibinfo  {journal} {Ann. Ec. Norm. Super.}\ }\textbf {\bibinfo
  {volume} {42}},\ \bibinfo {pages} {17} (\bibinfo {year} {1925})}\BibitemShut
  {NoStop}%
\bibitem [{\citenamefont {Veblen}(1933)}]{veblen33_projec_gener_relat}%
  \BibitemOpen
  \bibfield  {author} {\bibinfo {author} {\bibfnamefont {O.}~\bibnamefont
  {Veblen}},\ }\href@noop {} {\emph {\bibinfo {title} {Projective General
  Relativity}}},\ \bibinfo {series} {Advances in Mathematics and Related
  Topics}, Vol.~\bibinfo {volume} {2}\ (\bibinfo  {publisher} {Springer},\
  \bibinfo {year} {1933})\BibitemShut {NoStop}%
\bibitem [{\citenamefont
  {Einstein}(1923{\natexlab{a}})}]{einstein23_zur_affin_feldt}%
  \BibitemOpen
  \bibfield  {author} {\bibinfo {author} {\bibfnamefont {A.}~\bibnamefont
  {Einstein}},\ }\bibfield  {title} {\bibinfo {title} {Zur affinen
  feldtheorie},\ }\href@noop {} {\bibfield  {journal} {\bibinfo  {journal}
  {Sitzungsber. Preuss. Akad. Wiss.}\ ,\ \bibinfo {pages} {137}} (\bibinfo
  {year} {1923}{\natexlab{a}})}\BibitemShut {NoStop}%
\bibitem [{\citenamefont
  {Einstein}(1923{\natexlab{b}})}]{einstein23_theor_affin_field}%
  \BibitemOpen
  \bibfield  {author} {\bibinfo {author} {\bibfnamefont {A.}~\bibnamefont
  {Einstein}},\ }\bibfield  {title} {\bibinfo {title} {Theory of the affine
  field},\ }\href@noop {} {\bibfield  {journal} {\bibinfo  {journal} {Nature}\
  }\textbf {\bibinfo {volume} {112}},\ \bibinfo {pages} {448} (\bibinfo {year}
  {1923}{\natexlab{b}})}\BibitemShut {NoStop}%
\bibitem [{\citenamefont {Eddington}(1923)}]{eddington23}%
  \BibitemOpen
  \bibfield  {author} {\bibinfo {author} {\bibfnamefont {A.~S.}\ \bibnamefont
  {Eddington}},\ }\href@noop {} {\emph {\bibinfo {title} {The mathematical
  theory of relativity}}}\ (\bibinfo  {publisher} {Cambridge University
  Press},\ \bibinfo {address} {London},\ \bibinfo {year} {1923})\BibitemShut
  {NoStop}%
\bibitem [{\citenamefont
  {Schr\"odinger}(1948{\natexlab{a}})}]{schroedinger46_general_affin_field_laws}%
  \BibitemOpen
  \bibfield  {author} {\bibinfo {author} {\bibfnamefont {E.}~\bibnamefont
  {Schr\"odinger}},\ }\bibfield  {title} {\bibinfo {title} {{The General Affine
  Field Laws I}},\ }\bibfield  {journal} {\bibinfo  {journal} {Proc. R. Ir.
  Acad.}\ }\textbf {\bibinfo {volume} {51}},\ \href
  {https://doi.org/10.2307/20488470} {10.2307/20488470} (\bibinfo {year}
  {1945-1948}{\natexlab{a}})\BibitemShut {NoStop}%
\bibitem [{\citenamefont
  {Schr\"odinger}(1948{\natexlab{b}})}]{schroedinger47_final_affin_field_laws_i}%
  \BibitemOpen
  \bibfield  {author} {\bibinfo {author} {\bibfnamefont {E.}~\bibnamefont
  {Schr\"odinger}},\ }\bibfield  {title} {\bibinfo {title} {{The Final Affine
  Field Laws I}},\ }\bibfield  {journal} {\bibinfo  {journal} {Proc. R. Ir.
  Acad.}\ }\textbf {\bibinfo {volume} {51}},\ \href
  {https://doi.org/10.2307/20488479} {10.2307/20488479} (\bibinfo {year}
  {1945-1948}{\natexlab{b}})\BibitemShut {NoStop}%
\bibitem [{\citenamefont
  {Schr\"odinger}(1948{\natexlab{c}})}]{schroedinger48_final_affin_field_laws_ii}%
  \BibitemOpen
  \bibfield  {author} {\bibinfo {author} {\bibfnamefont {E.}~\bibnamefont
  {Schr\"odinger}},\ }\bibfield  {title} {\bibinfo {title} {{The Final Affine
  Field Laws II}},\ }\bibfield  {journal} {\bibinfo  {journal} {Proc. R. Ir.
  Acad.}\ }\textbf {\bibinfo {volume} {51}},\ \href
  {https://doi.org/10.2307/20488482} {10.2307/20488482} (\bibinfo {year}
  {1945-1948}{\natexlab{c}})\BibitemShut {NoStop}%
\bibitem [{\citenamefont
  {Schr\"odinger}(1948{\natexlab{d}})}]{schroedinger48_final_affin_field_laws_iii}%
  \BibitemOpen
  \bibfield  {author} {\bibinfo {author} {\bibfnamefont {E.}~\bibnamefont
  {Schr\"odinger}},\ }\bibfield  {title} {\bibinfo {title} {{The Final Affine
  Field Laws III}},\ }\bibfield  {journal} {\bibinfo  {journal} {Proc. R. Ir.
  Acad.}\ }\textbf {\bibinfo {volume} {52}},\ \href
  {https://doi.org/10.2307/20488487} {10.2307/20488487} (\bibinfo {year}
  {1945-1948}{\natexlab{d}})\BibitemShut {NoStop}%
\bibitem [{\citenamefont {Hlavat\'y}(1957)}]{hlavaty57_geomet_einst}%
  \BibitemOpen
  \bibfield  {author} {\bibinfo {author} {\bibfnamefont {V.}~\bibnamefont
  {Hlavat\'y}},\ }\href@noop {} {\emph {\bibinfo {title} {Geometry of
  Einstein's unified field theory}}}\ (\bibinfo  {publisher} {Noordhoff},\
  \bibinfo {year} {1957})\BibitemShut {NoStop}%
\bibitem [{\citenamefont {Tonnelat}(2014)}]{tonnelat14_einst}%
  \BibitemOpen
  \bibfield  {author} {\bibinfo {author} {\bibfnamefont {M.-A.}\ \bibnamefont
  {Tonnelat}},\ }\href@noop {} {\emph {\bibinfo {title} {Einstein's theory of
  unified fields}}},\ \bibinfo {edition} {2nd}\ ed.,\ \bibinfo {series}
  {Routledge library editions: 20th century science}, Vol.~\bibinfo {volume}
  {19}\ (\bibinfo  {publisher} {Routledge, Taylor \& Francis},\ \bibinfo {year}
  {2014})\BibitemShut {NoStop}%
\bibitem [{\citenamefont {Kaluza}(1921)}]{kaluza21_probl_unity_physic}%
  \BibitemOpen
  \bibfield  {author} {\bibinfo {author} {\bibfnamefont {T.}~\bibnamefont
  {Kaluza}},\ }\bibfield  {title} {\bibinfo {title} {{On the Problem of Unity
  in Physics}},\ }\href@noop {} {\bibfield  {journal} {\bibinfo  {journal}
  {Sitzungsber. Preuss. Akad. Wiss. Berlin (Math. Phys.)}\ }\textbf {\bibinfo
  {volume} {1921}},\ \bibinfo {pages} {966} (\bibinfo {year}
  {1921})}\BibitemShut {NoStop}%
\bibitem [{\citenamefont {Klein}(1926)}]{klein26_quant_theor_five_dimen_theor}%
  \BibitemOpen
  \bibfield  {author} {\bibinfo {author} {\bibfnamefont {O.}~\bibnamefont
  {Klein}},\ }\bibfield  {title} {\bibinfo {title} {{Quantum Theory and
  Five-Dimensional Theory of relativity}},\ }\href
  {https://doi.org/10.1007/BF01397481} {\bibfield  {journal} {\bibinfo
  {journal} {Z. Phys.}\ }\textbf {\bibinfo {volume} {37}},\ \bibinfo {pages}
  {895} (\bibinfo {year} {1926})}\BibitemShut {NoStop}%
\bibitem [{\citenamefont {Einstein}\ and\ \citenamefont
  {Bergmann}(1938)}]{einstein38_gener_kaluz_theor_elect}%
  \BibitemOpen
  \bibfield  {author} {\bibinfo {author} {\bibfnamefont {A.}~\bibnamefont
  {Einstein}}\ and\ \bibinfo {author} {\bibfnamefont {P.}~\bibnamefont
  {Bergmann}},\ }\bibfield  {title} {\bibinfo {title} {{On a Generalization of
  Kaluza's Theory of Electricity}},\ }\href {https://doi.org/10.2307/1968642}
  {\bibfield  {journal} {\bibinfo  {journal} {Annals Math.}\ }\textbf {\bibinfo
  {volume} {39}},\ \bibinfo {pages} {683} (\bibinfo {year} {1938})}\BibitemShut
  {NoStop}%
\bibitem [{\citenamefont {Clifton}\ \emph {et~al.}(2012)\citenamefont
  {Clifton}, \citenamefont {Ferreira}, \citenamefont {Padilla},\ and\
  \citenamefont {Skordis}}]{clifton12_modif_gravit_cosmol}%
  \BibitemOpen
  \bibfield  {author} {\bibinfo {author} {\bibfnamefont {T.}~\bibnamefont
  {Clifton}}, \bibinfo {author} {\bibfnamefont {P.~G.}\ \bibnamefont
  {Ferreira}}, \bibinfo {author} {\bibfnamefont {A.}~\bibnamefont {Padilla}},\
  and\ \bibinfo {author} {\bibfnamefont {C.}~\bibnamefont {Skordis}},\
  }\bibfield  {title} {\bibinfo {title} {Modified gravity and cosmology},\
  }\href {https://doi.org/10.1016/j.physrep.2012.01.001} {\bibfield  {journal}
  {\bibinfo  {journal} {Phys. Rep.}\ }\textbf {\bibinfo {volume} {513}},\
  \bibinfo {pages} {1} (\bibinfo {year} {2012})}\BibitemShut {NoStop}%
\bibitem [{\citenamefont {Saridakis}\ \emph {et~al.}(2021)\citenamefont
  {Saridakis}, \citenamefont {Lazkoz}, \citenamefont {Salzano}, \citenamefont
  {Moniz}, \citenamefont {Capozziello}, \citenamefont {Jim{\'e}nez},
  \citenamefont {Laurentis},\ and\ \citenamefont
  {Olmo}}]{saridakis21_modif_gravit_cosmol}%
  \BibitemOpen
  \bibfield  {author} {\bibinfo {author} {\bibfnamefont {E.~N.}\ \bibnamefont
  {Saridakis}}, \bibinfo {author} {\bibfnamefont {R.}~\bibnamefont {Lazkoz}},
  \bibinfo {author} {\bibfnamefont {V.}~\bibnamefont {Salzano}}, \bibinfo
  {author} {\bibfnamefont {P.~V.}\ \bibnamefont {Moniz}}, \bibinfo {author}
  {\bibfnamefont {S.}~\bibnamefont {Capozziello}}, \bibinfo {author}
  {\bibfnamefont {J.~B.}\ \bibnamefont {Jim{\'e}nez}}, \bibinfo {author}
  {\bibfnamefont {M.~D.}\ \bibnamefont {Laurentis}},\ and\ \bibinfo {author}
  {\bibfnamefont {G.~J.}\ \bibnamefont {Olmo}},\ }\href
  {http://arxiv.org/abs/2105.12582v2} {\emph {\bibinfo {title} {Modified
  Gravity and Cosmology: an Update By the Cantata Network}}}\ (\bibinfo
  {publisher} {Springer},\ \bibinfo {year} {2021})\ \Eprint
  {https://arxiv.org/abs/2105.12582} {arXiv:2105.12582 [gr-qc]} \BibitemShut
  {NoStop}%
\bibitem [{\citenamefont {Aldrovandi}\ and\ \citenamefont
  {Pereira}(2013)}]{aldrovandi13_telep_gravit}%
  \BibitemOpen
  \bibfield  {author} {\bibinfo {author} {\bibfnamefont {R.}~\bibnamefont
  {Aldrovandi}}\ and\ \bibinfo {author} {\bibfnamefont {J.~G.}\ \bibnamefont
  {Pereira}},\ }\href {https://doi.org/10.1007/978-94-007-5143-9} {\emph
  {\bibinfo {title} {Teleparallel Gravity}}},\ \bibinfo {series} {Fundamental
  Theories of Physics}, Vol.\ \bibinfo {volume} {173}\ (\bibinfo  {publisher}
  {Springer},\ \bibinfo {address} {London},\ \bibinfo {year}
  {2013})\BibitemShut {NoStop}%
\bibitem [{\citenamefont {Bahamonde}\ \emph {et~al.}(2023)\citenamefont
  {Bahamonde}, \citenamefont {Dialektopoulos}, \citenamefont
  {Escamilla-Rivera}, \citenamefont {Farrugia}, \citenamefont {Gakis},
  \citenamefont {Hendry}, \citenamefont {Hohmann}, \citenamefont {Said},
  \citenamefont {Mifsud},\ and\ \citenamefont
  {Valentino}}]{bahamonde21_telep_gravit}%
  \BibitemOpen
  \bibfield  {author} {\bibinfo {author} {\bibfnamefont {S.}~\bibnamefont
  {Bahamonde}}, \bibinfo {author} {\bibfnamefont {K.~F.}\ \bibnamefont
  {Dialektopoulos}}, \bibinfo {author} {\bibfnamefont {C.}~\bibnamefont
  {Escamilla-Rivera}}, \bibinfo {author} {\bibfnamefont {G.}~\bibnamefont
  {Farrugia}}, \bibinfo {author} {\bibfnamefont {V.}~\bibnamefont {Gakis}},
  \bibinfo {author} {\bibfnamefont {M.}~\bibnamefont {Hendry}}, \bibinfo
  {author} {\bibfnamefont {M.}~\bibnamefont {Hohmann}}, \bibinfo {author}
  {\bibfnamefont {J.~L.}\ \bibnamefont {Said}}, \bibinfo {author}
  {\bibfnamefont {J.}~\bibnamefont {Mifsud}},\ and\ \bibinfo {author}
  {\bibfnamefont {E.~D.}\ \bibnamefont {Valentino}},\ }\bibfield  {title}
  {\bibinfo {title} {Teleparallel gravity: From theory to cosmology},\ }\href
  {https://doi.org/10.1088/1361-6633/ac9cef} {\bibfield  {journal} {\bibinfo
  {journal} {Rep. Prog. Phys.}\ }\textbf {\bibinfo {volume} {86}},\ \bibinfo
  {pages} {026901} (\bibinfo {year} {2023})},\ \Eprint
  {https://arxiv.org/abs/2106.13793} {arXiv:2106.13793 [gr-qc]} \BibitemShut
  {NoStop}%
\bibitem [{\citenamefont {Jim{\'e}nez}\ \emph {et~al.}(2019)\citenamefont
  {Jim{\'e}nez}, \citenamefont {Heisenberg},\ and\ \citenamefont
  {Koivisto}}]{jimenez19_geomet_trinit_gravit}%
  \BibitemOpen
  \bibfield  {author} {\bibinfo {author} {\bibfnamefont {J.~B.}\ \bibnamefont
  {Jim{\'e}nez}}, \bibinfo {author} {\bibfnamefont {L.}~\bibnamefont
  {Heisenberg}},\ and\ \bibinfo {author} {\bibfnamefont {T.}~\bibnamefont
  {Koivisto}},\ }\bibfield  {title} {\bibinfo {title} {The geometrical trinity
  of gravity},\ }\href {https://doi.org/10.3390/universe5070173} {\bibfield
  {journal} {\bibinfo  {journal} {Universe}\ }\textbf {\bibinfo {volume} {5}},\
  \bibinfo {pages} {173} (\bibinfo {year} {2019})}\BibitemShut {NoStop}%
\bibitem [{Note2()}]{Note2}%
  \BibitemOpen
  \bibinfo {note} {In words of K. Krasnov \cite {krasnov20_formul_gener_relat},
  \begin {quote} There may be equivalent formulations of a theory, all leading
  to the same physical predictions. But such reformulations may be inequivalent
  if one decides to generalise. \end {quote}}\BibitemShut {NoStop}%
\bibitem [{\citenamefont {Sotiriou}\ and\ \citenamefont
  {Faraoni}(2010)}]{sotiriou10_r_theor_of_gravit}%
  \BibitemOpen
  \bibfield  {author} {\bibinfo {author} {\bibfnamefont {T.~P.}\ \bibnamefont
  {Sotiriou}}\ and\ \bibinfo {author} {\bibfnamefont {V.}~\bibnamefont
  {Faraoni}},\ }\bibfield  {title} {\bibinfo {title} {{\(f(R)\) Theories Of
  Gravity}},\ }\href {https://doi.org/10.1103/RevModPhys.82.451} {\bibfield
  {journal} {\bibinfo  {journal} {Rev. Mod. Phys.}\ }\textbf {\bibinfo {volume}
  {82}},\ \bibinfo {pages} {451} (\bibinfo {year} {2010})},\ \Eprint
  {https://arxiv.org/abs/0805.1726} {arXiv:0805.1726 [gr-qc]} \BibitemShut
  {NoStop}%
\bibitem [{\citenamefont {Olmo}(2011)}]{olmo11_palat_approac_to_modif_gravit}%
  \BibitemOpen
  \bibfield  {author} {\bibinfo {author} {\bibfnamefont {G.~J.}\ \bibnamefont
  {Olmo}},\ }\bibfield  {title} {\bibinfo {title} {{Palatini Approach To
  Modified Gravity: f(R) Theories and Beyond}},\ }\href
  {https://doi.org/10.1142/S0218271811018925} {\bibfield  {journal} {\bibinfo
  {journal} {Int. J. Mod. Phys. D}\ }\textbf {\bibinfo {volume} {20}},\
  \bibinfo {pages} {413} (\bibinfo {year} {2011})},\ \Eprint
  {https://arxiv.org/abs/1101.3864} {arXiv:1101.3864 [gr-qc]} \BibitemShut
  {NoStop}%
\bibitem [{\citenamefont {De~Felice}\ and\ \citenamefont
  {Tsujikawa}(2010)}]{de10_r}%
  \BibitemOpen
  \bibfield  {author} {\bibinfo {author} {\bibfnamefont {A.}~\bibnamefont
  {De~Felice}}\ and\ \bibinfo {author} {\bibfnamefont {S.}~\bibnamefont
  {Tsujikawa}},\ }\bibfield  {title} {\bibinfo {title} {{\(f(R)\) theories}},\
  }\href {https://doi.org/10.12942/lrr-2010-3} {\bibfield  {journal} {\bibinfo
  {journal} {Living Rev. Relativ.}\ }\textbf {\bibinfo {volume} {13}},\
  \bibinfo {pages} {3} (\bibinfo {year} {2010})},\ \Eprint
  {https://arxiv.org/abs/1002.4928} {arXiv:1002.4928 [gr-qc]} \BibitemShut
  {NoStop}%
\bibitem [{\citenamefont {Cai}\ \emph {et~al.}(2016)\citenamefont {Cai},
  \citenamefont {Capozziello}, \citenamefont {Laurentis},\ and\ \citenamefont
  {Saridakis}}]{cai16_t_telep_gravit_cosmol}%
  \BibitemOpen
  \bibfield  {author} {\bibinfo {author} {\bibfnamefont {Y.-F.}\ \bibnamefont
  {Cai}}, \bibinfo {author} {\bibfnamefont {S.}~\bibnamefont {Capozziello}},
  \bibinfo {author} {\bibfnamefont {M.~D.}\ \bibnamefont {Laurentis}},\ and\
  \bibinfo {author} {\bibfnamefont {E.~N.}\ \bibnamefont {Saridakis}},\
  }\bibfield  {title} {\bibinfo {title} {f(t) teleparallel gravity and
  cosmology},\ }\href {https://doi.org/10.1088/0034-4885/79/10/106901}
  {\bibfield  {journal} {\bibinfo  {journal} {Rep. Prog. Phys.}\ }\textbf
  {\bibinfo {volume} {79}},\ \bibinfo {pages} {106901} (\bibinfo {year}
  {2016})},\ \Eprint {https://arxiv.org/abs/1511.07586} {arXiv:1511.07586
  [gr-qc]} \BibitemShut {NoStop}%
\bibitem [{\citenamefont {Heisenberg}(2023)}]{heisenberg23_review_q_gravit}%
  \BibitemOpen
  \bibfield  {author} {\bibinfo {author} {\bibfnamefont {L.}~\bibnamefont
  {Heisenberg}},\ }\href {http://arxiv.org/abs/2309.15958v1} {\bibinfo {title}
  {Review on $f(q)$ gravity}} (\bibinfo {year} {2023}),\ \Eprint
  {https://arxiv.org/abs/2309.15958} {arXiv:2309.15958 [gr-qc]} \BibitemShut
  {NoStop}%
\bibitem [{\citenamefont {Jim{\'e}nez}\ \emph {et~al.}(2020)\citenamefont
  {Jim{\'e}nez}, \citenamefont {Heisenberg}, \citenamefont {Koivisto},\ and\
  \citenamefont {Pekar}}]{jimenez20_cosmol_q_geomet}%
  \BibitemOpen
  \bibfield  {author} {\bibinfo {author} {\bibfnamefont {J.~B.}\ \bibnamefont
  {Jim{\'e}nez}}, \bibinfo {author} {\bibfnamefont {L.}~\bibnamefont
  {Heisenberg}}, \bibinfo {author} {\bibfnamefont {T.}~\bibnamefont
  {Koivisto}},\ and\ \bibinfo {author} {\bibfnamefont {S.}~\bibnamefont
  {Pekar}},\ }\bibfield  {title} {\bibinfo {title} {Cosmology in f(q)
  geometry},\ }\href {https://doi.org/10.1103/physrevd.101.103507} {\bibfield
  {journal} {\bibinfo  {journal} {Phys. Rev. D}\ }\textbf {\bibinfo {volume}
  {101}},\ \bibinfo {pages} {103507} (\bibinfo {year} {2020})}\BibitemShut
  {NoStop}%
\bibitem [{\citenamefont {Myrzakulov}(2012)}]{myrzakulov12_dark_energ_f_r}%
  \BibitemOpen
  \bibfield  {author} {\bibinfo {author} {\bibfnamefont {R.}~\bibnamefont
  {Myrzakulov}},\ }\href {http://arxiv.org/abs/1205.5266v8} {\bibinfo {title}
  {Metric-affine myrzakulov gravity and its generalizations}} (\bibinfo {year}
  {2012}),\ \Eprint {https://arxiv.org/abs/1205.5266} {arXiv:1205.5266
  [physics.gen-ph]} \BibitemShut {NoStop}%
\bibitem [{\citenamefont {Harko}\ \emph {et~al.}(2011)\citenamefont {Harko},
  \citenamefont {Lobo}, \citenamefont {Nojiri},\ and\ \citenamefont
  {Odintsov}}]{harko11_r_t}%
  \BibitemOpen
  \bibfield  {author} {\bibinfo {author} {\bibfnamefont {T.}~\bibnamefont
  {Harko}}, \bibinfo {author} {\bibfnamefont {F.~S.~N.}\ \bibnamefont {Lobo}},
  \bibinfo {author} {\bibfnamefont {S.}~\bibnamefont {Nojiri}},\ and\ \bibinfo
  {author} {\bibfnamefont {S.~D.}\ \bibnamefont {Odintsov}},\ }\bibfield
  {title} {\bibinfo {title} {f(r,t)gravity},\ }\bibfield  {journal} {\bibinfo
  {journal} {Physical Review D}\ }\textbf {\bibinfo {volume} {84}},\ \href
  {https://doi.org/10.1103/physrevd.84.024020} {10.1103/physrevd.84.024020}
  (\bibinfo {year} {2011}),\ \Eprint {https://arxiv.org/abs/1104.2669}
  {arXiv:1104.2669 [gr-qc]} \BibitemShut {NoStop}%
\bibitem [{\citenamefont
  {Nakayama}(2023)}]{nakayama22_geomet_trinit_unimod_gravit}%
  \BibitemOpen
  \bibfield  {author} {\bibinfo {author} {\bibfnamefont {Y.}~\bibnamefont
  {Nakayama}},\ }\bibfield  {title} {\bibinfo {title} {Geometrical trinity of
  unimodular gravity},\ }\bibfield  {journal} {\bibinfo  {journal} {Class.
  Quant. Grav.}\ }\textbf {\bibinfo {volume} {40}},\ \href
  {https://doi.org/10.1088/1361-6382/acd100} {10.1088/1361-6382/acd100}
  (\bibinfo {year} {2023}),\ \Eprint {https://arxiv.org/abs/2209.09462}
  {arXiv:2209.09462 [gr-qc]} \BibitemShut {NoStop}%
\bibitem [{\citenamefont
  {Kijowski}(1978)}]{kijowski78_new_variat_princ_gener_relat}%
  \BibitemOpen
  \bibfield  {author} {\bibinfo {author} {\bibfnamefont {J.}~\bibnamefont
  {Kijowski}},\ }\bibfield  {title} {\bibinfo {title} {On a new variational
  principle in general relativity and the energy of the gravitational field},\
  }\href {https://doi.org/10.1007/bf00759646} {\bibfield  {journal} {\bibinfo
  {journal} {Gen. Rel. Grav.}\ }\textbf {\bibinfo {volume} {9}},\ \bibinfo
  {pages} {857} (\bibinfo {year} {1978})}\BibitemShut {NoStop}%
\bibitem [{\citenamefont {Kijowski}\ and\ \citenamefont
  {Tulczyjew}(1979)}]{kijowski79}%
  \BibitemOpen
  \bibfield  {author} {\bibinfo {author} {\bibfnamefont {J.}~\bibnamefont
  {Kijowski}}\ and\ \bibinfo {author} {\bibfnamefont {W.~M.}\ \bibnamefont
  {Tulczyjew}},\ }\href@noop {} {\emph {\bibinfo {title} {A symplectic
  framework for field theories}}}\ (\bibinfo  {publisher} {Springer-Verlag},\
  \bibinfo {year} {1979})\BibitemShut {NoStop}%
\bibitem [{\citenamefont {Ferraris}\ and\ \citenamefont
  {Kijowski}(1981)}]{ferraris81_gener_relat_is_gauge_type_theor}%
  \BibitemOpen
  \bibfield  {author} {\bibinfo {author} {\bibfnamefont {M.}~\bibnamefont
  {Ferraris}}\ and\ \bibinfo {author} {\bibfnamefont {J.}~\bibnamefont
  {Kijowski}},\ }\bibfield  {title} {\bibinfo {title} {General relativity is a
  gauge type theory},\ }\href {https://doi.org/10.1007/bf00403241} {\bibfield
  {journal} {\bibinfo  {journal} {Lett. Math. Phys.}\ }\textbf {\bibinfo
  {volume} {5}},\ \bibinfo {pages} {127} (\bibinfo {year} {1981})}\BibitemShut
  {NoStop}%
\bibitem [{\citenamefont {Ferraris}\ and\ \citenamefont
  {Kijowski}(1982)}]{ferraris82_equiv_relat_theor_gravit}%
  \BibitemOpen
  \bibfield  {author} {\bibinfo {author} {\bibfnamefont {M.}~\bibnamefont
  {Ferraris}}\ and\ \bibinfo {author} {\bibfnamefont {J.}~\bibnamefont
  {Kijowski}},\ }\bibfield  {title} {\bibinfo {title} {On the equivalence of
  the relativistic theories of gravitation},\ }\href
  {https://doi.org/10.1007/bf00756921} {\bibfield  {journal} {\bibinfo
  {journal} {Gen. Rel. Grav.}\ }\textbf {\bibinfo {volume} {14}},\ \bibinfo
  {pages} {165} (\bibinfo {year} {1982})}\BibitemShut {NoStop}%
\bibitem [{\citenamefont {Kijowski}\ and\ \citenamefont
  {Werpachowski}(2007)}]{kijowski07_univer_affin_formul_gener_relat}%
  \BibitemOpen
  \bibfield  {author} {\bibinfo {author} {\bibfnamefont {J.}~\bibnamefont
  {Kijowski}}\ and\ \bibinfo {author} {\bibfnamefont {R.}~\bibnamefont
  {Werpachowski}},\ }\bibfield  {title} {\bibinfo {title} {Universality of
  affine formulation in general relativity},\ }\href
  {https://doi.org/10.1016/s0034-4877(07)80001-2} {\bibfield  {journal}
  {\bibinfo  {journal} {Rept. Math. Phys.}\ }\textbf {\bibinfo {volume} {59}},\
  \bibinfo {pages} {1} (\bibinfo {year} {2007})},\ \Eprint
  {https://arxiv.org/abs/gr-qc/0406088} {arXiv:gr-qc/0406088 [gr-qc]}
  \BibitemShut {NoStop}%
\bibitem [{\citenamefont {Krasnov}(2007)}]{krasnov07_non_metric_gravit}%
  \BibitemOpen
  \bibfield  {author} {\bibinfo {author} {\bibfnamefont {K.}~\bibnamefont
  {Krasnov}},\ }\bibfield  {title} {\bibinfo {title} {{Non-Metric Gravity: A
  Status report}},\ }\href {https://doi.org/10.1142/S021773230702590X}
  {\bibfield  {journal} {\bibinfo  {journal} {Mod. Phys. Lett. A}\ }\textbf
  {\bibinfo {volume} {22}},\ \bibinfo {pages} {3013} (\bibinfo {year}
  {2007})},\ \Eprint {https://arxiv.org/abs/0711.0697} {arXiv:0711.0697
  [gr-qc]} \BibitemShut {NoStop}%
\bibitem [{\citenamefont {Krasnov}(2008)}]{krasnov08_non_metric_gravit_i}%
  \BibitemOpen
  \bibfield  {author} {\bibinfo {author} {\bibfnamefont {K.}~\bibnamefont
  {Krasnov}},\ }\bibfield  {title} {\bibinfo {title} {{Non-Metric Gravity. I.
  Field Equations}},\ }\href {https://doi.org/10.1088/0264-9381/25/2/025001}
  {\bibfield  {journal} {\bibinfo  {journal} {Class. Quant. Grav.}\ }\textbf
  {\bibinfo {volume} {25}},\ \bibinfo {pages} {025001} (\bibinfo {year}
  {2008})},\ \Eprint {https://arxiv.org/abs/gr-qc/0703002} {arXiv:gr-qc/0703002
  [gr-qc]} \BibitemShut {NoStop}%
\bibitem [{\citenamefont {Krasnov}\ and\ \citenamefont
  {Shtanov}(2008)}]{krasnov08_non_metric_gravit_ii}%
  \BibitemOpen
  \bibfield  {author} {\bibinfo {author} {\bibfnamefont {K.}~\bibnamefont
  {Krasnov}}\ and\ \bibinfo {author} {\bibfnamefont {Y.}~\bibnamefont
  {Shtanov}},\ }\bibfield  {title} {\bibinfo {title} {{Non-Metric Gravity. II.
  Spherically Symmetric Solution, Missing Mass and Redshifts of Quasars}},\
  }\href {https://doi.org/10.1088/0264-9381/25/2/025002} {\bibfield  {journal}
  {\bibinfo  {journal} {Class. Quant. Grav.}\ }\textbf {\bibinfo {volume}
  {25}},\ \bibinfo {pages} {025002} (\bibinfo {year} {2008})},\ \Eprint
  {https://arxiv.org/abs/0705.2047} {arXiv:0705.2047 [gr-qc]} \BibitemShut
  {NoStop}%
\bibitem [{\citenamefont
  {Krasnov}(2011)}]{krasnov11_pure_connec_action_princ_gener_relat}%
  \BibitemOpen
  \bibfield  {author} {\bibinfo {author} {\bibfnamefont {K.}~\bibnamefont
  {Krasnov}},\ }\bibfield  {title} {\bibinfo {title} {{Pure Connection Action
  Principle for General Relativity}},\ }\href
  {https://doi.org/10.1103/PhysRevLett.106.251103} {\bibfield  {journal}
  {\bibinfo  {journal} {Phys. Rev. Lett.}\ }\textbf {\bibinfo {volume} {106}},\
  \bibinfo {pages} {251103} (\bibinfo {year} {2011})},\ \Eprint
  {https://arxiv.org/abs/1103.4498} {arXiv:1103.4498 [gr-qc]} \BibitemShut
  {NoStop}%
\bibitem [{\citenamefont {Delfino}\ \emph
  {et~al.}(2015{\natexlab{a}})\citenamefont {Delfino}, \citenamefont
  {Krasnov},\ and\ \citenamefont
  {Scarinci}}]{delfino15_pure_connec_formal_gravit_lin}%
  \BibitemOpen
  \bibfield  {author} {\bibinfo {author} {\bibfnamefont {G.}~\bibnamefont
  {Delfino}}, \bibinfo {author} {\bibfnamefont {K.}~\bibnamefont {Krasnov}},\
  and\ \bibinfo {author} {\bibfnamefont {C.}~\bibnamefont {Scarinci}},\
  }\bibfield  {title} {\bibinfo {title} {{Pure Connection Formalism for
  Gravity: Linearized Theory}},\ }\href
  {https://doi.org/10.1007/JHEP03(2015)118} {\bibfield  {journal} {\bibinfo
  {journal} {J. High Energy Phys.}\ }\textbf {\bibinfo {volume} {03}},\
  \bibinfo {pages} {118}},\ \Eprint {https://arxiv.org/abs/1205.7045}
  {arXiv:1205.7045 [hep-th]} \BibitemShut {NoStop}%
\bibitem [{\citenamefont {Delfino}\ \emph
  {et~al.}(2015{\natexlab{b}})\citenamefont {Delfino}, \citenamefont
  {Krasnov},\ and\ \citenamefont
  {Scarinci}}]{delfino15_pure_connec_formal_gravit_feyn}%
  \BibitemOpen
  \bibfield  {author} {\bibinfo {author} {\bibfnamefont {G.}~\bibnamefont
  {Delfino}}, \bibinfo {author} {\bibfnamefont {K.}~\bibnamefont {Krasnov}},\
  and\ \bibinfo {author} {\bibfnamefont {C.}~\bibnamefont {Scarinci}},\
  }\bibfield  {title} {\bibinfo {title} {{Pure Connection Formalism for
  Gravity: Feynman Rules and the Graviton-Graviton scattering}},\ }\href
  {https://doi.org/10.1007/JHEP03(2015)119} {\bibfield  {journal} {\bibinfo
  {journal} {J. High Energy Phys.}\ }\textbf {\bibinfo {volume} {03}},\
  \bibinfo {pages} {119}},\ \Eprint {https://arxiv.org/abs/1210.6215}
  {arXiv:1210.6215 [hep-th]} \BibitemShut {NoStop}%
\bibitem [{\citenamefont
  {Pop{\l}awski}(2007{\natexlab{a}})}]{poplawski07_unified_purel_affin_theor_gravit_elect}%
  \BibitemOpen
  \bibfield  {author} {\bibinfo {author} {\bibfnamefont {N.~J.}\ \bibnamefont
  {Pop{\l}awski}},\ }\href {http://arxiv.org/abs/0705.0351} {\bibinfo {title}
  {A unified, purely affine theory of gravitation and electromagnetism}}
  (\bibinfo {year} {2007}{\natexlab{a}}),\ \Eprint
  {https://arxiv.org/abs/0705.0351} {arXiv:0705.0351 [gr-qc]} \BibitemShut
  {NoStop}%
\bibitem [{\citenamefont
  {Pop{\l}awski}(2007{\natexlab{b}})}]{poplawski07_nonsy_purel_affin}%
  \BibitemOpen
  \bibfield  {author} {\bibinfo {author} {\bibfnamefont {N.~J.}\ \bibnamefont
  {Pop{\l}awski}},\ }\bibfield  {title} {\bibinfo {title} {{On the Nonsymmetric
  Purely Affine gravity}},\ }\href {https://doi.org/10.1142/s0217732307025662}
  {\bibfield  {journal} {\bibinfo  {journal} {Mod. Phys. Lett. A}\ }\textbf
  {\bibinfo {volume} {22}},\ \bibinfo {pages} {2701} (\bibinfo {year}
  {2007}{\natexlab{b}})},\ \Eprint {https://arxiv.org/abs/gr-qc/0610132}
  {arXiv:gr-qc/0610132 [gr-qc]} \BibitemShut {NoStop}%
\bibitem [{\citenamefont
  {Pop{\l}awski}(2014)}]{poplawski14_affin_theor_gravit}%
  \BibitemOpen
  \bibfield  {author} {\bibinfo {author} {\bibfnamefont {N.~J.}\ \bibnamefont
  {Pop{\l}awski}},\ }\bibfield  {title} {\bibinfo {title} {{Affine Theory of
  gravitation}},\ }\href {https://doi.org/10.1007/s10714-013-1625-7} {\bibfield
   {journal} {\bibinfo  {journal} {Gen. Rel. Grav.}\ }\textbf {\bibinfo
  {volume} {46}},\ \bibinfo {pages} {1625} (\bibinfo {year} {2014})},\ \Eprint
  {https://arxiv.org/abs/1203.0294} {arXiv:1203.0294 [gr-qc]} \BibitemShut
  {NoStop}%
\bibitem [{\citenamefont {Castillo-Felisola}\ and\ \citenamefont
  {Skirzewski}(2015)}]{castillo-felisola15_polyn_model_purel_affin_gravit}%
  \BibitemOpen
  \bibfield  {author} {\bibinfo {author} {\bibfnamefont {O.}~\bibnamefont
  {Castillo-Felisola}}\ and\ \bibinfo {author} {\bibfnamefont {A.}~\bibnamefont
  {Skirzewski}},\ }\bibfield  {title} {\bibinfo {title} {{A Polynomial Model of
  Purely Affine Gravity}},\ }\href@noop {} {\bibfield  {journal} {\bibinfo
  {journal} {Rev. Mex. Fis.}\ }\textbf {\bibinfo {volume} {61}},\ \bibinfo
  {pages} {421} (\bibinfo {year} {2015})},\ \Eprint
  {https://arxiv.org/abs/1410.6183} {arXiv:1410.6183 [gr-qc]} \BibitemShut
  {NoStop}%
\bibitem [{\citenamefont {Castillo-Felisola}\ and\ \citenamefont
  {Skirzewski}(2018)}]{castillo-felisola18_einst_gravit_from_polyn_affin_model}%
  \BibitemOpen
  \bibfield  {author} {\bibinfo {author} {\bibfnamefont {O.}~\bibnamefont
  {Castillo-Felisola}}\ and\ \bibinfo {author} {\bibfnamefont {A.}~\bibnamefont
  {Skirzewski}},\ }\bibfield  {title} {\bibinfo {title} {Einstein's gravity
  from a polynomial affine model},\ }\href
  {https://doi.org/10.1088/1361-6382/aaa699} {\bibfield  {journal} {\bibinfo
  {journal} {Class. Quant. Grav.}\ }\textbf {\bibinfo {volume} {35}},\ \bibinfo
  {pages} {055012} (\bibinfo {year} {2018})},\ \Eprint
  {https://arxiv.org/abs/1505.04634} {arXiv:1505.04634 [gr-qc]} \BibitemShut
  {NoStop}%
\bibitem [{\citenamefont
  {Pop{\l}awski}(2010)}]{poplawski10_cosmol_with_torsion_alter_to_cosmic}%
  \BibitemOpen
  \bibfield  {author} {\bibinfo {author} {\bibfnamefont {N.~J.}\ \bibnamefont
  {Pop{\l}awski}},\ }\bibfield  {title} {\bibinfo {title} {{Cosmology With
  Torsion - an Alternative To Cosmic inflation}},\ }\href
  {https://doi.org/10.1016/j.physletb.2010.09.056} {\bibfield  {journal}
  {\bibinfo  {journal} {Phys. Lett. B}\ }\textbf {\bibinfo {volume} {694}},\
  \bibinfo {pages} {181} (\bibinfo {year} {2010})},\ \Eprint
  {https://arxiv.org/abs/1007.0587} {arXiv:1007.0587 [astro-ph.CO]}
  \BibitemShut {NoStop}%
\bibitem [{\citenamefont {Azri}\ and\ \citenamefont
  {Demir}(2017)}]{azri17_affin_inflat}%
  \BibitemOpen
  \bibfield  {author} {\bibinfo {author} {\bibfnamefont {H.}~\bibnamefont
  {Azri}}\ and\ \bibinfo {author} {\bibfnamefont {D.}~\bibnamefont {Demir}},\
  }\bibfield  {title} {\bibinfo {title} {Affine inflation},\ }\href
  {https://doi.org/10.1103/physrevd.95.124007} {\bibfield  {journal} {\bibinfo
  {journal} {Phys. Rev. D}\ }\textbf {\bibinfo {volume} {95}},\ \bibinfo
  {pages} {124007} (\bibinfo {year} {2017})}\BibitemShut {NoStop}%
\bibitem [{\citenamefont {Azri}(2018)}]{azri18_cosmol_implic_affin_gravit}%
  \BibitemOpen
  \bibfield  {author} {\bibinfo {author} {\bibfnamefont {H.}~\bibnamefont
  {Azri}},\ }\emph {\bibinfo {title} {Cosmological Implications of Affine
  Gravity}},\ \href@noop {} {Ph.D. thesis},\ \bibinfo  {school} {{{\.I}zm\.ir
  Institute of Technology}} (\bibinfo {year} {2018}),\ \Eprint
  {https://arxiv.org/abs/1805.03936} {arXiv:1805.03936 [gr-qc]} \BibitemShut
  {NoStop}%
\bibitem [{\citenamefont {Azri}\ and\ \citenamefont
  {Demir}(2018)}]{azri18_induc_affin_inflat}%
  \BibitemOpen
  \bibfield  {author} {\bibinfo {author} {\bibfnamefont {H.}~\bibnamefont
  {Azri}}\ and\ \bibinfo {author} {\bibfnamefont {D.}~\bibnamefont {Demir}},\
  }\bibfield  {title} {\bibinfo {title} {Induced affine inflation},\ }\href
  {https://doi.org/10.1103/physrevd.97.044025} {\bibfield  {journal} {\bibinfo
  {journal} {Phys. Rev. D}\ }\textbf {\bibinfo {volume} {97}},\ \bibinfo
  {pages} {044025} (\bibinfo {year} {2018})}\BibitemShut {NoStop}%
\bibitem [{\citenamefont {Castillo-Felisola}\ \emph {et~al.}(2019)\citenamefont
  {Castillo-Felisola}, \citenamefont {Perdiguero},\ and\ \citenamefont
  {Orellana}}]{castillo-felisola18_cosmol}%
  \BibitemOpen
  \bibfield  {author} {\bibinfo {author} {\bibfnamefont {O.}~\bibnamefont
  {Castillo-Felisola}}, \bibinfo {author} {\bibfnamefont {J.}~\bibnamefont
  {Perdiguero}},\ and\ \bibinfo {author} {\bibfnamefont {O.}~\bibnamefont
  {Orellana}},\ }\bibinfo {title} {Redefining standard model cosmology}\
  (\bibinfo  {publisher} {IntechOpen},\ \bibinfo {address} {London},\ \bibinfo
  {year} {2019})\ Chap.\ \bibinfo {chapter} {Cosmological Solutions to
  Polynomial Affine Gravity in the Torsion-Free Sector}, p.~\bibinfo {pages}
  {NA},\ \Eprint {https://arxiv.org/abs/1808.05970} {arXiv:1808.05970 [gr-qc]}
  \BibitemShut {NoStop}%
\bibitem [{\citenamefont {Castillo-Felisola}\ \emph {et~al.}(2020)\citenamefont
  {Castillo-Felisola}, \citenamefont {Perdiguero}, \citenamefont {Orellana},\
  and\ \citenamefont
  {Zerwekh}}]{castillo-felisola20_emerg_metric_geodes_analy_cosmol}%
  \BibitemOpen
  \bibfield  {author} {\bibinfo {author} {\bibfnamefont {O.}~\bibnamefont
  {Castillo-Felisola}}, \bibinfo {author} {\bibfnamefont {J.}~\bibnamefont
  {Perdiguero}}, \bibinfo {author} {\bibfnamefont {O.}~\bibnamefont
  {Orellana}},\ and\ \bibinfo {author} {\bibfnamefont {A.~R.}\ \bibnamefont
  {Zerwekh}},\ }\bibfield  {title} {\bibinfo {title} {Emergent metric and
  geodesic analysis in cosmological solutions of (torsion-free) polynomial
  affine gravity},\ }\href {https://doi.org/10.1088/1361-6382/ab58ef}
  {\bibfield  {journal} {\bibinfo  {journal} {Class. Quant. Grav.}\ }\textbf
  {\bibinfo {volume} {37}},\ \bibinfo {pages} {075013} (\bibinfo {year}
  {2020})},\ \Eprint {https://arxiv.org/abs/1908.06654} {arXiv:1908.06654
  [gr-qc]} \BibitemShut {NoStop}%
\bibitem [{\citenamefont {Castillo-Felisola}\ \emph
  {et~al.}(2024{\natexlab{a}})\citenamefont {Castillo-Felisola}, \citenamefont
  {Grez}, \citenamefont {Olmo}, \citenamefont {Orellana},\ and\ \citenamefont
  {{Perdiguero G{\'a}rate}}}]{castillo-felisola24_cosmol_solut_polyn_affin}%
  \BibitemOpen
  \bibfield  {author} {\bibinfo {author} {\bibfnamefont {O.}~\bibnamefont
  {Castillo-Felisola}}, \bibinfo {author} {\bibfnamefont {B.}~\bibnamefont
  {Grez}}, \bibinfo {author} {\bibfnamefont {G.~J.}\ \bibnamefont {Olmo}},
  \bibinfo {author} {\bibfnamefont {O.}~\bibnamefont {Orellana}},\ and\
  \bibinfo {author} {\bibfnamefont {J.}~\bibnamefont {{Perdiguero
  G{\'a}rate}}},\ }\bibfield  {title} {\bibinfo {title} {Cosmological solutions
  in polynomial affine gravity with torsion},\ }\Eprint
  {https://arxiv.org/abs/2404.11703} {arXiv:2404.11703 [gr-qc]}  (\bibinfo
  {year} {2024}{\natexlab{a}})\BibitemShut {NoStop}%
\bibitem [{\citenamefont {Aalbers}(2013)}]{aalbers13_confor}%
  \BibitemOpen
  \bibfield  {author} {\bibinfo {author} {\bibfnamefont {J.}~\bibnamefont
  {Aalbers}},\ }\emph {\bibinfo {title} {Conformal symmetry in classical
  gravity}},\ \href@noop {} {Master's thesis},\ \bibinfo  {school} {Utrecht
  University} (\bibinfo {year} {2013})\BibitemShut {NoStop}%
\bibitem [{\citenamefont {Bailey}\ \emph {et~al.}(1994)\citenamefont {Bailey},
  \citenamefont {Eastwood},\ and\ \citenamefont
  {Gover}}]{bailey94_thomas_struc_bundl_confor_projec_relat_struc}%
  \BibitemOpen
  \bibfield  {author} {\bibinfo {author} {\bibfnamefont {T.~N.}\ \bibnamefont
  {Bailey}}, \bibinfo {author} {\bibfnamefont {M.~G.}\ \bibnamefont
  {Eastwood}},\ and\ \bibinfo {author} {\bibfnamefont {A.~R.}\ \bibnamefont
  {Gover}},\ }\bibfield  {title} {\bibinfo {title} {Thomas's structure bundle
  for conformal, projective and related structures},\ }\href
  {https://doi.org/10.1216/rmjm/1181072333} {\bibfield  {journal} {\bibinfo
  {journal} {Rocky Mountain Journal of Mathematics}\ }\textbf {\bibinfo
  {volume} {24}},\ \bibinfo {pages} {1191} (\bibinfo {year}
  {1994})}\BibitemShut {NoStop}%
\bibitem [{\citenamefont
  {Borthwick}(2022)}]{borthwick21_projec_differ_geomet_asymp_analy_gener_relat}%
  \BibitemOpen
  \bibfield  {author} {\bibinfo {author} {\bibfnamefont {J.}~\bibnamefont
  {Borthwick}},\ }\href {http://arxiv.org/abs/2109.05834v2} {\bibinfo {title}
  {Projective differential geometry and asymptotic analysis in general
  relativity}} (\bibinfo {year} {2022}),\ \Eprint
  {https://arxiv.org/abs/2109.05834} {arXiv:2109.05834 [math.DG]} \BibitemShut
  {NoStop}%
\bibitem [{\citenamefont {Synge}\ and\ \citenamefont
  {Schild}(1978)}]{synge78_tensor}%
  \BibitemOpen
  \bibfield  {author} {\bibinfo {author} {\bibfnamefont {J.~L.}\ \bibnamefont
  {Synge}}\ and\ \bibinfo {author} {\bibfnamefont {A.}~\bibnamefont {Schild}},\
  }\href@noop {} {\emph {\bibinfo {title} {Tensor calculus}}},\ Vol.~\bibinfo
  {volume} {5}\ (\bibinfo  {publisher} {Courier Corporation},\ \bibinfo {year}
  {1978})\BibitemShut {NoStop}%
\bibitem [{\citenamefont {Schouten}(2013)}]{schouten13_ricci}%
  \BibitemOpen
  \bibfield  {author} {\bibinfo {author} {\bibfnamefont {J.~A.}\ \bibnamefont
  {Schouten}},\ }\href@noop {} {\emph {\bibinfo {title} {Ricci-calculus: an
  introduction to tensor analysis and its geometrical applications}}},\
  Vol.~\bibinfo {volume} {10}\ (\bibinfo  {publisher} {Springer},\ \bibinfo
  {address} {Berlin},\ \bibinfo {year} {2013})\BibitemShut {NoStop}%
\bibitem [{Note3()}]{Note3}%
  \BibitemOpen
  \bibinfo {note} {In our convention, the torsion is defined as the difference
  between the components of the connection, \(\protect \tilde {\protect
  \mathcal {T}}_{\mu }{}^{\lambda }{}_{\nu } = \protect \tilde {\Gamma }_{\mu
  }{}^{\lambda }{}_{\nu } - \protect \tilde {\Gamma }_{\nu }{}^{\lambda
  }{}_{\mu }\). In order to avoid the half factor in the second line of Eq.
  \protect \eqref {eq:conn-decomp}, it is usually referred to as the tensor
  \(\protect \tilde {\protect \mathcal {S}}_{\mu }{}^{\lambda }{}_{\nu } =
  \protect \tilde {\Gamma }_{[\mu }{}^{\lambda }{}_{\nu ]}\).}\BibitemShut
  {Stop}%
\bibitem [{\citenamefont {Zanelli}(2000)}]{zanelli00_chern_simon_gravit}%
  \BibitemOpen
  \bibfield  {author} {\bibinfo {author} {\bibfnamefont {J.}~\bibnamefont
  {Zanelli}},\ }\bibfield  {title} {\bibinfo {title} {{Chern-Simons Gravity:
  From (2+1)-dimensions To (2n+1)- dimensions}},\ }\href
  {https://doi.org/10.1590/S0103-97332000000200006} {\bibfield  {journal}
  {\bibinfo  {journal} {Braz. J. Phys.}\ }\textbf {\bibinfo {volume} {30}},\
  \bibinfo {pages} {251} (\bibinfo {year} {2000})},\ \Eprint
  {https://arxiv.org/abs/hep-th/0010049} {arXiv:hep-th/0010049 [hep-th]}
  \BibitemShut {NoStop}%
\bibitem [{\citenamefont {Zanelli}(2005)}]{zanelli05_lectur_notes_chern_simon}%
  \BibitemOpen
  \bibfield  {author} {\bibinfo {author} {\bibfnamefont {J.}~\bibnamefont
  {Zanelli}},\ }\href {http://arxiv.org/abs/hep-th/0502193v4} {\bibinfo {title}
  {Lecture notes on chern-simons (super-)gravities. second edition (february
  2008)}} (\bibinfo {year} {2005}),\ \Eprint
  {https://arxiv.org/abs/hep-th/0502193} {arXiv:hep-th/0502193 [hep-th]}
  \BibitemShut {NoStop}%
\bibitem [{\citenamefont {Zanelli}\ and\ \citenamefont
  {Hassaine}(2016)}]{zanelli16_chern}%
  \BibitemOpen
  \bibfield  {author} {\bibinfo {author} {\bibfnamefont {J.}~\bibnamefont
  {Zanelli}}\ and\ \bibinfo {author} {\bibfnamefont {M.}~\bibnamefont
  {Hassaine}},\ }\href@noop {} {\emph {\bibinfo {title} {{C}hern--{S}imons
  (super)gravity}}},\ \bibinfo {series} {100 years of general relativity},
  Vol.~\bibinfo {volume} {2}\ (\bibinfo  {publisher} {World Scientific},\
  \bibinfo {year} {2016})\BibitemShut {NoStop}%
\bibitem [{Note4()}]{Note4}%
  \BibitemOpen
  \bibinfo {note} {Note that when the connection is metric the Pontryagin
  \(P_2\) vanishes explicitly, since the anti-symmetric part of the Ricci
  tensor is zero, \(\protect \mathcal {R}_{[\mu \nu ]} = 0\).}\BibitemShut
  {Stop}%
\bibitem [{\citenamefont {McGady}\ and\ \citenamefont
  {Rodina}(2014)}]{mcgady14_higher_spin_massl_s_dimen}%
  \BibitemOpen
  \bibfield  {author} {\bibinfo {author} {\bibfnamefont {D.~A.}\ \bibnamefont
  {McGady}}\ and\ \bibinfo {author} {\bibfnamefont {L.}~\bibnamefont
  {Rodina}},\ }\bibfield  {title} {\bibinfo {title} {{Higher-Spin Massless
  $S$-matrices in four-Dimensions}},\ }\href
  {https://doi.org/10.1103/PhysRevD.90.084048} {\bibfield  {journal} {\bibinfo
  {journal} {Phys. Rev. D}\ }\textbf {\bibinfo {volume} {90}},\ \bibinfo
  {pages} {084048} (\bibinfo {year} {2014})},\ \Eprint
  {https://arxiv.org/abs/1311.2938} {arXiv:1311.2938 [hep-th]} \BibitemShut
  {NoStop}%
\bibitem [{\citenamefont {Camanho}\ \emph {et~al.}(2016)\citenamefont
  {Camanho}, \citenamefont {Edelstein}, \citenamefont {Maldacena},\ and\
  \citenamefont {Zhiboedov}}]{camanho16_causal_const_correc_to_gravit}%
  \BibitemOpen
  \bibfield  {author} {\bibinfo {author} {\bibfnamefont {X.~O.}\ \bibnamefont
  {Camanho}}, \bibinfo {author} {\bibfnamefont {J.~D.}\ \bibnamefont
  {Edelstein}}, \bibinfo {author} {\bibfnamefont {J.}~\bibnamefont
  {Maldacena}},\ and\ \bibinfo {author} {\bibfnamefont {A.}~\bibnamefont
  {Zhiboedov}},\ }\bibfield  {title} {\bibinfo {title} {{Causality Constraints
  on Corrections To the Graviton Three-Point Coupling}},\ }\href
  {https://doi.org/10.1007/JHEP02(2016)020} {\bibfield  {journal} {\bibinfo
  {journal} {J. High Energy Phys.}\ }\textbf {\bibinfo {volume} {02}},\
  \bibinfo {pages} {020}},\ \Eprint {https://arxiv.org/abs/1407.5597}
  {arXiv:1407.5597 [hep-th]} \BibitemShut {NoStop}%
\bibitem [{Note5()}]{Note5}%
  \BibitemOpen
  \bibinfo {note} {This is similar to what happens in the unimodular model of
  gravity \cite
  {ng91_unimod_theor_gravit_cosmol_const,jirousek23_unimod_approac_to_cosmol}.}\BibitemShut
  {Stop}%
\bibitem [{\citenamefont {Hojman}(2015)}]{hojman_privat}%
  \BibitemOpen
  \bibfield  {author} {\bibinfo {author} {\bibfnamefont {S.}~\bibnamefont
  {Hojman}},\ }\href@noop {} {\bibinfo {title} {Private communication}}
  (\bibinfo {year} {2015}),\ \bibinfo {note} {we thank S.~Hojman for
  clarifications in this respect.}\BibitemShut {Stop}%
\bibitem [{\citenamefont {Nomizu}\ and\ \citenamefont
  {Sasaki}(1994)}]{nomizu94_affin}%
  \BibitemOpen
  \bibfield  {author} {\bibinfo {author} {\bibfnamefont {K.}~\bibnamefont
  {Nomizu}}\ and\ \bibinfo {author} {\bibfnamefont {T.}~\bibnamefont
  {Sasaki}},\ }\href@noop {} {\emph {\bibinfo {title} {Affine differential
  geometry}}}\ (\bibinfo  {publisher} {Cambridge University Press},\ \bibinfo
  {year} {1994})\BibitemShut {NoStop}%
\bibitem [{\citenamefont {Derdzi{\'n}ski}\ and\ \citenamefont
  {Shen}(1983)}]{derdzinski83_codaz_tensor_field_curvat_pontr_forms}%
  \BibitemOpen
  \bibfield  {author} {\bibinfo {author} {\bibfnamefont {A.}~\bibnamefont
  {Derdzi{\'n}ski}}\ and\ \bibinfo {author} {\bibfnamefont {C.-L.}\
  \bibnamefont {Shen}},\ }\bibfield  {title} {\bibinfo {title} {{C}odazzi
  tensor fields, curvature and {P}ontryagin forms},\ }\href
  {https://doi.org/10.1112/plms/s3-47.1.15} {\bibfield  {journal} {\bibinfo
  {journal} {Proc. London Math. Soc.}\ }\textbf {\bibinfo {volume} {s3-47}},\
  \bibinfo {pages} {15} (\bibinfo {year} {1983})}\BibitemShut {NoStop}%
\bibitem [{\citenamefont {Besse}(2007)}]{besse07_einst}%
  \BibitemOpen
  \bibfield  {author} {\bibinfo {author} {\bibfnamefont {A.~L.}\ \bibnamefont
  {Besse}},\ }\href@noop {} {\emph {\bibinfo {title} {Einstein manifolds}}}\
  (\bibinfo  {publisher} {Springer},\ \bibinfo {address} {Berlin},\ \bibinfo
  {year} {2007})\BibitemShut {NoStop}%
\bibitem [{\citenamefont
  {Derdzi{\'n}ski}(1980)}]{derdzinski80_class_certain_compac_rieman_manif}%
  \BibitemOpen
  \bibfield  {author} {\bibinfo {author} {\bibfnamefont {A.}~\bibnamefont
  {Derdzi{\'n}ski}},\ }\bibfield  {title} {\bibinfo {title} {{Classification of
  Certain Compact Riemannian Manifolds With Harmonic Curvature and Non-Parallel
  Ricci tensor}},\ }\href@noop {} {\bibfield  {journal} {\bibinfo  {journal}
  {Math. Z.}\ }\textbf {\bibinfo {volume} {172}},\ \bibinfo {pages} {273}
  (\bibinfo {year} {1980})}\BibitemShut {NoStop}%
\bibitem [{\citenamefont
  {Derdzi{\'n}ski}(1982)}]{derdzinski82_compac_rieman_manif_with_harmon_curvat}%
  \BibitemOpen
  \bibfield  {author} {\bibinfo {author} {\bibfnamefont {A.}~\bibnamefont
  {Derdzi{\'n}ski}},\ }\bibfield  {title} {\bibinfo {title} {On compact
  riemannian manifolds with harmonic curvature},\ }\href@noop {} {\bibfield
  {journal} {\bibinfo  {journal} {Mathematische Annalen}\ }\textbf {\bibinfo
  {volume} {259}},\ \bibinfo {pages} {145} (\bibinfo {year}
  {1982})}\BibitemShut {NoStop}%
\bibitem [{\citenamefont {Derdzi\'nski}(1985)}]{derdzinski85_rieman}%
  \BibitemOpen
  \bibfield  {author} {\bibinfo {author} {\bibfnamefont {A.}~\bibnamefont
  {Derdzi\'nski}},\ }\bibfield  {title} {\bibinfo {title} {Riemannian manifolds
  with harmonic curvature},\ }in\ \href {https://doi.org/10.1007/BFb0075087}
  {\emph {\bibinfo {booktitle} {Global Differential Geometry and Global
  Analysis 1984}}},\ \bibinfo {series} {Lecture Notes in Mathematics}, Vol.\
  \bibinfo {volume} {1156},\ \bibinfo {editor} {edited by\ \bibinfo {editor}
  {\bibfnamefont {D.}~\bibnamefont {Ferus}}, \bibinfo {editor} {\bibfnamefont
  {R.~B.}\ \bibnamefont {Gardner}}, \bibinfo {editor} {\bibfnamefont
  {S.}~\bibnamefont {Helgason}},\ and\ \bibinfo {editor} {\bibfnamefont
  {U.}~\bibnamefont {Simon}}}\ (\bibinfo  {publisher} {Springer},\ \bibinfo
  {address} {Berlin},\ \bibinfo {year} {1985})\ p.~\bibinfo {pages}
  {74}\BibitemShut {NoStop}%
\bibitem [{\citenamefont
  {Stephenson}(1958)}]{stephenson58_quadr_lagran_gener_relat}%
  \BibitemOpen
  \bibfield  {author} {\bibinfo {author} {\bibfnamefont {G.}~\bibnamefont
  {Stephenson}},\ }\bibfield  {title} {\bibinfo {title} {Quadratic lagrangians
  and general relativity},\ }\href {https://doi.org/10.1007/bf02724929}
  {\bibfield  {journal} {\bibinfo  {journal} {Nuovo Cimento}\ }\textbf
  {\bibinfo {volume} {9}},\ \bibinfo {pages} {263} (\bibinfo {year}
  {1958})}\BibitemShut {NoStop}%
\bibitem [{\citenamefont {Kilmister}\ and\ \citenamefont
  {Newman}(1961)}]{kilmister61_use_alg_struct_phys}%
  \BibitemOpen
  \bibfield  {author} {\bibinfo {author} {\bibfnamefont {C.~W.}\ \bibnamefont
  {Kilmister}}\ and\ \bibinfo {author} {\bibfnamefont {D.~J.}\ \bibnamefont
  {Newman}},\ }\bibfield  {title} {\bibinfo {title} {The use of algebraic
  structures in physics},\ }in\ \href
  {https://doi.org/10.1017/S0305004100036008} {\emph {\bibinfo {booktitle}
  {Mathematical Proceedings of the Cambridge Philosophical Society}}},\
  Vol.~\bibinfo {volume} {57}\ (\bibinfo  {publisher} {Cambridge University
  Press},\ \bibinfo {year} {1961})\ p.\ \bibinfo {pages} {851}\BibitemShut
  {NoStop}%
\bibitem [{\citenamefont {Yang}(1974)}]{yang74_integ_formal_gauge_field}%
  \BibitemOpen
  \bibfield  {author} {\bibinfo {author} {\bibfnamefont {C.~N.}\ \bibnamefont
  {Yang}},\ }\bibfield  {title} {\bibinfo {title} {Integral formalism for gauge
  fields},\ }\href {https://doi.org/10.1103/PhysRevLett.33.445} {\bibfield
  {journal} {\bibinfo  {journal} {Phys. Rev. Lett.}\ }\textbf {\bibinfo
  {volume} {33}},\ \bibinfo {pages} {445} (\bibinfo {year} {1974})}\BibitemShut
  {NoStop}%
\bibitem [{\citenamefont
  {Pavelle}(1975)}]{pavelle75_unphy_solut_yangs_gravit_field_equat}%
  \BibitemOpen
  \bibfield  {author} {\bibinfo {author} {\bibfnamefont {R.}~\bibnamefont
  {Pavelle}},\ }\bibfield  {title} {\bibinfo {title} {Unphysical solutions of
  yang's gravitational-field equations},\ }\href
  {https://doi.org/10.1103/physrevlett.34.1114} {\bibfield  {journal} {\bibinfo
   {journal} {Phys. Rev. Lett.}\ }\textbf {\bibinfo {volume} {34}},\ \bibinfo
  {pages} {1114} (\bibinfo {year} {1975})}\BibitemShut {NoStop}%
\bibitem [{\citenamefont
  {Thompson}(1975)}]{thompson75_geomet_degen_solut_kilmis_yang_equat}%
  \BibitemOpen
  \bibfield  {author} {\bibinfo {author} {\bibfnamefont {A.~H.}\ \bibnamefont
  {Thompson}},\ }\bibfield  {title} {\bibinfo {title} {Geometrically degenerate
  solutions of the kilmister-yang equations},\ }\href
  {https://doi.org/10.1103/PhysRevLett.35.320} {\bibfield  {journal} {\bibinfo
  {journal} {Phys. Rev. Lett.}\ }\textbf {\bibinfo {volume} {35}},\ \bibinfo
  {pages} {320} (\bibinfo {year} {1975})}\BibitemShut {NoStop}%
\bibitem [{\citenamefont {Zanelli}(2015)}]{zanelli_privat}%
  \BibitemOpen
  \bibfield  {author} {\bibinfo {author} {\bibfnamefont {J.}~\bibnamefont
  {Zanelli}},\ }\href@noop {} {\bibinfo {title} {Private communication}}
  (\bibinfo {year} {2015}),\ \bibinfo {note} {we thank J.~Zanelli for sharing
  his personal communications with C.~N.~Yang, and clarify this aspect of the
  effective model.}\BibitemShut {Stop}%
\bibitem [{\citenamefont
  {Harada}(2021)}]{harada21_emerg_cotton_tensor_descr_gravit}%
  \BibitemOpen
  \bibfield  {author} {\bibinfo {author} {\bibfnamefont {J.}~\bibnamefont
  {Harada}},\ }\bibfield  {title} {\bibinfo {title} {Emergence of the cotton
  tensor for describing gravity},\ }\href
  {https://doi.org/10.1103/PhysRevD.103.L121502} {\bibfield  {journal}
  {\bibinfo  {journal} {Phys. Rev. D}\ }\textbf {\bibinfo {volume} {103}}
  (\bibinfo {year} {2021})},\ \Eprint {https://arxiv.org/abs/2105.09304}
  {arXiv:2105.09304 [gr-qc]} \BibitemShut {NoStop}%
\bibitem [{\citenamefont
  {Harada}(2022)}]{harada22_cotton_gravit_galax_rotat_curves}%
  \BibitemOpen
  \bibfield  {author} {\bibinfo {author} {\bibfnamefont {J.}~\bibnamefont
  {Harada}},\ }\bibfield  {title} {\bibinfo {title} {Cotton gravity and 84
  galaxy rotation curves},\ }\href
  {https://doi.org/10.1103/PhysRevD.106.064044} {\bibfield  {journal} {\bibinfo
   {journal} {Phys. Rev. D}\ }\textbf {\bibinfo {volume} {106}},\ \bibinfo
  {pages} {064044} (\bibinfo {year} {2022})},\ \Eprint
  {https://arxiv.org/abs/2209.04055} {arXiv:2209.04055 [gr-qc]} \BibitemShut
  {NoStop}%
\bibitem [{\citenamefont {Mantica}\ and\ \citenamefont
  {Molinari}(2023)}]{mantica23_note_harad_confor_killin_gravit}%
  \BibitemOpen
  \bibfield  {author} {\bibinfo {author} {\bibfnamefont {C.~A.}\ \bibnamefont
  {Mantica}}\ and\ \bibinfo {author} {\bibfnamefont {L.~G.}\ \bibnamefont
  {Molinari}},\ }\bibfield  {title} {\bibinfo {title} {A note on harada's
  conformal killing gravity},\ }\href
  {https://doi.org/10.1103/PhysRevD.108.124029} {\bibfield  {journal} {\bibinfo
   {journal} {Phys. Rev. D}\ }\textbf {\bibinfo {volume} {108}},\ \bibinfo
  {pages} {124029} (\bibinfo {year} {2023})},\ \Eprint
  {https://arxiv.org/abs/2308.06803} {arXiv:2308.06803 [gr-qc]} \BibitemShut
  {NoStop}%
\bibitem [{\citenamefont
  {Harada}(2023)}]{harada23_dark_energ_confor_killin_gravit}%
  \BibitemOpen
  \bibfield  {author} {\bibinfo {author} {\bibfnamefont {J.}~\bibnamefont
  {Harada}},\ }\bibfield  {title} {\bibinfo {title} {Dark energy in conformal
  killing gravity},\ }\href {10.1103/PhysRevD.108.104037} {\bibfield  {journal}
  {\bibinfo  {journal} {Phys. Rev. D}\ }\textbf {\bibinfo {volume} {108}},\
  \bibinfo {pages} {104037} (\bibinfo {year} {2023})},\ \Eprint
  {https://arxiv.org/abs/2308.07634} {arXiv:2308.07634 [gr-qc]} \BibitemShut
  {NoStop}%
\bibitem [{\citenamefont {Cotton}(1899)}]{cotton99_sur}%
  \BibitemOpen
  \bibfield  {author} {\bibinfo {author} {\bibfnamefont {{\'E}.}~\bibnamefont
  {Cotton}},\ }\bibfield  {title} {\bibinfo {title} {Sur les vari{\'e}t{\'e}s
  {\`a} trois dimensions},\ }in\ \href@noop {} {\emph {\bibinfo {booktitle}
  {Annales de la Facult{\'e} des sciences de Toulouse: Math{\'e}matiques}}},\
  Vol.~\bibinfo {volume} {1}\ (\bibinfo {year} {1899})\ pp.\ \bibinfo {pages}
  {385--438}\BibitemShut {NoStop}%
\bibitem [{Note6()}]{Note6}%
  \BibitemOpen
  \bibinfo {note} {Operationally, a projective equivalent quantity is defined
  by a similar expression where the terms containing the metric tensor field
  are dropped.}\BibitemShut {Stop}%
\bibitem [{\citenamefont {Weyl}(1918)}]{weyl18_reine_infin}%
  \BibitemOpen
  \bibfield  {author} {\bibinfo {author} {\bibfnamefont {H.}~\bibnamefont
  {Weyl}},\ }\bibfield  {title} {\bibinfo {title} {Reine
  infinitesimalgeometrie},\ }\href@noop {} {\bibfield  {journal} {\bibinfo
  {journal} {Mathematische Zeitschrift}\ }\textbf {\bibinfo {volume} {2}},\
  \bibinfo {pages} {384} (\bibinfo {year} {1918})}\BibitemShut {NoStop}%
\bibitem [{\citenamefont {Eisenhart}(1927)}]{eisenhart27_non_rieman}%
  \BibitemOpen
  \bibfield  {author} {\bibinfo {author} {\bibfnamefont {L.}~\bibnamefont
  {Eisenhart}},\ }\href@noop {} {\emph {\bibinfo {title} {Non-Riemannian
  geometry}}}\ (\bibinfo  {publisher} {American Mathematical Society},\
  \bibinfo {address} {New York},\ \bibinfo {year} {1927})\BibitemShut {NoStop}%
\bibitem [{\citenamefont {ichi
  Amari}(2010)}]{amari10_infor_geomet_optim_machin}%
  \BibitemOpen
  \bibfield  {author} {\bibinfo {author} {\bibfnamefont {S.}~\bibnamefont {ichi
  Amari}},\ }\bibfield  {title} {\bibinfo {title} {Information geometry in
  optimization, machine learning and statistical inference},\ }\href
  {https://doi.org/10.1007/s11460-010-0101-3} {\bibfield  {journal} {\bibinfo
  {journal} {Frontiers of Electrical and Electronic Engineering in China}\
  }\textbf {\bibinfo {volume} {5}},\ \bibinfo {pages} {241} (\bibinfo {year}
  {2010})}\BibitemShut {NoStop}%
\bibitem [{\citenamefont {Suzuki}(2014)}]{suzuki14_infor_geomet_statis_manif}%
  \BibitemOpen
  \bibfield  {author} {\bibinfo {author} {\bibfnamefont {M.}~\bibnamefont
  {Suzuki}},\ }\href {http://arxiv.org/abs/1410.3369v1} {\bibinfo {title}
  {Information geometry and statistical manifold}} (\bibinfo {year} {2014}),\
  \Eprint {https://arxiv.org/abs/1410.3369} {arXiv:1410.3369 [math.ST]}
  \BibitemShut {NoStop}%
\bibitem [{\citenamefont {ichi Amari}(2016)}]{amari16_infor_geomet_its_applic}%
  \BibitemOpen
  \bibfield  {author} {\bibinfo {author} {\bibfnamefont {S.}~\bibnamefont {ichi
  Amari}},\ }\href@noop {} {\emph {\bibinfo {title} {Information Geometry and
  Its Applications}}},\ \bibinfo {edition} {1st}\ ed.,\ Applied Mathematical
  Sciences\ (\bibinfo  {publisher} {Springer},\ \bibinfo {year}
  {2016})\BibitemShut {NoStop}%
\bibitem [{Note7()}]{Note7}%
  \BibitemOpen
  \bibinfo {note} {Note that there are sixty-four of these functions in four
  dimensions.}\BibitemShut {Stop}%
\bibitem [{\citenamefont
  {Castillo-Felisola}(2018)}]{castillo-felisola18_beyond_einstein}%
  \BibitemOpen
  \bibfield  {author} {\bibinfo {author} {\bibfnamefont {O.}~\bibnamefont
  {Castillo-Felisola}},\ }\bibinfo {title} {Gravity}\ (\bibinfo  {publisher}
  {IntechOpen},\ \bibinfo {address} {London},\ \bibinfo {year} {2018})\ Chap.\
  \bibinfo {chapter} {Beyond Einstein: A Polynomial Affine Model of Gravity},
  pp.\ \bibinfo {pages} {183--201},\ \Eprint {https://arxiv.org/abs/1902.09131}
  {arXiv:1902.09131 [gr-qc]} \BibitemShut {NoStop}%
\bibitem [{Note8()}]{Note8}%
  \BibitemOpen
  \bibinfo {note} {The Lie derivatives of the affine connection were computed
  using the free and open mathematical software \protect \texttt {SAGE} \cite
  {stein18_sage_mathem_softw_version}, through its module \protect \texttt
  {sagemanifolds} \cite
  {gourgoulhon18_sagem_version,gourgoulhon15_tensor_calcul_with_open_sourc_softw,gourgoulhon18_symbol_tensor_calcul_manif}.}\BibitemShut
  {Stop}%
\bibitem [{\citenamefont {Toloza}\ and\ \citenamefont
  {Zanelli}(2013)}]{toloza13_cosmol_with_scalar_euler_form_coupl}%
  \BibitemOpen
  \bibfield  {author} {\bibinfo {author} {\bibfnamefont {A.}~\bibnamefont
  {Toloza}}\ and\ \bibinfo {author} {\bibfnamefont {J.}~\bibnamefont
  {Zanelli}},\ }\bibfield  {title} {\bibinfo {title} {Cosmology with
  scalar-euler form coupling},\ }\href
  {https://doi.org/10.1088/0264-9381/30/13/135003} {\bibfield  {journal}
  {\bibinfo  {journal} {Class. Quant. Grav.}\ }\textbf {\bibinfo {volume}
  {30}},\ \bibinfo {pages} {135003} (\bibinfo {year} {2013})}\BibitemShut
  {NoStop}%
\bibitem [{\citenamefont {Castillo-Felisola}\ \emph {et~al.}(2022)\citenamefont
  {Castillo-Felisola}, \citenamefont {Orellana}, \citenamefont {Perdiguero},
  \citenamefont {Ram\'\i{}rez}, \citenamefont {Skirzewski},\ and\ \citenamefont
  {Zerwekh}}]{castillo-felisola21_aspec_polyn_affin_model_gravit_three}%
  \BibitemOpen
  \bibfield  {author} {\bibinfo {author} {\bibfnamefont {O.}~\bibnamefont
  {Castillo-Felisola}}, \bibinfo {author} {\bibfnamefont {O.}~\bibnamefont
  {Orellana}}, \bibinfo {author} {\bibfnamefont {J.}~\bibnamefont
  {Perdiguero}}, \bibinfo {author} {\bibfnamefont {F.}~\bibnamefont
  {Ram\'\i{}rez}}, \bibinfo {author} {\bibfnamefont {A.}~\bibnamefont
  {Skirzewski}},\ and\ \bibinfo {author} {\bibfnamefont {A.~R.}\ \bibnamefont
  {Zerwekh}},\ }\bibfield  {title} {\bibinfo {title} {{Aspects of the
  Polynomial Affine Model of Gravity in Three dimensions}},\ }\href
  {https://doi.org/10.1140/epjc/s10052-021-09938-4} {\bibfield  {journal}
  {\bibinfo  {journal} {Eur. Phys. J. C}\ }\textbf {\bibinfo {volume} {82}},\
  \bibinfo {pages} {8} (\bibinfo {year} {2022})},\ \Eprint
  {https://arxiv.org/abs/2107.07209} {arXiv:2107.07209 [gr-qc]} \BibitemShut
  {NoStop}%
\bibitem [{Note9()}]{Note9}%
  \BibitemOpen
  \bibinfo {note} {Assuming that $D_1 - 2D_2 + D_3 \protect \neq 0, F_3
  \protect \neq 0, B_3 \protect \neq 0$}\BibitemShut {NoStop}%
\bibitem [{\citenamefont {Mukhanov}\ \emph {et~al.}(1992)\citenamefont
  {Mukhanov}, \citenamefont {Feldman},\ and\ \citenamefont
  {Brandenberger}}]{mukhanov92_theor_cosmol_pertur}%
  \BibitemOpen
  \bibfield  {author} {\bibinfo {author} {\bibfnamefont {V.~F.}\ \bibnamefont
  {Mukhanov}}, \bibinfo {author} {\bibfnamefont {H.~A.}\ \bibnamefont
  {Feldman}},\ and\ \bibinfo {author} {\bibfnamefont {R.~H.}\ \bibnamefont
  {Brandenberger}},\ }\bibfield  {title} {\bibinfo {title} {{Theory of
  Cosmological Perturbations}},\ }\href
  {https://doi.org/10.1016/0370-1573(92)90044-Z} {\bibfield  {journal}
  {\bibinfo  {journal} {Phys. Rep.}\ }\textbf {\bibinfo {volume} {215}},\
  \bibinfo {pages} {203} (\bibinfo {year} {1992})}\BibitemShut {NoStop}%
\bibitem [{\citenamefont {Poisson}\ and\ \citenamefont
  {Will}(2014)}]{poisson14_gravit}%
  \BibitemOpen
  \bibfield  {author} {\bibinfo {author} {\bibfnamefont {E.}~\bibnamefont
  {Poisson}}\ and\ \bibinfo {author} {\bibfnamefont {C.~M.}\ \bibnamefont
  {Will}},\ }\href@noop {} {\emph {\bibinfo {title} {Gravity: Newtonian,
  post-newtonian, relativistic}}}\ (\bibinfo  {publisher} {Cambridge University
  Press},\ \bibinfo {address} {London},\ \bibinfo {year} {2014})\BibitemShut
  {NoStop}%
\bibitem [{Note10()}]{Note10}%
  \BibitemOpen
  \bibinfo {note} {It should be noted as the metric \(g\) has to be symmetric,
  the perturbation field \(h\) is symmetric too.}\BibitemShut {Stop}%
\bibitem [{\citenamefont
  {Straumann}(2008)}]{straumann08_proof_decom_theor_symmet}%
  \BibitemOpen
  \bibfield  {author} {\bibinfo {author} {\bibfnamefont {N.}~\bibnamefont
  {Straumann}},\ }\bibfield  {title} {\bibinfo {title} {Proof of a
  decomposition theorem for symmetric tensors on spaces with constant
  curvature},\ }\href {https://doi.org/10.1002/andp.20085200805} {\bibfield
  {journal} {\bibinfo  {journal} {Annalen der Physik}\ }\textbf {\bibinfo
  {volume} {520}},\ \bibinfo {pages} {609} (\bibinfo {year}
  {2008})}\BibitemShut {NoStop}%
\bibitem [{\citenamefont {Castillo-Felisola}\ \emph
  {et~al.}(2024{\natexlab{b}})\citenamefont {Castillo-Felisola}, \citenamefont
  {Gannouji}, \citenamefont {Morocho-L{\'o}pez},\ and\ \citenamefont
  {Rozas-Rojas}}]{castillo-felisola24_model_cosmol_pertur_affin_gravit}%
  \BibitemOpen
  \bibfield  {author} {\bibinfo {author} {\bibfnamefont {O.}~\bibnamefont
  {Castillo-Felisola}}, \bibinfo {author} {\bibfnamefont {R.}~\bibnamefont
  {Gannouji}}, \bibinfo {author} {\bibfnamefont {M.}~\bibnamefont
  {Morocho-L{\'o}pez}},\ and\ \bibinfo {author} {\bibfnamefont
  {M.}~\bibnamefont {Rozas-Rojas}},\ }\bibfield  {title} {\bibinfo {title} {A
  model for cosmological perturbations in affine gravity},\ }\href
  {http://arxiv.org/abs/2401.08198v1} {\  (\bibinfo {year}
  {2024}{\natexlab{b}})},\ \bibinfo {note} {accepted for publication in
  \emph{Phys. Rev. D}},\ \Eprint {https://arxiv.org/abs/2401.08198}
  {arXiv:2401.08198 [gr-qc]} \BibitemShut {NoStop}%
\bibitem [{Note11()}]{Note11}%
  \BibitemOpen
  \bibinfo {note} {A cosmological perturbation technique for metric-affine
  theories has been proposed in Ref. \cite
  {aoki23_cosmol_pertur_theor_metric}.}\BibitemShut {Stop}%
\bibitem [{Note12()}]{Note12}%
  \BibitemOpen
  \bibinfo {note} {In Eq. \protect \eqref {eq:relation-c-indices} the \protect
  \emph {lowering} of the \(t\)-index should be considered an identification to
  simplify the presentation of the forthcoming expression, and not as the
  action of a metric.}\BibitemShut {Stop}%
\bibitem [{\citenamefont {Landsberg}(2011)}]{landsberg11_tensor}%
  \BibitemOpen
  \bibfield  {author} {\bibinfo {author} {\bibfnamefont {J.~M.}\ \bibnamefont
  {Landsberg}},\ }\href@noop {} {\emph {\bibinfo {title} {Tensors: Geometry and
  Applications}}},\ \bibinfo {edition} {1st}\ ed.,\ \bibinfo {series} {Graduate
  studies in mathematics}, Vol.\ \bibinfo {volume} {128}\ (\bibinfo
  {publisher} {American Mathematical Society},\ \bibinfo {year}
  {2011})\BibitemShut {NoStop}%
\bibitem [{\citenamefont {Itin}\ and\ \citenamefont
  {Reches}(2021)}]{itin21_decom_third_order_const_tensor}%
  \BibitemOpen
  \bibfield  {author} {\bibinfo {author} {\bibfnamefont {Y.}~\bibnamefont
  {Itin}}\ and\ \bibinfo {author} {\bibfnamefont {S.}~\bibnamefont {Reches}},\
  }\bibfield  {title} {\bibinfo {title} {Decomposition of third-order
  constitutive tensors},\ }\href {https://doi.org/10.1177/10812865211016530}
  {\bibfield  {journal} {\bibinfo  {journal} {Math. Mech. Solids}\ }\textbf
  {\bibinfo {volume} {27}},\ \bibinfo {pages} {222} (\bibinfo {year}
  {2021})}\BibitemShut {NoStop}%
\bibitem [{Note13()}]{Note13}%
  \BibitemOpen
  \bibinfo {note} {Generically the independence of the affine connection under
  a change of metric would require also a transformation of the torsion
  tensor.}\BibitemShut {Stop}%
\bibitem [{Note14()}]{Note14}%
  \BibitemOpen
  \bibinfo {note} {In fact, making a functional derivative of the action with
  respect to the metric is going to return a trivial output.}\BibitemShut
  {Stop}%
\bibitem [{\citenamefont {Ehlers}\ \emph {et~al.}(2012)\citenamefont {Ehlers},
  \citenamefont {Pirani},\ and\ \citenamefont {Schild}}]{ehlers12_repub_of}%
  \BibitemOpen
  \bibfield  {author} {\bibinfo {author} {\bibfnamefont {J.}~\bibnamefont
  {Ehlers}}, \bibinfo {author} {\bibfnamefont {F.~A.~E.}\ \bibnamefont
  {Pirani}},\ and\ \bibinfo {author} {\bibfnamefont {A.}~\bibnamefont
  {Schild}},\ }\bibfield  {title} {\bibinfo {title} {Republication of: The
  geometry of free fall and light propagation},\ }\href
  {https://doi.org/10.1007/s10714-012-1353-4} {\bibfield  {journal} {\bibinfo
  {journal} {Gen. Rel. Grav.}\ }\textbf {\bibinfo {volume} {44}},\ \bibinfo
  {pages} {1587} (\bibinfo {year} {2012})}\BibitemShut {NoStop}%
\bibitem [{\citenamefont {Matveev}\ and\ \citenamefont
  {Trautman}(2013)}]{matveev13_criter_compat_confor_projec_struc}%
  \BibitemOpen
  \bibfield  {author} {\bibinfo {author} {\bibfnamefont {V.~S.}\ \bibnamefont
  {Matveev}}\ and\ \bibinfo {author} {\bibfnamefont {A.}~\bibnamefont
  {Trautman}},\ }\bibfield  {title} {\bibinfo {title} {A criterion for
  compatibility of conformal and projective structures},\ }\href
  {https://doi.org/10.1007/s00220-013-1850-7} {\bibfield  {journal} {\bibinfo
  {journal} {Commun. Math. Phys.}\ }\textbf {\bibinfo {volume} {329}},\
  \bibinfo {pages} {821} (\bibinfo {year} {2013})}\BibitemShut {NoStop}%
\bibitem [{\citenamefont {Castillo-Felisola}\ \emph
  {et~al.}(2023{\natexlab{a}})\citenamefont {Castillo-Felisola}, \citenamefont
  {Grez}, \citenamefont {Orellana}, \citenamefont {Perdiguero}, \citenamefont
  {Skirzewski},\ and\ \citenamefont {Zerwekh}}]{castillo-felisola23_corrig}%
  \BibitemOpen
  \bibfield  {author} {\bibinfo {author} {\bibfnamefont {O.}~\bibnamefont
  {Castillo-Felisola}}, \bibinfo {author} {\bibfnamefont {B.}~\bibnamefont
  {Grez}}, \bibinfo {author} {\bibfnamefont {O.}~\bibnamefont {Orellana}},
  \bibinfo {author} {\bibfnamefont {J.}~\bibnamefont {Perdiguero}}, \bibinfo
  {author} {\bibfnamefont {A.}~\bibnamefont {Skirzewski}},\ and\ \bibinfo
  {author} {\bibfnamefont {A.~R.}\ \bibnamefont {Zerwekh}},\ }\bibfield
  {title} {\bibinfo {title} {{Corrigendum: Emergent Metric and Geodesic
  Analysis in Cosmological Solutions of (torsion-Free) Polynomial Affine
  Gravity (2020 Class. Quantum Grav. 37 075013)}},\ }\href
  {https://doi.org/10.1088/1361-6382/ad0356} {\bibfield  {journal} {\bibinfo
  {journal} {Class. Quant. Grav.}\ }\textbf {\bibinfo {volume} {40}},\ \bibinfo
  {pages} {249501} (\bibinfo {year} {2023}{\natexlab{a}})}\BibitemShut
  {NoStop}%
\bibitem [{Note15()}]{Note15}%
  \BibitemOpen
  \bibinfo {note} {Note that one could restrict to the traceless part of the
  torsion, and obtain another metric as the trace of the squared \(\protect
  \mathcal {B}\)-tensor.}\BibitemShut {Stop}%
\bibitem [{Note16()}]{Note16}%
  \BibitemOpen
  \bibinfo {note} {In the literature on Horndeski gravity (see for example Ref.
  \cite {horndeski24_years_hornd_gravit}), this tensor is usually denoted by
  \(X_{\mu \nu }\).}\BibitemShut {Stop}%
\bibitem [{Note17()}]{Note17}%
  \BibitemOpen
  \bibinfo {note} {Notice that in general one could \protect \emph {scale} each
  term of the action by different functions of the scalar field.}\BibitemShut
  {Stop}%
\bibitem [{Note18()}]{Note18}%
  \BibitemOpen
  \bibinfo {note} {In metric Friedmann--Robertson--Walker scenario, the Hubble
  function is expressed in terms of the scale factor as \(H(t) = \protect \dot
  {a}/a\).}\BibitemShut {Stop}%
\bibitem [{\citenamefont {Adame}\ \emph {et~al.}(2024)\citenamefont {Adame}
  \emph {et~al.}}]{collaboration24_desi_vi}%
  \BibitemOpen
  \bibfield  {author} {\bibinfo {author} {\bibfnamefont {A.~G.}\ \bibnamefont
  {Adame}} \emph {et~al.} (\bibinfo {collaboration} {DESI}),\ }\bibfield
  {title} {\bibinfo {title} {{DESI} 2024 {VI}: Cosmological constraints from
  the measurements of baryon acoustic oscillations},\ }\href
  {http://arxiv.org/abs/2404.03002v3} {\bibfield  {journal} {\bibinfo
  {journal} {Accepted for publication in JCAP}\ } (\bibinfo {year} {2024})},\
  \Eprint {https://arxiv.org/abs/2404.03002} {arXiv:2404.03002 [astro-ph.CO]}
  \BibitemShut {NoStop}%
\bibitem [{\citenamefont {Roy}(2024)}]{roy24_dynam_dark_energ_light}%
  \BibitemOpen
  \bibfield  {author} {\bibinfo {author} {\bibfnamefont {N.}~\bibnamefont
  {Roy}},\ }\href {http://arxiv.org/abs/2406.00634v2} {\bibinfo {title}
  {{Dynamical Dark Energy in the Light of Desi 2024 Data}}} (\bibinfo {year}
  {2024}),\ \Eprint {https://arxiv.org/abs/2406.00634} {arXiv:2406.00634
  [astro-ph.CO]} \BibitemShut {NoStop}%
\bibitem [{\citenamefont {Castillo-Felisola}\ \emph
  {et~al.}(2023{\natexlab{b}})\citenamefont {Castillo-Felisola}, \citenamefont
  {Grez}, \citenamefont {Perdiguero},\ and\ \citenamefont
  {Skirzewski}}]{castillo-felisola23_inflat_scenar_effec_polyn}%
  \BibitemOpen
  \bibfield  {author} {\bibinfo {author} {\bibfnamefont {O.}~\bibnamefont
  {Castillo-Felisola}}, \bibinfo {author} {\bibfnamefont {B.}~\bibnamefont
  {Grez}}, \bibinfo {author} {\bibfnamefont {J.}~\bibnamefont {Perdiguero}},\
  and\ \bibinfo {author} {\bibfnamefont {A.}~\bibnamefont {Skirzewski}},\
  }\bibfield  {title} {\bibinfo {title} {Inflationary scenarios in an effective
  polynomial affine model of gravity},\ }\Eprint
  {https://arxiv.org/abs/2312.07312} {arXiv:2312.07312 [gr-qc]}  (\bibinfo
  {year} {2023}{\natexlab{b}})\BibitemShut {NoStop}%
\bibitem [{\citenamefont {Bekenstein}\ and\ \citenamefont
  {Majhi}(2015)}]{bekenstein15_is_princ_least_action}%
  \BibitemOpen
  \bibfield  {author} {\bibinfo {author} {\bibfnamefont {J.~D.}\ \bibnamefont
  {Bekenstein}}\ and\ \bibinfo {author} {\bibfnamefont {B.~R.}\ \bibnamefont
  {Majhi}},\ }\bibfield  {title} {\bibinfo {title} {{Is the Principle of Least
  Action a must?}},\ }\href {https://doi.org/10.1016/j.nuclphysb.2015.01.015}
  {\bibfield  {journal} {\bibinfo  {journal} {Nucl. Phys. B}\ }\textbf
  {\bibinfo {volume} {892}},\ \bibinfo {pages} {337} (\bibinfo {year}
  {2015})},\ \Eprint {https://arxiv.org/abs/1411.2424} {arXiv:1411.2424
  [hep-th]} \BibitemShut {NoStop}%
\bibitem [{Note19()}]{Note19}%
  \BibitemOpen
  \bibinfo {note} {This strategy is similar to the utilised in thermodynamics,
  where one supplements the thermodynamic equations with the equation of
  states.}\BibitemShut {Stop}%
\bibitem [{\citenamefont {Percacci}\ and\ \citenamefont
  {Randjbar-Daemi}(1983)}]{percacci83_kaluz_klein_theor_bundl_with_homog_fiber}%
  \BibitemOpen
  \bibfield  {author} {\bibinfo {author} {\bibfnamefont {R.}~\bibnamefont
  {Percacci}}\ and\ \bibinfo {author} {\bibfnamefont {S.}~\bibnamefont
  {Randjbar-Daemi}},\ }\bibfield  {title} {\bibinfo {title} {Kaluza-klein
  theories on bundles with homogeneous fibers. i},\ }\href
  {https://doi.org/10.1063/1.525753} {\bibfield  {journal} {\bibinfo  {journal}
  {J. Math. Phys.}\ }\textbf {\bibinfo {volume} {24}},\ \bibinfo {pages} {807}
  (\bibinfo {year} {1983})}\BibitemShut {NoStop}%
\bibitem [{\citenamefont {Choquet-Bruhat}\ \emph {et~al.}(1989)\citenamefont
  {Choquet-Bruhat}, \citenamefont {DeWitt-Morette},\ and\ \citenamefont
  {Dillard-Bleick}}]{choquet-bruhat89_analy}%
  \BibitemOpen
  \bibfield  {author} {\bibinfo {author} {\bibfnamefont {Y.}~\bibnamefont
  {Choquet-Bruhat}}, \bibinfo {author} {\bibfnamefont {C.}~\bibnamefont
  {DeWitt-Morette}},\ and\ \bibinfo {author} {\bibfnamefont {M.}~\bibnamefont
  {Dillard-Bleick}},\ }\href@noop {} {\emph {\bibinfo {title} {Analysis,
  manifolds and physics}}},\ Vol.\ \bibinfo {volume} {1 \& 2}\ (\bibinfo
  {publisher} {North-Holland},\ \bibinfo {year} {1989})\BibitemShut {NoStop}%
\bibitem [{\citenamefont {Kol{\'a}{\v{r}}}\ \emph {et~al.}(1993)\citenamefont
  {Kol{\'a}{\v{r}}}, \citenamefont {Michor},\ and\ \citenamefont
  {Slov{\'a}k}}]{ivan93_natur_operat_differ_geomet}%
  \BibitemOpen
  \bibfield  {author} {\bibinfo {author} {\bibfnamefont {I.}~\bibnamefont
  {Kol{\'a}{\v{r}}}}, \bibinfo {author} {\bibfnamefont {P.~W.}\ \bibnamefont
  {Michor}},\ and\ \bibinfo {author} {\bibfnamefont {J.}~\bibnamefont
  {Slov{\'a}k}},\ }\href@noop {} {\emph {\bibinfo {title} {Natural Operations
  in Differential Geometry}}},\ \bibinfo {edition} {corrected second printing}\
  ed.\ (\bibinfo  {publisher} {Springer},\ \bibinfo {year} {1993})\BibitemShut
  {NoStop}%
\bibitem [{\citenamefont {Nakahara}(2005)}]{nakahara05_geomet_topol_physic}%
  \BibitemOpen
  \bibfield  {author} {\bibinfo {author} {\bibfnamefont {M.}~\bibnamefont
  {Nakahara}},\ }\href@noop {} {\emph {\bibinfo {title} {Geometry, Topology and
  Physics}}}\ (\bibinfo  {publisher} {Institute Of Physics},\ \bibinfo {year}
  {2005})\BibitemShut {NoStop}%
\bibitem [{\citenamefont {Dudek}\ and\ \citenamefont
  {Garecki}(2018)}]{dudek18_ehres_theor_connec_princ_bundl_compen_physic}%
  \BibitemOpen
  \bibfield  {author} {\bibinfo {author} {\bibfnamefont {M.}~\bibnamefont
  {Dudek}}\ and\ \bibinfo {author} {\bibfnamefont {J.}~\bibnamefont
  {Garecki}},\ }\bibfield  {title} {\bibinfo {title} {Ehreshmann theory of
  connection in a principal bundle - compendium for physicists},\ }\Eprint
  {https://arxiv.org/abs/1810.03447} {arXiv:1810.03447 [gr-qc]}  (\bibinfo
  {year} {2018}),\ \bibinfo {note} {lecture delivered at the Conference
  "Hypercomplex Seminar 2018" in July 2018, Bedlewo, Poland}\BibitemShut
  {NoStop}%
\bibitem [{Note20()}]{Note20}%
  \BibitemOpen
  \bibinfo {note} {The notation \(T\pi \) refers to the induced map in any
  \protect \emph {tangent} bundle (including tensor bundles) \cite
  {ivancevic07_applied_differ_geomet}. If we would like to refer to an specific
  induced map, e.g over the tangent or cotangent bundles, we would use the more
  standard notation \(\pi _{*}\) and \(\pi ^{*}\) respectively.}\BibitemShut
  {Stop}%
\bibitem [{Note21()}]{Note21}%
  \BibitemOpen
  \bibinfo {note} {It worth to clarify that the quantity \(\protect \tilde
  {g}(\protect \tilde {X},\protect \tilde {Y}) = \protect \hat {g}(\protect
  \tilde {X},\protect \tilde {Y})\) would be, according with Eq. \protect
  \eqref {eq:identification-operation-bundles}, the effective metric in the
  manifold \(\protect \mathcal {M}\).}\BibitemShut {Stop}%
\bibitem [{Note22()}]{Note22}%
  \BibitemOpen
  \bibinfo {note} {When we considered the Schwarzschild-like metric with two
  unknown functions, we were unable to solve the field equations exactly.
  However, a power series expansion of the unknown functions point to the
  necessity of three conditions to determine the solution, we hypothesise that
  those conditions are related to three different scales in gravity, which
  might be thought as celestial scale (Newtonian gravity plus corrections),
  galactic scale (which might explain the rotation curves of galaxies, i.e.
  dark matter), and cosmological scale (related to the cosmological constant,
  i.e. dark energy).}\BibitemShut {Stop}%
\bibitem [{Note23()}]{Note23}%
  \BibitemOpen
  \bibinfo {note} {OCF wants to dedicate this article to the memory of Ivan,
  who has passed away recently.}\BibitemShut {Stop}%
\bibitem [{\citenamefont {Stein}\ \emph {et~al.}(2024)\citenamefont {Stein}
  \emph {et~al.}}]{stein18_sage_mathem_softw_version}%
  \BibitemOpen
  \bibfield  {author} {\bibinfo {author} {\bibfnamefont {W.~A.}\ \bibnamefont
  {Stein}} \emph {et~al.},\ }\href {http://www.sagemath.org} {\emph {\bibinfo
  {title} {{S}age {M}athematics {S}oftware ({V}ersion 10.5)}}},\ \bibinfo
  {organization} {The Sage Development Team} (\bibinfo {year}
  {2024})\BibitemShut {NoStop}%
\bibitem [{\citenamefont {Gourgoulhon}\ \emph {et~al.}(2024)\citenamefont
  {Gourgoulhon}, \citenamefont {Bejger} \emph
  {et~al.}}]{gourgoulhon18_sagem_version}%
  \BibitemOpen
  \bibfield  {author} {\bibinfo {author} {\bibfnamefont {E.}~\bibnamefont
  {Gourgoulhon}}, \bibinfo {author} {\bibfnamefont {M.}~\bibnamefont {Bejger}},
  \emph {et~al.},\ }\href {http://sagemanifolds.obspm.fr/index.html} {\emph
  {\bibinfo {title} {{S}age{M}anifolds ({V}ersion 10.5)}}},\ \bibinfo
  {organization} {SageManifolds Development Team} (\bibinfo {year}
  {2024})\BibitemShut {NoStop}%
\bibitem [{\citenamefont
  {Peeters}(2007{\natexlab{a}})}]{peeters07_symbol_field_theor_with_cadab}%
  \BibitemOpen
  \bibfield  {author} {\bibinfo {author} {\bibfnamefont {K.}~\bibnamefont
  {Peeters}},\ }\bibfield  {title} {\bibinfo {title} {Symbolic field theory
  with cadabra},\ }\href
  {http://www.fachgruppe-computeralgebra.de/CA-Rundbrief/car41.pdf} {\bibfield
  {journal} {\bibinfo  {journal} {Computeralgebra Rundbrief}\ }\textbf
  {\bibinfo {volume} {41}},\ \bibinfo {pages} {16} (\bibinfo {year}
  {2007}{\natexlab{a}})}\BibitemShut {NoStop}%
\bibitem [{\citenamefont
  {Peeters}(2007{\natexlab{b}})}]{peeters07_introd_cadab}%
  \BibitemOpen
  \bibfield  {author} {\bibinfo {author} {\bibfnamefont {K.}~\bibnamefont
  {Peeters}},\ }\bibfield  {title} {\bibinfo {title} {{Introducing Cadabra: A
  Symbolic Computer Algebra System for Field Theory problems}},\ }\Eprint
  {https://arxiv.org/abs/hep-th/0701238} {arXiv:hep-th/0701238 [hep-th]}
  (\bibinfo {year} {2007}{\natexlab{b}})\BibitemShut {NoStop}%
\bibitem [{\citenamefont {Peeters}(2007{\natexlab{c}})}]{peeters07_cadab}%
  \BibitemOpen
  \bibfield  {author} {\bibinfo {author} {\bibfnamefont {K.}~\bibnamefont
  {Peeters}},\ }\bibfield  {title} {\bibinfo {title} {Cadabra: a field-theory
  motivated symbolic computer algebra system},\ }\href
  {https://doi.org/10.1016/j.cpc.2007.01.003} {\bibfield  {journal} {\bibinfo
  {journal} {Comput. Phys. Commun.}\ }\textbf {\bibinfo {volume} {176}},\
  \bibinfo {pages} {550} (\bibinfo {year} {2007}{\natexlab{c}})}\BibitemShut
  {NoStop}%
\bibitem [{\citenamefont
  {Brewin}(2019)}]{brewin19_using_cadab_tensor_comput_gener_relat}%
  \BibitemOpen
  \bibfield  {author} {\bibinfo {author} {\bibfnamefont {L.}~\bibnamefont
  {Brewin}},\ }\bibfield  {title} {\bibinfo {title} {Using cadabra for tensor
  computations in general relativity},\ }\Eprint
  {https://arxiv.org/abs/1912.08839} {arXiv:1912.08839 [gr-qc]}  (\bibinfo
  {year} {2019})\BibitemShut {NoStop}%
\bibitem [{\citenamefont {Kulyabov}\ \emph {et~al.}(2019)\citenamefont
  {Kulyabov}, \citenamefont {Korol'kova},\ and\ \citenamefont
  {Sevast'yanov}}]{kulyabov19_new_featur_secon_version_cadab}%
  \BibitemOpen
  \bibfield  {author} {\bibinfo {author} {\bibfnamefont {D.~S.}\ \bibnamefont
  {Kulyabov}}, \bibinfo {author} {\bibfnamefont {A.~V.}\ \bibnamefont
  {Korol'kova}},\ and\ \bibinfo {author} {\bibfnamefont {L.~A.}\ \bibnamefont
  {Sevast'yanov}},\ }\bibfield  {title} {\bibinfo {title} {New features in the
  second version of the cadabra computer algebra system},\ }\href
  {https://doi.org/10.1134/s0361768819020063} {\bibfield  {journal} {\bibinfo
  {journal} {Programming and Computer Software}\ }\textbf {\bibinfo {volume}
  {45}},\ \bibinfo {pages} {58} (\bibinfo {year} {2019})}\BibitemShut {NoStop}%
\bibitem [{\citenamefont
  {Sakharov}(2000)}]{sakharov00_vacuum_quant_fluct_curved}%
  \BibitemOpen
  \bibfield  {author} {\bibinfo {author} {\bibfnamefont {A.~D.}\ \bibnamefont
  {Sakharov}},\ }\bibfield  {title} {\bibinfo {title} {Vacuum quantum
  fluctuations in curved space and the theory of gravitation},\ }\href
  {https://doi.org/10.1023/a:1001947813563} {\bibfield  {journal} {\bibinfo
  {journal} {Gen. Rel. Grav.}\ }\textbf {\bibinfo {volume} {32}},\ \bibinfo
  {pages} {365} (\bibinfo {year} {2000})},\ \bibinfo {note} {translated from
  Doklady Akademii Nauk SSSR, vol. 177, No. 1, pp. 70–71, November 1967.
  Original article submitted August 28, 1967.}\BibitemShut {Stop}%
\bibitem [{\citenamefont {Krasnov}(2020)}]{krasnov20_formul_gener_relat}%
  \BibitemOpen
  \bibfield  {author} {\bibinfo {author} {\bibfnamefont {K.}~\bibnamefont
  {Krasnov}},\ }\href@noop {} {\emph {\bibinfo {title} {{Formulations of
  General Relativity : Gravity, Spinors and Differential Forms}}}},\ \bibinfo
  {edition} {1st}\ ed.,\ Cambridge Monographs on Mathematical Physics\
  (\bibinfo  {publisher} {Cambridge University Press},\ \bibinfo {year}
  {2020})\BibitemShut {NoStop}%
\bibitem [{\citenamefont {Ng}\ and\ \citenamefont {van
  Dam}(1991)}]{ng91_unimod_theor_gravit_cosmol_const}%
  \BibitemOpen
  \bibfield  {author} {\bibinfo {author} {\bibfnamefont {Y.~J.}\ \bibnamefont
  {Ng}}\ and\ \bibinfo {author} {\bibfnamefont {H.}~\bibnamefont {van Dam}},\
  }\bibfield  {title} {\bibinfo {title} {Unimodular theory of gravity and the
  cosmological constant},\ }\href {https://doi.org/10.1063/1.529283} {\bibfield
   {journal} {\bibinfo  {journal} {J. Math. Phys.}\ }\textbf {\bibinfo {volume}
  {32}},\ \bibinfo {pages} {1337} (\bibinfo {year} {1991})}\BibitemShut
  {NoStop}%
\bibitem [{\citenamefont
  {Jirou{\v{s}}ek}(2023)}]{jirousek23_unimod_approac_to_cosmol}%
  \BibitemOpen
  \bibfield  {author} {\bibinfo {author} {\bibfnamefont {P.}~\bibnamefont
  {Jirou{\v{s}}ek}},\ }\bibfield  {title} {\bibinfo {title} {Unimodular
  approaches to the cosmological constant problem},\ }\href
  {https://doi.org/10.3390/universe9030131} {\bibfield  {journal} {\bibinfo
  {journal} {Universe}\ }\textbf {\bibinfo {volume} {9}},\ \bibinfo {pages}
  {131} (\bibinfo {year} {2023})},\ \Eprint {https://arxiv.org/abs/2301.01662}
  {arXiv:2301.01662 [gr-qc]} \BibitemShut {NoStop}%
\bibitem [{\citenamefont {Gourgoulhon}\ \emph {et~al.}(2015)\citenamefont
  {Gourgoulhon}, \citenamefont {Bejger},\ and\ \citenamefont
  {Mancini}}]{gourgoulhon15_tensor_calcul_with_open_sourc_softw}%
  \BibitemOpen
  \bibfield  {author} {\bibinfo {author} {\bibfnamefont {E.}~\bibnamefont
  {Gourgoulhon}}, \bibinfo {author} {\bibfnamefont {M.}~\bibnamefont
  {Bejger}},\ and\ \bibinfo {author} {\bibfnamefont {M.}~\bibnamefont
  {Mancini}},\ }\bibfield  {title} {\bibinfo {title} {Tensor calculus with
  open-source software: the sagemanifolds project},\ }\href
  {https://doi.org/10.1088/1742-6596/600/1/012002} {\bibfield  {journal}
  {\bibinfo  {journal} {Journal of Physics: Conference Series}\ }\textbf
  {\bibinfo {volume} {600}},\ \bibinfo {pages} {012002} (\bibinfo {year}
  {2015})},\ \Eprint {https://arxiv.org/abs/1412.4765} {arXiv:1412.4765
  [gr-qc]} \BibitemShut {NoStop}%
\bibitem [{\citenamefont {Gourgoulhon}\ and\ \citenamefont
  {Mancini}(2018)}]{gourgoulhon18_symbol_tensor_calcul_manif}%
  \BibitemOpen
  \bibfield  {author} {\bibinfo {author} {\bibfnamefont {E.}~\bibnamefont
  {Gourgoulhon}}\ and\ \bibinfo {author} {\bibfnamefont {M.}~\bibnamefont
  {Mancini}},\ }\bibfield  {title} {\bibinfo {title} {Symbolic tensor calculus
  on manifolds: a sagemath implementation},\ }\href
  {https://doi.org/10.5802/ccirm.26} {\bibfield  {journal} {\bibinfo  {journal}
  {Les cours du CIRM}\ }\textbf {\bibinfo {volume} {6}},\ \bibinfo {pages} {1}
  (\bibinfo {year} {2018})},\ \Eprint {https://arxiv.org/abs/1804.07346}
  {arXiv:1804.07346 [gr-qc]} \BibitemShut {NoStop}%
\bibitem [{\citenamefont {Aoki}\ \emph {et~al.}(2024)\citenamefont {Aoki},
  \citenamefont {Bahamonde}, \citenamefont {Valcarcel},\ and\ \citenamefont
  {Gorji}}]{aoki23_cosmol_pertur_theor_metric}%
  \BibitemOpen
  \bibfield  {author} {\bibinfo {author} {\bibfnamefont {K.}~\bibnamefont
  {Aoki}}, \bibinfo {author} {\bibfnamefont {S.}~\bibnamefont {Bahamonde}},
  \bibinfo {author} {\bibfnamefont {J.~G.}\ \bibnamefont {Valcarcel}},\ and\
  \bibinfo {author} {\bibfnamefont {M.~A.}\ \bibnamefont {Gorji}},\ }\bibfield
  {title} {\bibinfo {title} {Cosmological perturbation theory in metric-affine
  gravity},\ }\href {https://doi.org/10.1103/PhysRevD.110.024017} {\bibfield
  {journal} {\bibinfo  {journal} {Phys. Rev. D}\ }\textbf {\bibinfo {volume}
  {110}},\ \bibinfo {pages} {024017} (\bibinfo {year} {2024})},\ \Eprint
  {https://arxiv.org/abs/2310.16007} {arXiv:2310.16007 [gr-qc]} \BibitemShut
  {NoStop}%
\bibitem [{\citenamefont {Horndeski}\ and\ \citenamefont
  {Silvestri}(2024)}]{horndeski24_years_hornd_gravit}%
  \BibitemOpen
  \bibfield  {author} {\bibinfo {author} {\bibfnamefont {G.~W.}\ \bibnamefont
  {Horndeski}}\ and\ \bibinfo {author} {\bibfnamefont {A.}~\bibnamefont
  {Silvestri}},\ }\bibfield  {title} {\bibinfo {title} {50 years of horndeski
  gravity: Past, present and future},\ }\bibfield  {journal} {\bibinfo
  {journal} {Int. J. Theor. Phys.}\ }\textbf {\bibinfo {volume} {63}},\ \href
  {https://doi.org/10.1007/s10773-024-05558-2} {10.1007/s10773-024-05558-2}
  (\bibinfo {year} {2024})\BibitemShut {NoStop}%
\bibitem [{\citenamefont {Ivancevic}\ and\ \citenamefont
  {Ivancevic}(2007)}]{ivancevic07_applied_differ_geomet}%
  \BibitemOpen
  \bibfield  {author} {\bibinfo {author} {\bibfnamefont {V.~G.}\ \bibnamefont
  {Ivancevic}}\ and\ \bibinfo {author} {\bibfnamefont {T.~T.}\ \bibnamefont
  {Ivancevic}},\ }\href@noop {} {\emph {\bibinfo {title} {Applied Differential
  Geometry: A Modern Introduction}}}\ (\bibinfo  {publisher} {World
  Scientific},\ \bibinfo {year} {2007})\BibitemShut {NoStop}%
\end{thebibliography}
%

\end{document}